\documentclass{aa}  
\usepackage{algorithm}
\usepackage{float}
\usepackage{graphicx}
\usepackage{float}
\usepackage{indentfirst}
\usepackage{lscape}
\usepackage{longtable}
\usepackage{amsmath}
\usepackage{tablefootnote}
\usepackage{amssymb}
\usepackage{graphicx}
\usepackage{lastpage}
\usepackage{natbib}
\usepackage{longtable}
\usepackage{sidecap}
\usepackage{txfonts}
\begin{document}
   \title{Evolution of Chemistry in the envelope of Hot Corinos (ECHOS)}

   \subtitle{I. Extremely young sulphur chemistry in the isolated Class 0 object B\,335}

   \author{G. Esplugues
          \inst{1},   
          M. Rodr\'iguez-Baras
          \inst{1},  
          D. San Andr\'es
          \inst{2},     
          D. Navarro-Almaida
          \inst{3},  
          A. Fuente  
          \inst{1,2},
          P. Rivière-Marichalar  
          \inst{1}, 
          Á. Sánchez-Monge 
          \inst{4,5},
          M. N. Drozdovskaya 
          \inst{6}, 
          S. Spezzano 
          \inst{7}, and
          P. Caselli 
          \inst{7}
         }

   \institute{Observatorio Astron\'omico Nacional (OAN), Alfonso XII, 3, 28014. Madrid. Spain\\
              \email{g.esplugues@oan.es}
        \and
Centro de Astrobiolog\'ia (CAB), INTA-CSIC, Carretera de Ajalvir Km. 4, Torrej\'on de Ardoz, 28850, Madrid, Spain       
        \and
D\'epartement d’Astrophysique (DAp), Commissariat à l’\'Energie Atomique et aux \'Energies Alternatives (CEA), Orme des Merisiers, Bât. 709, 91191 Gif sur Yvette, Paris-Saclay, France
		\and
Institut de Ciències de l'Espai (ICE, CSIC), Can Magrans s/n, E-08193, Bellaterra, Barcelona, Spain
		\and
Institut d'Estudis Espacials de Catalunya (IEEC), Barcelona, Spain
		\and
Center for Space and Habitability, Universität Bern, Gesellschaftsstrasse 6, CH-3012 Bern, Switzerland    
		\and
Max-Planck-Institut für extraterrestrische Physik, 85748 Garching, Germany
}


 
  \abstract   
   {Within the project Evolution of Chemistry in the envelope of HOt corinoS (ECHOS), we present a study of sulphur chemistry in the envelope of the Class 0 source B\,335 through observations in the spectral range $\lambda$ = 7, 3, and 2 mm.   
   } 
   {Our goal is to characterise the sulphur chemistry in this isolated protostellar source and compare it with other Class 0 objects to determine the environmental and evolutionary effects on the sulphur chemistry in these young sources. 
   }
   {We have modelled observations and computed column densities assuming local thermodynamic equilibrium (LTE) and large velocity gradient (LVG) approximation. We have also used the code Nautilus to study the time evolution of sulphur species, as well as of several sulphur molecular ratios.    
   }
   {We have detected 20 sulphur species in B\,335 with a total gas-phase S abundance similar to that found in the envelopes of other Class 0 objects, but with significant differences in the abundances between sulphur carbon chains and sulphur molecules containing oxygen and nitrogen. Our results highlight the nature of B\,335 as a source especially rich in sulphur carbon chains unlike other Class 0 sources. The low presence or absence of some molecules, such as SO and SO$^+$, suggests a chemistry not particularly influenced by shocks. We, however, detect a large presence of HCS$^+$ that, together with the low rotational temperatures obtained for all the S species ($<$15 K), reveals the moderate or low density of the envelope of B\,335. Model results also show the large influence of the cosmic ray ionisation rate and density variations on the abundances of some S species (e.g. SO, SO$_2$, CCS, and CCCS) with differences of up to $\sim$4 orders of magnitude. We also find that observations are better reproduced by models with a sulphur depletion factor of 10 with respect to the sulphur cosmic elemental abundance. 
}   
   {The comparison between our model and observational results for B\,335 reveals an age of 10$^4$$<$$t$$<$10$^5$ yr, which highlights the particularly early evolutionary stage of this source. B\,335 presents a different chemistry compared to other young protostars that have formed in dense molecular clouds, which could be the result of accretion of surrounding material from the diffuse cloud onto the protostellar envelope of B\,335. In addition, the theoretical analysis and comparison with observations of the SO$_2$/C$_2$S, SO/CS, and HCS$^+$/CS ratios within a sample of prestellar cores and Class 0 objects show that they could be used as good chemical evolutionary indicators of the prestellar to protostellar transition.        
}

   \keywords{survey-stars: formation - ISM: abundances - ISM: clouds - ISM: molecules - Radio lines: ISM}
   \titlerunning{Extremely young sulphur chemistry in the isolated Class 0 object B\,335}
   \authorrunning{G. Esplugues et al.}
   \maketitle
%

\section{Introduction}
\label{section:Introduction}

Hot corinos are regions that arise during the formation of solar-type protostars. The progressive collapse of a prestellar core leads to the heating of the infalling envelope of dust and gas to temperatures of hundreds of Kelvin, while the densities increase up to $\sim$10$^8$-10$^9$ cm$^{-3}$ in the inner $\sim$100 au around the central protostar. 
These are also the scales at which protoplanetary disks are expected to arise owing to the conservation of angular momentum. As the temperatures increase, the water-rich ice mantles sublimate, injecting molecules into the gas phase. Hot corinos are therefore chemically rich regions in Solar-like young protostars \citep[]{Ceccarelli2004, Ceccarelli2007b}. In particular, these regions are relatively rich in interstellar complex organic molecules \citep{Herbst2009}. Studying, therefore, the inventory and abundances of molecules in the earliest stages of star formation, that is in hot corinos, is essential to probe the initial conditions of planet and star formation processes. In spite of the detection of various hot corinos in the last decades \citep[e.g.][]{Cazaux2003, Okoda2023}, just a few of them \citep[such as IRAS\,16293-2422, NGC\,1333, and IRAS\,4A][]{vanDishoeck1995, Ceccarelli2007, Jaber2014, Caselli2012, Jorgensen2016} have been investigated in detail until now. Many questions, therefore, still remain open about their nature and molecular composition: for instance, chemical differences between the earliest stages of star formation (Class 0 versus Class I objects), the role of the chemical characteristics inherited from the parent cloud, as well as the environmental effect on the chemical composition of hot corinos. 

One way to understand the dynamics and evolution of the earliest stages of star formation is through molecular depletion. At temperatures of roughly 10 K and densities above 10$^{4}$ cm$^{-3}$, several molecules, such as CO and CS, condense out onto dust grain surfaces \citep[e.g.][]{Caselli1999, Crapsi2005}, and depletion increases with time \citep[e.g.][]{Bergin1997, Aikawa2003}. Therefore, depletion can be used as a clock, since evolved cores should be more depleted of certain species than younger cores. Regarding depletion, sulphur plays a fundamental role since, although sulphur is one of the most abundant species in the Universe and plays a crucial role in biological systems on Earth \citep[e.g.][]{Leustek2002, Francioso2020}, S-bearing molecules are not as abundant as expected in the interstellar medium (ISM). In particular, unlike the diffuse ISM and photon-dominated regions where the observed gaseous sulphur accounts \citep[e.g.][]{Goicoechea2006, Howk2006} for its total solar abundance \citep[S/H$\sim$1.5$\times$10$^{-5}$;][]{Asplund2009}, sulphur is strongly depleted in more dense molecular gas, since the sum of the detectable S-bearing molecules only accounts for a very small fraction of the elemental S abundance \citep{Tieftrunk1994}. In fact, to reproduce observations in hot corinos and hot cores, one needs to assume a significant sulphur depletion of at least one order of magnitude lower than the solar elemental sulphur abundance \citep[e.g.][]{Wakelam2004, Esplugues2014, Crockett2014, Vastel2018, Bulut2021, Navarro-Almaida2021, Esplugues2022, Hily-Blant2022, Fuente2023}. 
Most of the sulphur seems to be locked on the icy grain mantles \citep[e.g.][]{Millar1990, Ruffle1999, Vidal2017, Laas2019} in the form of organo-sulphur molecules, also present in the comet 67P \citep[]{Calmonte2016} and in meteoritic material \citep[e.g.][]{Naraoka2022, Ruf2021}. However, OCS and SO$_2$ are the only S-bearing molecules unambiguously detected in ice mantles to date \citep[]{Geballe1985, Palumbo1995, McClure2023}, while H$_2$S has not been detected \citep[]{Yang2022, McClure2023}. Theoretical models also suggest that sulphur could be locked into pure allotropes of sulphur, especially S$_8$ \citep[]{Shingledecker2020}. In any case, sulphur reservoirs in dense regions remain as an open question, as well as the time at which sulphur depletion occurs.

Given the low number of discovered hot corinos and, therefore, the limited knowledge about them, the project Evolution of Chemistry in the envelope of HOt corinoS (ECHOS) aims to provide a complete inventory of the chemistry of hot corinos (with a sample including IRAS16293-2422, NGC 1333 IRAS 4A, HH 212, and B\,335). ECHOS is carried out through homogeneous and systematic spectral surveys between 30 and 200 GHz using the Yebes-40m and IRAM-30m telescopes. In this paper, we focus on the analysis of B\,335 and present a study of sulphur-bearing molecules in this region. Observations and a description of the source are presented in Sects. \ref{section:observations} and \ref{section:source}. In Sect. \ref{section:results} we present the data and calculate column densities and abundances using different methods, depending on the number of detected transitions for each species, and also depending on the availability of collisional coefficients. A discussion of the results (including a comparison with other similar objects, as well as the use of a chemical code to study the fractional abundance evolution of sulphur species) is presented in Sect. \ref{discussion}. We finally summarise our conclusions in Sect. \ref{section:summary}.

\section{Observations}
\label{section:observations}

The observations were carried out with the Yebes-40m and IRAM-30m telescopes pointing to the position with coordinates $\alpha$$_{J2000.0}$ = 19$^{\mathrm{h}}$37$^{\mathrm{m}}$0.93$^{\mathrm{s}}$, $\delta$$_{J2000.0}$ = +07$^{\mathrm{o}}$34$\arcmin$09.9$\arcsec$. For both telescopes, data were reduced and processed by using the CLASS and GREG packages provided within the GILDAS software\footnote{http://www.iram.fr/IRAMFR/GILDAS}, developed by the IRAM institute.

\subsection{Yebes-40m telescope}

The observations from the Yebes-40m radiotelescope located in Yebes (Guadalajara, Spain) were obtained using a receiver that consists of two cold high electron mobility transistor amplifiers covering the 31.0-50.3 GHz Q-band with horizontal and vertical polarisations. The backends are 2$\times$8$\times$2.5 GHz fast Fourier transform spectrometers with a spectral resolution of 38.15 kHz, providing the whole coverage of the Q-band in both polarisations. The observations were performed in the position-switching mode with (-400$\arcsec$, 0$\arcsec$) as the reference position. The main beam efficiency varies from 0.6 at 32 GHz to 0.43 at 50 GHz. Pointing corrections were derived from nearby quasars and SiO masers, and errors remained within 2-3$\arcsec$. 

\begin{table}[b!]
\caption{Yebes 40 m and IRAM 30 m telescope efficiency data along the covered frequency range.}               
\centering          
\label{table:tablaeficiencias}
\begin{tabular}{c c c c}     
\hline\hline                    
Telescope  & Frequency  &  $\eta$$_{\mathrm{MB}}$ & HPBW \\
           & (GHz)      &                         & ($\arcsec$)\\
\hline                    
Yebes 40 m &  32.4 & 0.61 & 54.4  \\
           & 34.6 & 0.58 & 51.0  \\
           & 36.9 & 0.57 & 47.8  \\
           & 39.2 & 0.55 & 45.0  \\
           & 41.5 & 0.53 & 42.5  \\
           & 43.8 & 0.52 & 40.3  \\
           & 46.1 & 0.49 & 38.3   \\
           & 48.4 & 0.47 & 36.4   \\
\hline   
IRAM 30 m  & 86  & 0.82 & 29.0  \\
           & 100 & 0.79 & 22.0  \\
           & 145 & 0.74 & 17.0  \\
           & 170 & 0.70 & 14.5  \\         
\hline
\end{tabular}
\end{table}

\subsection{IRAM-30m telescope}

The observations from the IRAM-30m radio telescope located in Pico Veleta (Granada, Spain) cover the 73-118 GHz and 130-175 GHz spectral ranges. These observations were carried out in one session in July 2021 with the Wobbler-Switching mode ($\pm$120$\arcsec$), using the broad-band EMIR (E090 and E150) receivers and the fast Fourier transform spectrometer in its 200 kHz of spectral resolution. 
The intensity scale in antenna temperature ($T$$^{\star}_{\mathrm{A}}$, which is corrected for atmospheric absorption and for antenna ohmic and spillover losses) was calibrated using two absorbers at different temperatures and the atmospheric transmission model \citep[ATM;][]{Cernicharo1985, Pardo2001}. Calibration uncertainties were adopted to be 10$\%$. 

To convert to main beam brightness temperature ($T$$_{\mathrm{MB}}$), we used the expression 

\begin{equation}
T_{\mathrm{MB}}=\left(F_{\mathrm{eff}}/B_{\mathrm{eff}}\right)\times{T^{\star}_{\mathrm{A}}}=\left(T^{\star}_{\mathrm{A}}/\eta_{\mathrm{MB}}\right),
\end{equation}

\noindent where $F$$_{\mathrm{eff}}$ is the telescope forward efficiency and $B$$_{\mathrm{eff}}$ is the main beam efficiency\footnote{https://rt40m.oan.es/rayo/index.php}$^,$\footnote{http://www.iram.es/IRAMES/mainWiki/Iram30mEfficiencies}. 
See Table \ref{table:tablaeficiencias} for a summary of the telescope beam sizes depending on frequency.

\section{The source}
\label{section:source}

B\,335 is an isolated dense globule associated with the embedded far-infrared source IRAS\,19347+0727 \citep[]{Keene1980, Keene1983}, located at a distance of $\sim$100 pc \citep[]{Olofsson2009}. The central source is a Class 0 source with a luminosity of 0.72 L$_{\odot}$ \citep[]{Evans2015}, whose mass is estimated to be 0.08 M$_{\odot}$ \citep[]{Yen2015}. Complex organic molecules (COMs), such as acetaldehyde (CH$_3$CHO), methyl formate (HCOOCH$_3$), and formamide (NH$_2$CHO), have been detected \citep[]{Imai2016, Okoda2022} in the vicinity of the protostar ($r$$\sim$10 au). It reveals that B\,335 harbors a hot corino \citep[]{Cazaux2003, Herbst2009}, which warms up the surrounding regions leading to the sublimation of grain mantles and injecting complex organic molecules into the gas-phase \citep[e.g.][]{Ceccarelli2004, Caselli2012
}. Emission of CCH and c-C$_3$H$_2$ has also been detected around the protostar, which indicates the presence of carbon-chain molecules in this source \citep[]{Sakai2008, Sakai2013}. This carbon-chain molecules could suggest the presence of a so-called warm carbon-chain chemistry \citep[WCCC,][]{Sakai2008}, characterised by various carbon-chain molecules being produced by reactions of CH$_4$ (evaporated from grain mantles in a lukewarm region) and where the formation of HC$_n$N is slower than that of C$_n$H$_m$ molecules \citep[]{Sakai2008, Sakai2009, Oya2017}. 

Particularly interesting are the single-dish observations of CS and H$_2$CO in B\,335 \citep[]{Zhou1993, Choi1995}, since their optically thick lines show asymmetric profiles. This implies the presence of infalling motions in the envelope around B335. A more detailed analysis with C$^{18}$O, SO, HCN, and HCO$^+$ ALMA observations \citep[]{Evans2015, Yen2015, Evans2023} confirms an infalling gas motion of the inner part of the protostellar core, but rotation motion has not been detected at a spatial resolution of 0.3$\arcsec$ ($\sim$30 au). A rotation structure is only found at a scale of about 10 au from CH$_3$OH and HCOOH emission \citep[]{Imai2019}. Single-dish and interferometric $^{12}$CO (1–0) and $^{13}$CO (1– 0) observations \citep[e.g.][]{Hirano1988, Hirano1992, Stutz2008, Bjerkeli2019} show also the presence of an east-west molecular outflow on $\sim$0.2 pc \citep[]{Cabrit1988, Moriarty1989} with an inclination between 3$^{\mathrm{o}}$-10$^{\mathrm{o}}$ on the plane of the sky and an outflow opening angle of 45$^{\mathrm{o}}$ \citep[]{Hirano1988, Hirano1992, Stutz2008, Yen2010}. A dynamical analysis of the outflow molecular structure \citep[]{Yildiz2015} implies that dynamical timescales for the CO emitting gas are of the order of 10$^4$ yr. This result, along with the absence of a rotationally supported disk on scales greater than $\sim$10 au in size \citep[]{Evans2015, Yen2015}, suggest that the B\,335 system is a particularly young source compared to other Class 0 objects, such as L\,1527 which contains an embedded IRAS source and has a dynamical age of 2$\times$10$^4$-10$^5$ yr \citep[]{Tamura1996, Andre2000b, Agundez2019}. The detection of abundant CCS (see Sect. \ref{column_densities}) and other carbon-chain molecules in B\,335 also confirms the early stage of this source since it is generally recognised that CCS and other carbon-chain molecules are abundant in the early stage of chemical evolution and become deficient at the advanced stage \citep[]{Sakai2008}.

So far, B\,335 is one of the few hot corinos identified in a Bok globule, isolated from a large molecular cloud complex. 
This is therefore an ideal source for the study of the earliest stages of star formation with a chemical composition that could be regarded as a standard template for isolated protostellar cores.

\section{Data analysis and results}
\label{section:results}

We have detected 81 emission lines from 21 different sulphur-bearing species (including isotopologs) in B\,335. In particular, we have detected three transitions of CS, three of $^{13}$CS, three of C$^{34}$S, three of C$^{33}$S, nineteen of CCS, three of CC$^{34}$S, seven of CCCS, three of CCC$^{34}$S, five of OCS, three of HCS$^+$, two of HC$^{34}$S$^+$, seven of SO, one of SO$_2$, one of $^{34}$SO, one of H$_2$S, one of H$_2$$^{34}$S, ten of H$_2$CS, one of HSCN, one of HNCS, and six of NS (see Table \ref{table:line_parameters} and Figs. \ref{figure:CS_lines}-\ref{figure:HSCN_lines} in Appendix). These transitions span an energy range of $E_{\mathrm{up}}$=2.0-65.3 K.

\subsection{Line profiles}
\label{Line_profiles}

All the lines are observed in emission. We first fitted the observed lines with Gaussian profiles using the CLASS software to derive the radial velocity (v$_{\mathrm{LSR}}$), the line width, and the intensity for each line. Results are given in Table \ref{table:line_parameters}. We observed that most of the detected molecules show narrow line profiles ($\Delta$v$\lesssim$1.5 km s$^{-1}$) that can be adjusted by a single velocity component. 

It is well known that the gravitational collapse of a molecular cloud when a new star is formed is accompanied by the development of highly supersonic outflows \citep[e.g.][]{Bachiller1997, Holdship2019}. The material ejected through these jets collides with the surrounding cloud, compressing and heating the gas, which leads to a drastic alteration of the chemistry as endothermic reactions become efficient and dust grains are partially destroyed. Many observations have shown that several sulphur-bearing species are enhanced in the shocked regions with respect to other areas of the cloud \citep[e.g.][]{PineauDesForets1993, Bachiller1997, Codella1999, Wakelam2004, Wakelam2005, Esplugues2013}. However, the sulphur lines observed in B\,335 present very narrow line widths (Figs. \ref{figure:CS_lines}-\ref{figure:HSCN_lines}), which is more consistent with the emission arising from the ambient quiescent cloud. Only for H$_2$S, formed through the hydrogenation of sticked atomic sulphur on grains \citep[][]{Hatchell1998, Garrod2007, Esplugues2014, Oba2019} and thought to be the main sulphur reservoir in ices \citep[][]{Vidal2017, Navarro-Almaida2020}, it is necessary to also consider a wider component (with $\Delta$v=2.5 km s$^{-1}$) to fit its only detected line, since the line profile shows the presence of faint wings. This could indicate that part of the H$_2$S emission arises from the outflow, however, the line width of this component is much smaller ($\sim$2.5 km s$^{-1}$) than the ones observed for $^{12}$CO ($\sim$10-20 km s$^{-1}$) in the outflow \citep[]{Yen2010}. Therefore, the wider H$_2$S emission may be likely related to infall motions.

On the other hand, there are some species (CCS, CCCS, and H$_2$CS) whose low transition lines present a double peak emission. The peak with the lowest intensity is located for all the cases at lower velocity (v$_{\mathrm{LSR}}$=7.39-8.1 km s$^{-1}$, blue peak) than for the more intense peak (v$_{\mathrm{LSR}}$=8.80-8.97 km s$^{-1}$, red peak). Line widths are slightly larger ($\Delta$v$\lesssim$0.83-1.63 km s$^{-1}$) in the blue peaks than in the red ones ($\Delta$v$\lesssim$0.55-0.83 km s$^{-1}$). The comparison of these profiles with those from the less abundant isotopologues shows that these two-peak profiles are due to self-absorption due to the large optical depths. 
All the detected lines present 0.005$<$$T_{\mathrm{MB}}$$<$1.5 K.

\subsection{Rotational diagrams}
\label{Rotational_diagrams}

\begin{figure*}
\centering
\includegraphics[scale=0.39, angle=0]{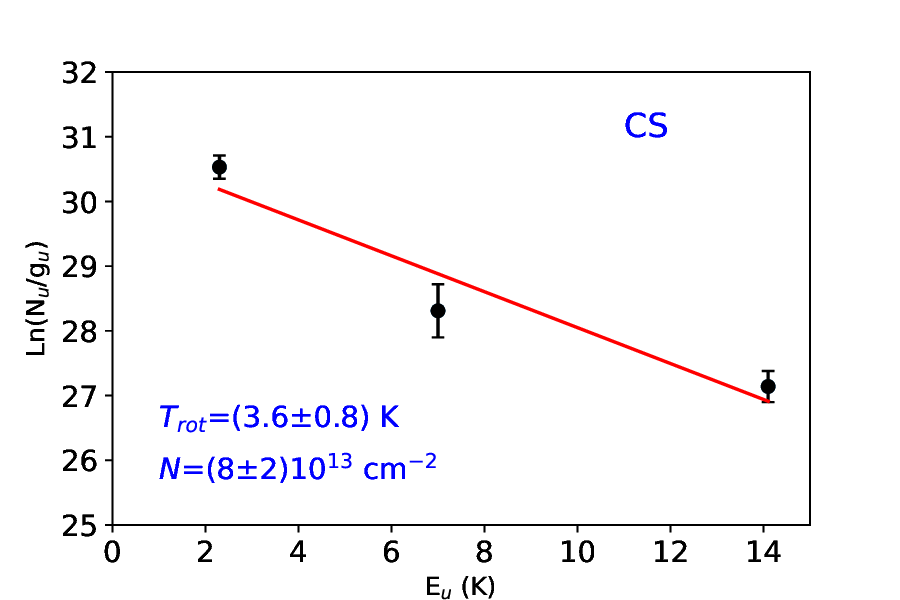}  
\includegraphics[scale=0.39, angle=0]{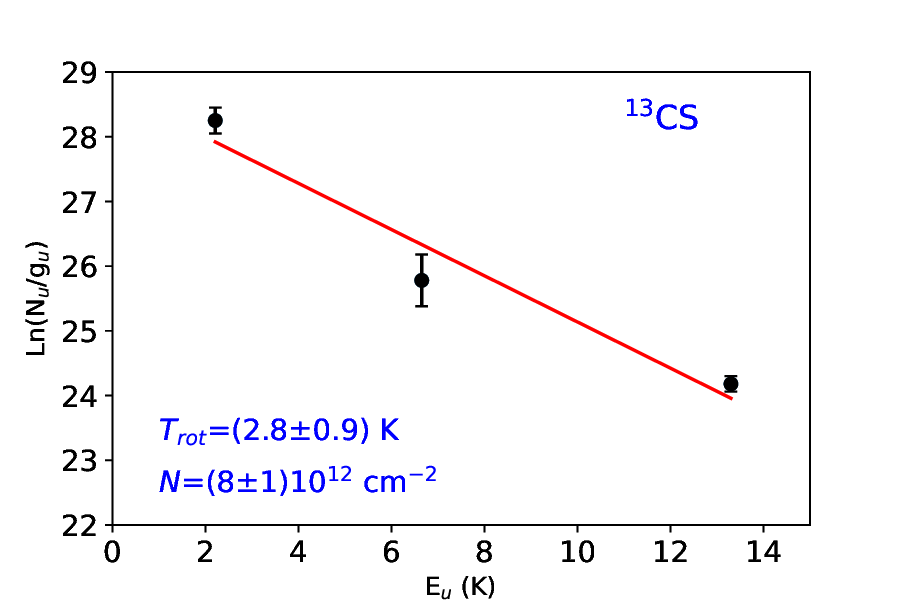} 
\includegraphics[scale=0.39, angle=0]{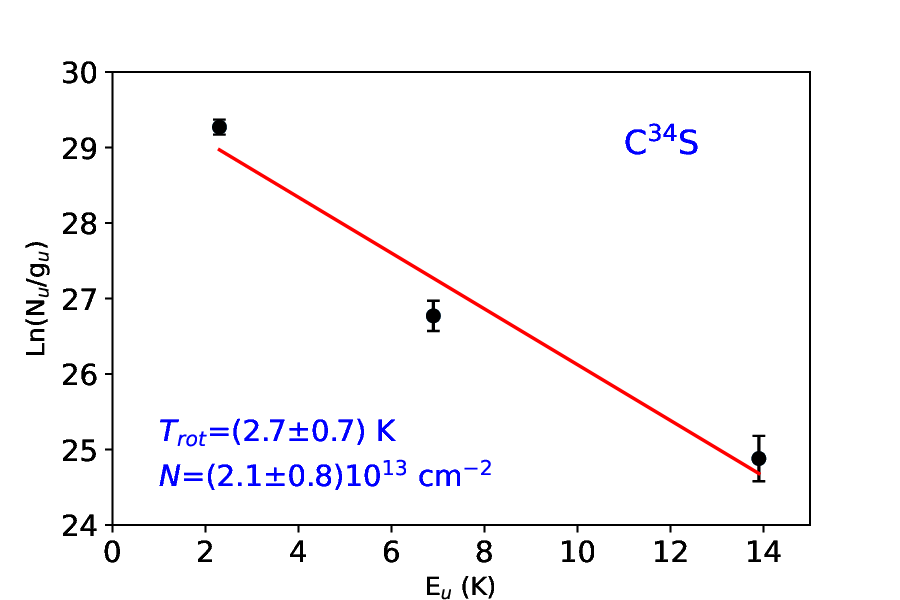} 

\includegraphics[scale=0.39, angle=0]{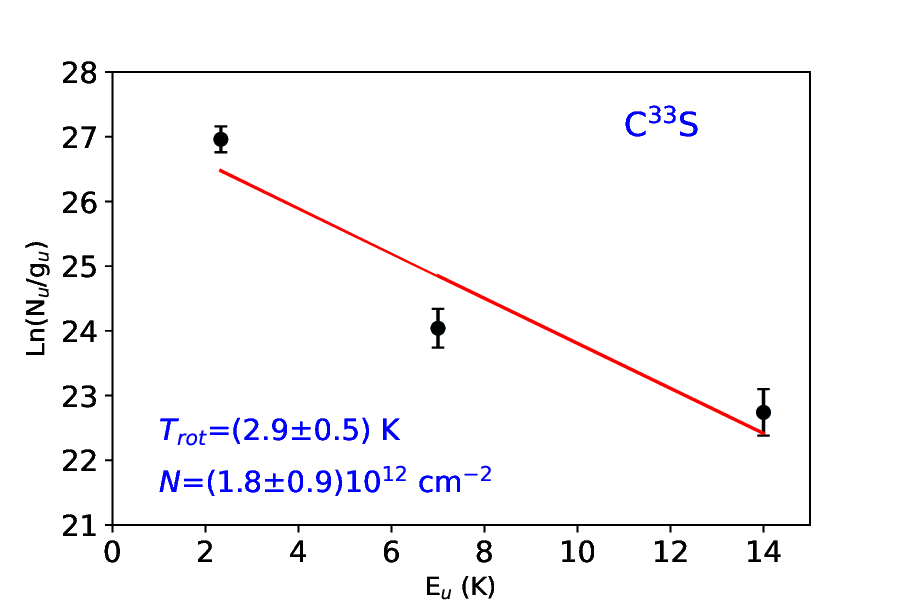}  
\includegraphics[scale=0.39, angle=0]{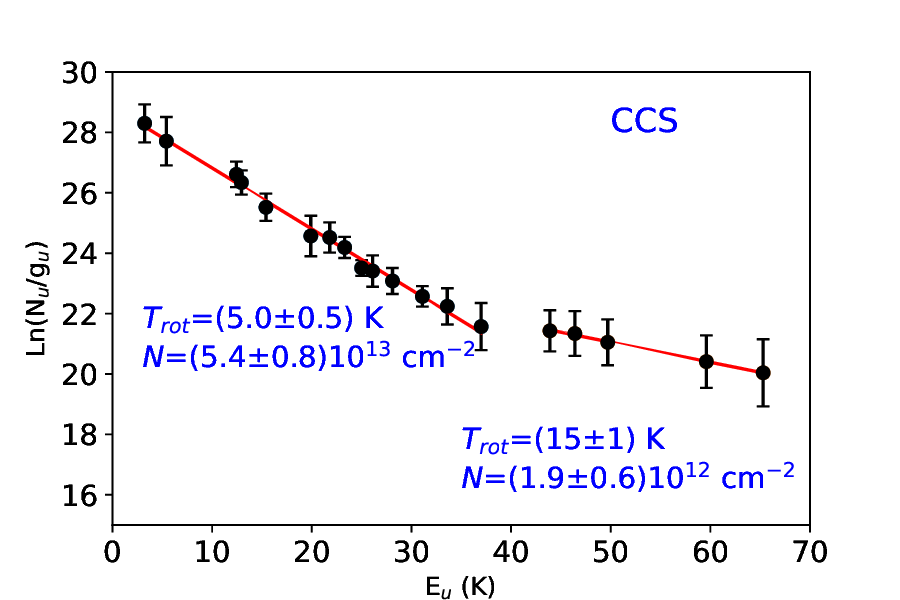}  
\includegraphics[scale=0.39, angle=0]{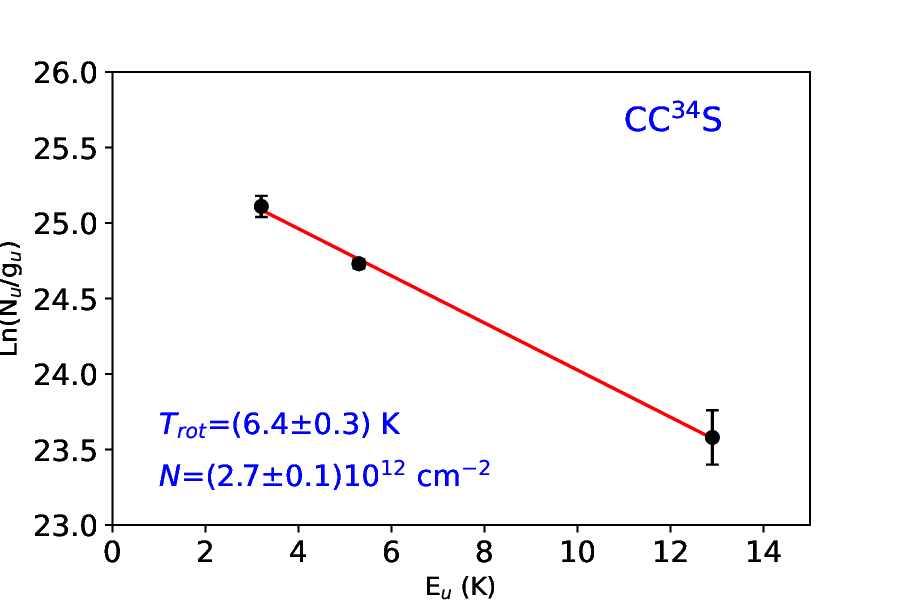} 

\vspace{0.1cm}

\includegraphics[scale=0.39, angle=0]{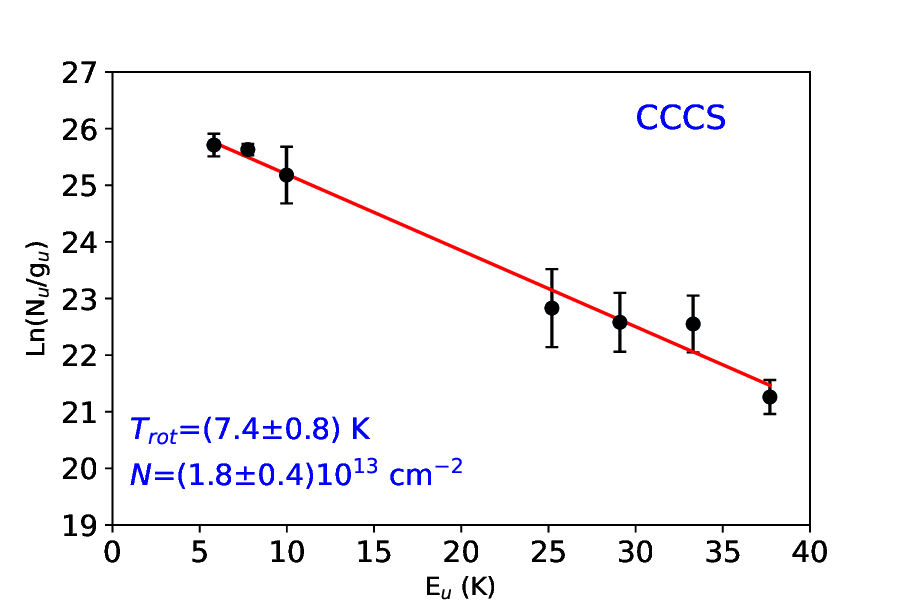} 
\includegraphics[scale=0.39, angle=0]{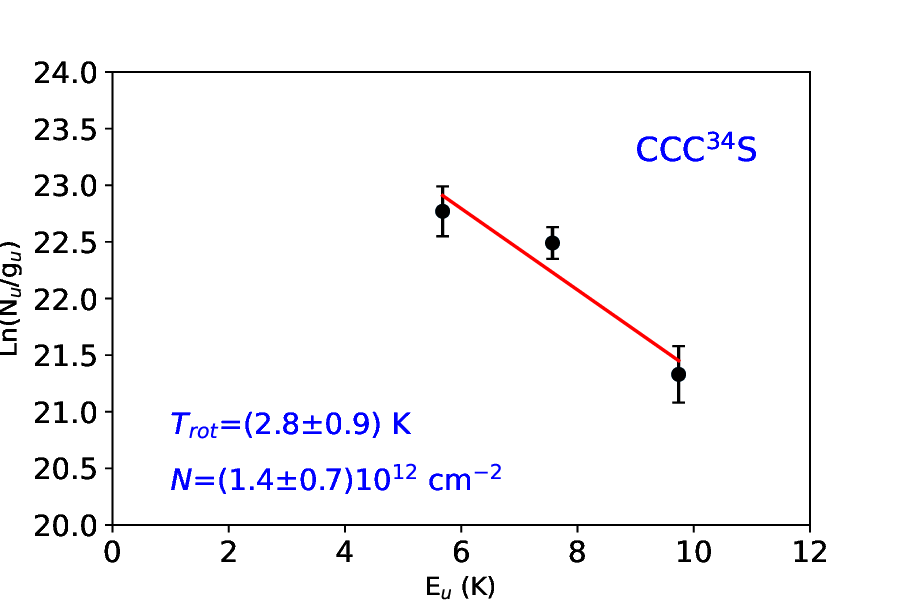} 
\includegraphics[scale=0.39, angle=0]{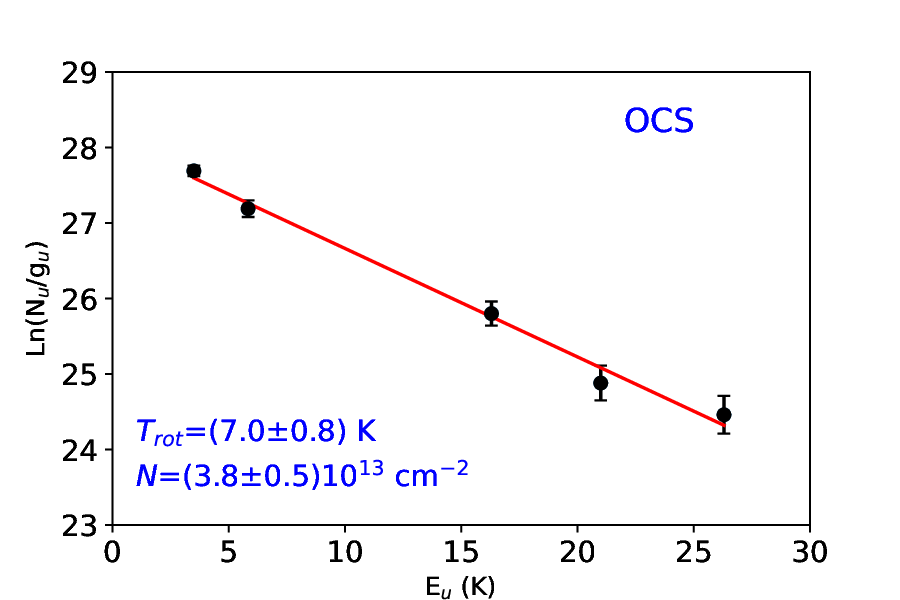} 

\vspace{0.1cm}
 
\includegraphics[scale=0.39, angle=0]{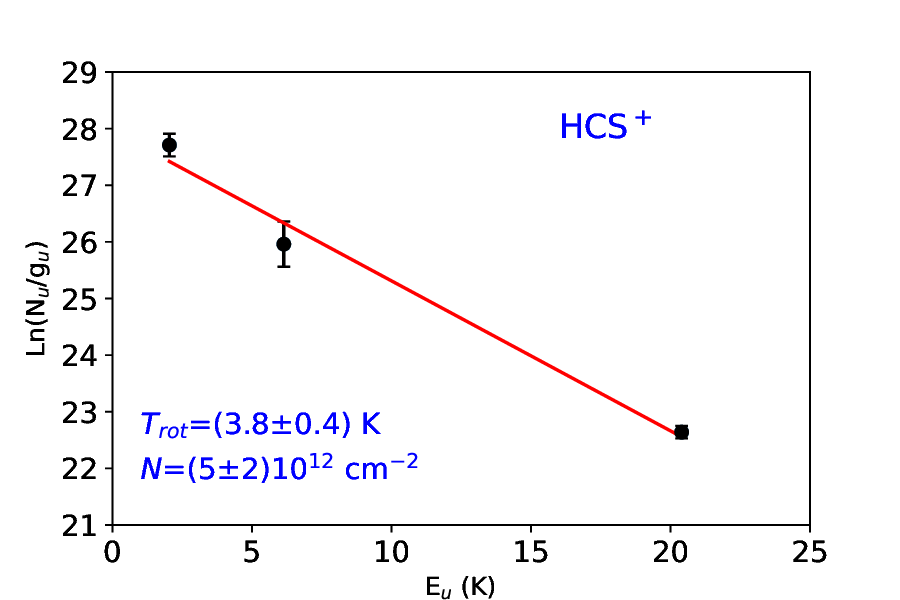} 
\includegraphics[scale=0.39, angle=0]{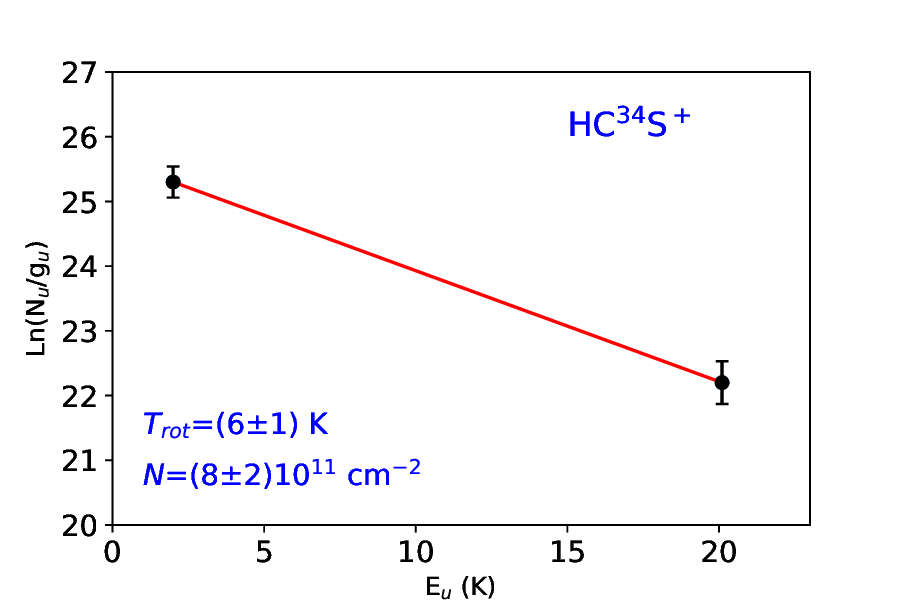} 
\includegraphics[scale=0.39, angle=0]{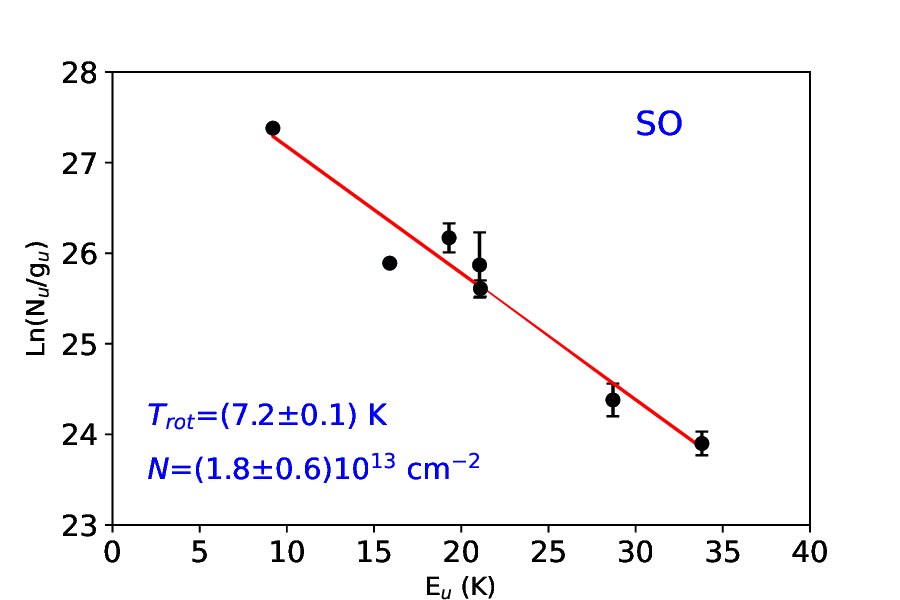}  

\vspace{0.1cm}

\includegraphics[scale=0.39, angle=0]{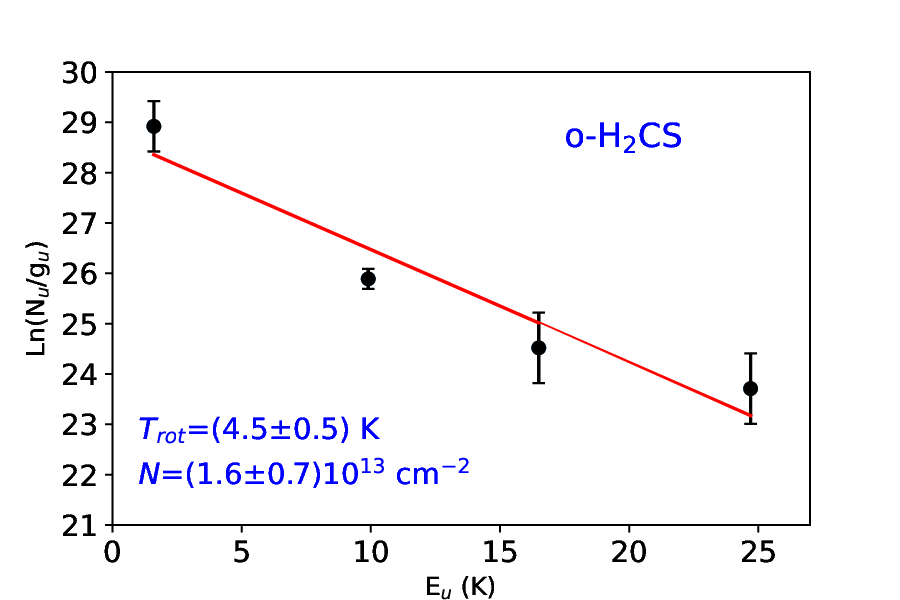} 
 \includegraphics[scale=0.39, angle=0]{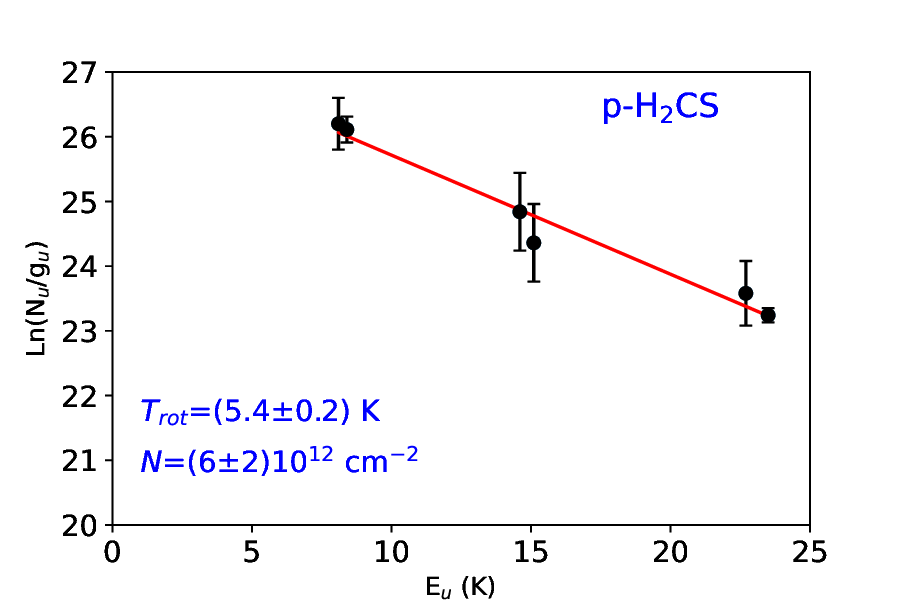}

\hspace{1cm}
\\
\caption{Rotational diagrams of the detected sulphur-bearing molecules in B\,335. Fitted values of the rotational temperature, $T$$_{\mathrm{rot}}$, column density, $N$, and their respective uncertainties are also indicated for each molecule.}
\label{figure:RD}
\end{figure*}

For each molecule, we have computed a representative rotational temperature ($T$$_{\mathrm{rot}}$) and a column density ($N$) by constructing a rotational diagram, assuming a single rotational temperature for all energy levels \citep[]{Goldsmith1999}. Rotational diagrams represent a technique to analyse cloud properties from molecular line emission assuming local thermodynamic equilibrium
(LTE). 
The standard relation for the rotational diagram analysis is

\textbf{
\begin{equation}
\mathrm{ln}(\frac{N_{\mathrm{u}}}{g_{\mathrm{u}}})=\mathrm{ln} N -\mathrm{ln}Z-\frac{E_{\mathrm{u}}}{kT_{\mathrm{rot}}},
\end{equation}
}

\noindent with $N$$_{\mathrm{u}}$/$g$$_{\mathrm{u}}$ given by

\textbf{
\begin{equation}
\frac{N_{\mathrm{u}}}{g_{\mathrm{u}}} = \frac{8k\pi}{hc^3} \times \frac{\nu^{2}_{\mathrm{ul}}}{A_{\mathrm{ul}}g_{\mathrm{u}}} \times {f}^{-1} \times \int T_{\mathrm{MB}} d$v$,
\end{equation}
}

\noindent where $N$$_{\mathrm{u}}$ is the column density of the upper level in the optically thin limit, $N$ is the total column density, $g$$_{\mathrm{u}}$ is the statistical weight of the upper state of each level, $Z$ is the partition function evaluated at a rotational temperature $T$$_{\mathrm{rot}}$, $E$$_{\mathrm{u}}$/k is the energy of the upper level of the transition, $\nu$$_{\mathrm{ul}}$ is the frequency of the $u$→$l$ transition, $\int$ $T$$_{\mathrm{MB}}$ dv is the velocity-integrated line intensity corrected from beam efficiency, and $f$ is the beam filling factor. Assuming that the emission source has a 2D Gaussian shape, $f$ is equal to $f$$_{\mathrm{bf}}$=$\theta$$^{2}_{\mathrm{s}}$/($\theta$$^{2}_{\mathrm{s}}$ + $\theta$$^{2}_{\mathrm{b}}$) being $\theta$$_{\mathrm{b}}$ the HPBW of the telescope in arcsec and $\theta$$_{\mathrm{s}}$ the diameter of the Gaussian source in arcsec. In order to derive the values of $f$ in each case, we have used the HPBW values listed in Table \ref{table:tablaeficiencias}, and we have assumed a source size of 10$\arcsec$ according to interferometric maps of C$^{17}$O emission in B\,335 \citep[]{Cabedo2021}.

The resulting rotational diagrams are shown in Fig. \ref{figure:RD}, while the column densities and rotational temperatures obtained 
are listed in Table \ref{table:results_from_RD}. The uncertainties shown in Table \ref{table:results_from_RD} indicate the uncertainty obtained in the least squares fit of the rotational diagrams. These uncertainties include the 10$\%$ error in the observed line intensities due to calibration. The uncertainty obtained in the determination of the line parameters with the Gaussian fitting programme is included within the error bars at each point of the rotational diagram. 
We did not applied the rotational diagram method to NS since its four observed lines are close to the detection limit and/or blended with other species, which would introduce a high uncertainty in the results. The analysis of NS and others molecules with only one emission line detected (e.g. SO$_2$) has been done only with a LVG code (see next Section) instead.

The obtained rotational temperatures are $\lesssim$15 K. In the case of CCS, the molecule with the largest number of detected transitions, we observe a trend variation (Fig. \ref{figure:RD}) between the low and the high transition energies. In particular, the transitions with $E$$_{\mathrm{u}}$<40 K are well adjusted for a $T$$_{\mathrm{rot}}$$\sim$5 K, while transitions with $E$$_{\mathrm{u}}$>40 K are better adjusted for a $T$$_{\mathrm{rot}}$$\sim$15 K. This suggests that CCS arises from different regions characterised by distinct excitation temperatures. The rotational temperatures for the rest of molecules with several detected transitions (e.g. C$_3$S, OCS, SO, H$_2$CS) have 4.5$\lesssim$$T$$_{\mathrm{rot}}$$\lesssim$7.4 K. These low temperatures together with their narrow line widths are consistent with emission arising from a cold and extended part of the dense core. Gas density at these scales is probably not too high as also indicated by the low $T$$_{\mathrm{rot}}$ obtained for most of the S-molecules. All this suggests that the gas is sub-thermally excited with a moderate or low ($<$10$^5$ cm$^{-3}$) density, in agreement with \cite{Frerking1987}. In addition, the low number of transitions detected for several species (CS, $^{13}$CS, C$^{34}$S, C$^{33}$S, HCS$^+$, and HC$^{34}$S$^+$) and the optically thick emission of some transitions of CCS, CCCS, and H$_2$CS, also indicate that the rotational diagram method not suitable to derive column densities.

\begin{table}
\caption{Column densities ($N$), rotational temperatures, and abundances ($X$) relative to H$_2$ of the sulphur-bearing molecules detected in B\,335.}             
\resizebox{\columnwidth}{!}{%
\begin{tabular}{lllll}     
\hline\hline       
Species & $N$$_{\mathrm{rot}}$$\times$10$^{13}$ & $T$$_{\mathrm{rot}}$  & $^a$$N$$_{\mathrm{LVG}}$$\times$10$^{13}$  & $^b$$X$$\times$10$^{-9}$\\ 
        & (cm$^{-2}$)           & (K)             &    (cm$^{-2}$)  &          \\
\hline 
CS               & 8$\pm$2            & 3.6$\pm$0.8    & 12.00 (34.2)  & 11.0\ \\
$^{13}$CS        & 0.8$\pm$0.1        & 2.8$\pm$0.9    & 0.73          & 0.24\ \\
C$^{34}$S        & 2.1$\pm$0.8        & 2.7$\pm$0.7    & 1.40          & 0.45\ \\
C$^{33}$S        & 0.18$\pm$0.09      & 2.9$\pm$0.5    & 0.13          & 0.04\ \\
CCS              & 5.4$\pm$0.8        & 5$\pm$0.5      & 4.40          & 1.42\ \\
                 & 0.19$\pm$0.06      & 15$\pm$1       &               &  \\
CC$^{34}$S       & 0.27$\pm$0.01      & 6.4$\pm$0.3    & 0.90          & 0.29 \\
CCCS             & 1.8$\pm$0.4        & 7.4$\pm$0.8    & 0.65          & 0.02 \\
CCC$^{34}$S      & 0.14$\pm$0.07      & 2.8$\pm$0.9    & 0.12          & 0.04 \\
OCS              & 3.8$\pm$0.5        & 7.0$\pm$0.8    & 2.00          & 0.65 \\ 
HCS$^+$          & 0.5$\pm$0.2        & 3.8$\pm$0.4    & 0.55 (1.5)    & 0.05 \\
HC$^{34}$S$^+$   & 0.08$\pm$0.02      & 6$\pm$1        & 0.06          & 0.01 \\
SO               & 1.8$\pm$0.6        & 7.2$\pm$0.1    & 6.10          & 1.97 \\
$^{34}$SO        & -                  & -              & 0.25          & 0.08 \\
o-H$_2$CS        & 1.6$\pm$0.7        & 4.5$\pm$0.5    & 1.00          & 0.32 \\
p-H$_2$CS        & 0.6$\pm$0.2        & 5.4$\pm$0.2    & 0.70          & 0.22 \\
NS               & -                  & -              & 0.13          & 0.04 \\
SO$_2$           & -                  & -              & 0.47          & 0.21  \\
H$_2$S           & -                  & -              & 6.50          & 2.10   \\
H$_2$$^{34}$S    & -                  & -              & 0.11          & 0.04   \\
HSCN             & -                  & -              & 0.10          & 0.03\\
HNCS             & -                  & -              & 0.47          & 0.15\\

\hline
Total $N$        &                    &                & 6.2$\times$10$^{14}$  &  \\
\hline 
Total $X$  &                    &                &         & 2.0$\times$10$^{-8}$\\
\hline 
\label{table:results_from_RD}   
\end{tabular} 
}
Numbers in parentheses are obtained from its corresponding $^{34}$S isotopologue considering $^{32}$S/$^{34}$S=24.4 \citep[]{Mauersberger2004}. \\
$^a$ LVG calculations assuming $n$$_{\mathrm{H_2}}$=5$\times$10$^4$ cm$^{-3}$ and $T$$_{\mathrm{K}}$=15 K \citep[]{Shirley2011, Cabedo2021}.\\            
$^b$ Considering $N$$_{\mathrm{H_2}}$=3.1$\times$10$^{22}$ cm$^{-2}$ \citep[]{Cabedo2021}.\\

\end{table}

\subsection{Column densities and abundances}
\label{column_densities}

We have therefore calculated molecular column densities for the detected S-species using a large velocity gradient (LVG) approximation. In particular, we have derived them by fitting the emission line profiles using the LVG code MADEX \citep[]{Cernicharo2012}, which assumes that the radiative coupling between two relatively close points is negligible, and the excitation problem is local. The LVG models are based on the \cite{Goldreich1974} formalism. MADEX also includes corrections for the beam dilution of each line, depending on the different beam sizes at different frequencies. The final considered fit is the one that reproduces more line profiles better from all the observed transitions
. To carry out the fits, we assume uniform physical conditions (kinetic temperature, density, line width, radial velocity).
In particular, we have assumed a gas density of 5$\times$10$^4$ cm$^{-3}$ according to results from Sect. \ref{Rotational_diagrams} and to results from \cite{Cabedo2023}, who deduced a density of $\sim$10$^4$ cm$^{-3}$ for the outer regions of B\,335 based on interferometric observations. We have also assumed  a source size of 10$\arcsec$ according to interferometric maps \citep[]{Cabedo2021}, and a kinetic temperature of the gas $T$$_{\mathrm{K}}$$\sim$15 K estimated by \cite{Shirley2011} from the formula for dust temperature considering only central heating by the B\,335 protostar, and that dust and gas are coupled.

In the case of H$_2$CS and H$_2$S, we have used the code RADEX \citep[]{vanderTak2007} since there are not available collisional rates in MADEX for these two molecules. The results are shown in Table \ref{table:results_from_RD}. For those species for which one transition is observed ($^{34}$SO, SO$_2$, H$_2$S, H$_2$$^{34}$S, HSCN, and HNCS), we have derived their column densities using only the LVG approximation and not the rotational diagram technique described in Section \ref{Rotational_diagrams}. For NS, since there are not available collisional rates, we have considered LTE approximation. 

Column densities derived using the rotational diagram technique and LVG calculations agree within a factor $\lesssim$3. 
Sources of uncertainty are the low angular resolution of the used telescopes (with beam sizes between $\sim$36$\arcsec$ and $\sim$54$\arcsec$ for the 40m-Yebes telescope, and $\sim$14$\arcsec$ and $\sim$29$\arcsec$ for the 30m-IRAM telescope depending on frequency), which implies that the emission from the inner region of the B\,335 is blended with the outer envelope, the possible overlap with the emission from other species (e.g. the case of H$_2$$^{34}$S), the limited number of detected transitions (one transition in many cases, such as for H$_2$S, H$_2$$^{34}$S, SO$_2$, HSCN, HNCS, and $^{34}$SO), the lack of collisional rates for some species (e.g. NS), and the assumed source size derived from C$^{17}$O interferometric observations. This is one of the most important uncertainties due to the lack of interferometric observations of S-species in B\,335, since the emission from the different S-molecules may arise from distinct regions with slightly different sizes leading to a significant impact on the fits. In order to derive the uncertainty associated with the LVG results, we have run the LVG code varying the considered values for density, temperature, and source size. In particular, we have run models with density 5$\times$10$^4$ cm$^{-3}$ and 2.5$\times$10$^4$ cm$^{-3}$ (variation of a factor 2), temperatures of 15 K and 12 K, and source sizes of 10$\arcsec$ and 8 $\arcsec$ (reduction of 20$\%$). We found variations in the S-column densities $\lesssim$10$\%$ between these two $T$$_{\mathrm{K}}$ which is comparable to the calibration errors, while the variations when changing the density and the source size were $\lesssim$20$\%$ and $\sim$30$\%$, respectively. Taking all this into account, we estimate the uncertainties in the column densities derived with LVG approximation to be of the order of $\sim$50$\%$. 

The total column density of the sulphur-bearing species observed in B\,335 is 6.2$\times$10$^{14}$ cm$^{-2}$, with the highest contribution coming from CS \citep[calculated from C$^{34}$S considering $^{32}$S/$^{34}$S=24.4 since CS is optically thick,][]{Mauersberger2004}, H$_2$S, SO, and CCS. The lowest values are found for NS and HSCN. We have also observed several isotopologues relative to $^{13}$C, $^{34}$S, and $^{33}$S. In particular, from CS, we obtain $^{34}$S/$^{33}$S=10.7 \citep[which is slightly higher than the solar abundance value $^{34}$S/$^{33}$S=5.6-6.3,][]{AndersGrevesse1989, Mauersberger2004}, and $^{12}$C/$^{13}$C=46.8, which is similar to that found in dark clouds by \cite{Cernicharo1987}.

\begin{figure}[h!]
\includegraphics[scale=0.65, angle=0]{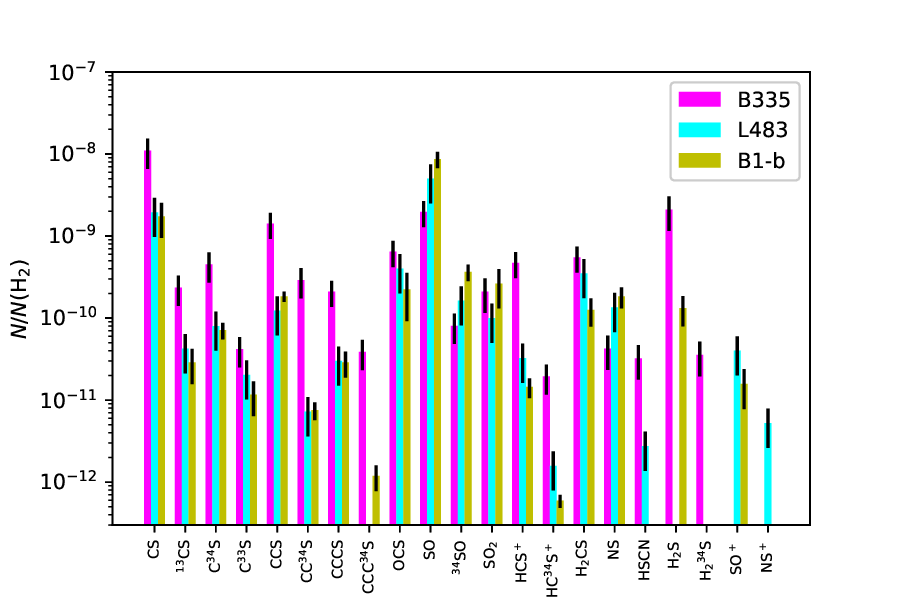}  \hspace{0.0cm}
\caption{Comparison of fractional abundances of sulphur-bearing species between three Class 0 objects B\,335 (this work), L\,483 \citep[]{Agundez2019}, and B1-b \citep[]{Fuente2016}.}
\label{figure:abundances_histograms}
\end{figure}

Table \ref{table:results_from_RD} also shows the fractional abundances relative to H$_2$
\citep[with $N$$_{\mathrm{H_2}}$ obtained from][]{Cabedo2021} of the sulphur molecules detected in the range $\lambda$ = 2, 3, and 7 mm. Considering the detected S-molecules, we have obtained a total molecular sulphur abundance of 2.0$\times$10$^{-8}$, which is alike to the one found in similar sources, such as L\,1544 \citep[1.1$\times$10$^{-8}$,][]{Vastel2018}, L\,483 \citep[9.1$\times$10$^{-9}$,][]{Agundez2019}, and B1-b \citep[1.2$\times$10$^{-8}$,][]{Fuente2016}. The highest S-abundances are $>$10$^{-9}$ (CS, H$_2$S, CCS, and SO), with also several relatively abundant ($>$10$^{-10}$)
species, such as OCS and H$_2$CS. In general, B\,335 presents a great variety of S-bearing molecules, but less than other similar sources. In particular, although the total sulphur abundance in B\,335 is similar to other sources, such as L\,483, we detect a significant lower number of sulphur molecules in B\,335 than in L\,483. We are also missing some other S-species already reported towards L\,483, including SO$^+$, HCS, and CH$_3$SH, for which we have derived upper limits for their column densities of $N$(SO$^+$)$<$1.5$\times$10$^{11}$ cm$^{-2}$, $N$(HCS)$<$3$\times$10$^{12}$ cm$^{-2}$, and $N$(CH$_3$SH)$<$2$\times$10$^{12}$ cm$^{-2}$. To derive these upper limits, once the physical parameters are fixed, we vary (for each species) the column density until the model fit reaches the observed intensity peak of any of the observed lines. We do not allow the model fit to be greater than any observed line.

\section{discussion}
\label{discussion}

\subsection{Chemical comparison}
\label{Comparison}

Even though the total sulphur abundance in B\,335 is similar to that from other Class 0 sources, the abundances of some specific molecules significantly vary among them. Figure \ref{figure:abundances_histograms} shows a comparison between abundances of different sulphur species in three Class 0 sources, B\,335, L\,483, and B1-b. The first thing that stands out is that sulphur carbon chains (CS, CCS, and CCCS) in B\,335 are about one order of magnitude higher than the ones observed in L\,483 and B1-b. Although high abundances of carbon-chain molecules have already been found in other cold clouds, such as TMC1 \citep[]{Agundez2013}, it is not a universal characteristic of this type of early regions, since many of them present different degrees of carbon chain richness \citep[e.g.][]{Suzuki1992, Hirota2009}. This therefore highlights the nature of
B\,335 as a source especially rich in sulphur carbon chains. 

\begin{figure}
\includegraphics[scale=0.65, angle=0]{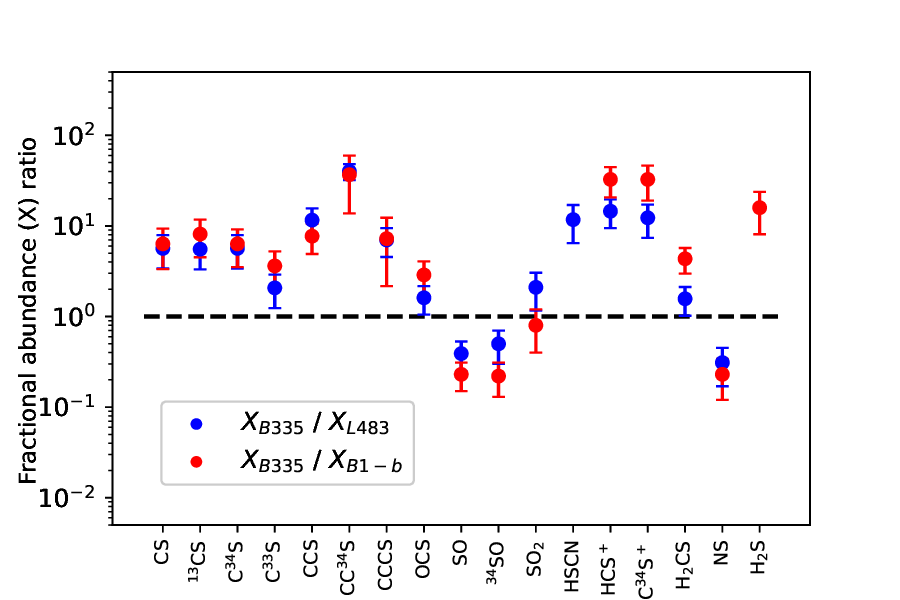}  \hspace{0.0cm}
\caption{Fractional abundance ($X$) ratios between B\,335 and L\,483, and B\,335 and B1-b for all the detected sulphur-bearing species in B\,335.}
\label{figure:abundances_ratios_other_sources}
\end{figure}

Unlike carbon chains, sulphur molecules containing oxygen in B\,335 have a similar (OCS) or even lower (SO and SO$_2$) abundance by up to one order of magnitude than those found in L\,483 and B1-b (see also Fig. \ref{figure:abundances_ratios_other_sources}). These oxygen-sulphur molecules are usually detected in cold dense regions with abundances between 10$^{-10}$-10$^{-8}$, while in hot cores and hot corinos they are significantly enhanced \citep[e.g.][]{Tercero2010, Esplugues2013}. In particular, SO$_2$ is a better tracer of warm gas than SO \citep[][]{Esplugues2013}, while SO is a well-known outflow tracer \citep[e.g.][]{Codella2003, Lee2010, Tafalla2010}. SO seems to be more enhanced than SO$_2$ in shocks for timescales $\sim$10$^4$ yr \citep[e.g.][]{Codella1999, Viti2004, Jimenez-Serra2005}. The low SO abundance found in B\,335 with respect to the other two sources L\,483 and B1-b suggests that its chemistry may not be significantly affected by shocks associated with the bipolar outflow \citep[e.g.][]{Hirano1988, Hirano1992, Stutz2008, Bjerkeli2019}.  

Other sulphur molecules detected in B\,335 with low abundances (Fig. \ref{figure:abundances_ratios_other_sources}) are those containing nitrogen, \text{that is} NS and HSCN (4$\times$10$^{-11}$ and 3$\times$10$^{-11}$, respectively). HSCN, unlike HNCS which is the most stable isomer, is metastable and only detected on Earth under specific conditions (through UV-photolysis of HNCS \citep[]{Wierzejewska2001} or through a low-pressure discharge \citep[]{Brunken2009}). In contrast, HSCN is more easily observed in the interstellar medium. For instance, in Sgr B2 (where HSCN was first time detected) the ratio HNCS/HSCN$\sim$3 \citep[]{Halfen2009} and in TMC1 this ratio is $\sim$1 \citep[]{Adande2010}, while in L\,1544 \cite{Vastel2018} only detect HSCN, but not HNCS. This shows the important role of non-equilibrium chemistry in dense clouds. In B\,335, we only detect one transition of HNCS with a signal to noise ratio $\sigma$$\sim$3.1, so we have considered its column density ($N_{\mathrm{HNCS}}$$\leq$4.7$\times$10$^{12}$ cm$^{-2}$) as an upper limit. This leads to a ratio HNCS/HSCN$<$4.7 consistent with the one found in Sgr B2.

\subsection{Sulphur ions}
\label{Ions}

Apart from neutral sulphur species, we have also observed the presence of some ions in B\,335. Molecular ions are commonly observed in dense prestellar or protostellar regions \citep[e.g.][]{Caselli1998, Hogerheijde1998, Caselli2002, vanderTak2005}. To a lesser extent, ions (such as HCO$^+$ and N$_2$H$^+$) have also been observed in protostellar shocks  \citep[e.g.][]{Bachiller1997, Hogerheijde1998, Caselli2002, vanderTak2005}. The presence of ions in shocked regions is a balance between their enhancement by the sputtering of dust grain mantles, which ejects molecules into the gas phase that undergo reactions forming ions, and their destruction by electronic recombination \citep[e.g.][]{Neufeld1989, Viti2002}. For this reason, those ions that usually decrease in shocks (such as N$_2$H$^+$) may indicate the presence of a pre-shock chemistry \citep[]{Codella2013}, while the ions that are enhanced by shock chemistry will be effective tracers of shocked gas. This is the case of HOCO$^+$ and SO$^+$ \citep[]{Podio2014} since their abundances are predicted to be very low in the quiescent gas, but significantly enhanced in shocks following the release of CO$_2$ and S-species from dust grain mantles \citep[]{Minh1991, Turner1992, Turner1994, Deguchi2006}.

Among the sulphur ions, in B\,335 we do not detect SO$^+$ \citep[strong tracer of shocked gas which forms from S$^+$ and OH][]{Neufeld1989}, but we detect HCS$^+$ and HC$^{34}$S$^+$ with abundances of 4.7$\times$10$^{-10}$ and 1.9$\times$10$^{-11}$, respectively. The value we have obtained for HCS$^+$ is similar to the one observed in the bow shock L\,1157-B1 \citep[]{Podio2014} and one order of magnitude higher than the ones obtained by \cite{Agundez2019} and \cite{Fuente2016} in the Class 0 objects L\,483 and B1-b, respectively, as also found for CS. 
The abundances of HCS$^+$ and CS are strongly related, since this ion forms through the reaction of CS with HCO$^+$, H$^{+}_{3}$, H$_3$O$^{+}$ \citep[]{Millar1985}. In addition, large abundances of CS and HCS$^+$ may also indicate that OCS is one of the main sulphur carriers on dust grains upon sputtering from the grain mantles 
according to previous theoretical \citep[]{Podio2014} and observational \citep[]{Wakelam2005, Codella2005} results.

Regarding the ratio $R$=HCS$^+$/CS, we obtain from observations a value of 0.043. \cite{Thaddeus1981} determined an abundance ratio $R$$\sim$0.01-0.03 in ten molecular sources located in Sgr B2 and Ori A, while \cite{Millar1983} found that $R$ could be as large as 0.1 in cold dark clouds. This result was also verified by \cite{Irvine1983}. A similar $R$ value to that from B\,335 is found as well in the pre-stellar core L\,1544 \citep[0.03,][]{Vastel2018}, while in the protostellar sources L\,483 and B1-b the HCS$^+$/CS ratio is 0.017 and 0.008, respectively \citep[]{Agundez2019, Fuente2016}. According to \cite{Clary1985}, rate coefficients for ion-CS reactions rapidly increase at low temperatures suggesting that the sources B\,335 and L\,1544 are characterised by lower temperatures than L\,483 and B1-b, which can be due to an earlier evolutionary stage of B\,335 and L\,1544 compared to L\,483 and B1-b.

\subsection{Time evolution of fractional abundances}
\label{fractional_abundances_evolution}

\begin{SCfigure*}
  \centering
  \caption{Evolution of fractional abundances of CS, CCS, C$_3$S, OCS, SO, SO$_2$, HCS$^+$, H$_2$CS, and NS as a function of time for an initial sulphur abundance $S$$^{+}_{\mathrm{init}}$=1.5$\times$10$^{-6}$, a hydrogen number density $n$$_{\mathrm{H}}$=2$\times$10$^4$ cm$^{-3}$, $T$$_{\mathrm{gas}}$=7 K, and two values of CR ionisation rate $\zeta$=1.3$\times$10$^{-17}$ s$^{-1}$ (solid line) and $\zeta$=1.3$\times$10$^{-16}$ s$^{-1}$ (dashed line).}
  \includegraphics[width=0.4\textwidth]%
    {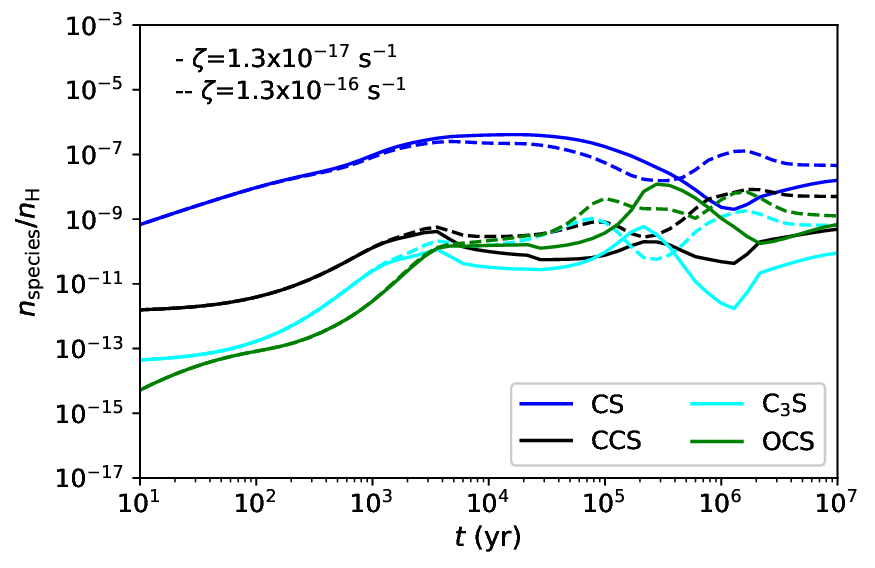}
  \includegraphics[width=0.4\textwidth]%
    {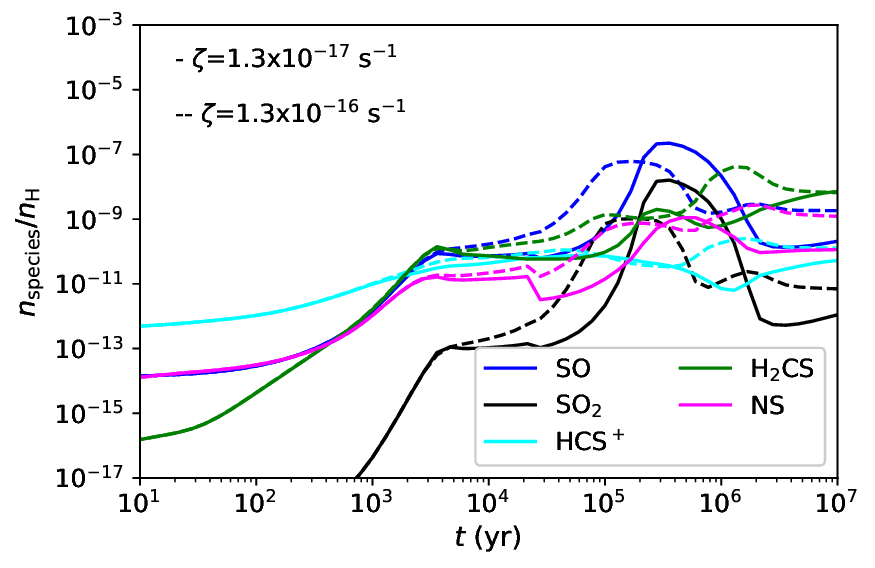}
   \label{figure:CR_effect_plot1}
\end{SCfigure*}

\begin{SCfigure*}
  \centering
  \caption{Evolution of fractional abundances of CS, CCS, C$_3$S, OCS, SO, SO$_2$, HCS$^+$, H$_2$CS, and NS as a function of time for an initial sulphur abundance $S$$^{+}_{\mathrm{init}}$=1.5$\times$10$^{-6}$, a hydrogen number density $n$$_{\mathrm{H}}$=2$\times$10$^4$ cm$^{-3}$, $T$$_{\mathrm{gas}}$=7 K, and two values of CR ionisation rate $\zeta$=1.3$\times$10$^{-17}$ s$^{-1}$ (solid line) and $\zeta$=1.3$\times$10$^{-16}$ s$^{-1}$ (dashed line).}
  \includegraphics[width=0.4\textwidth]%
    {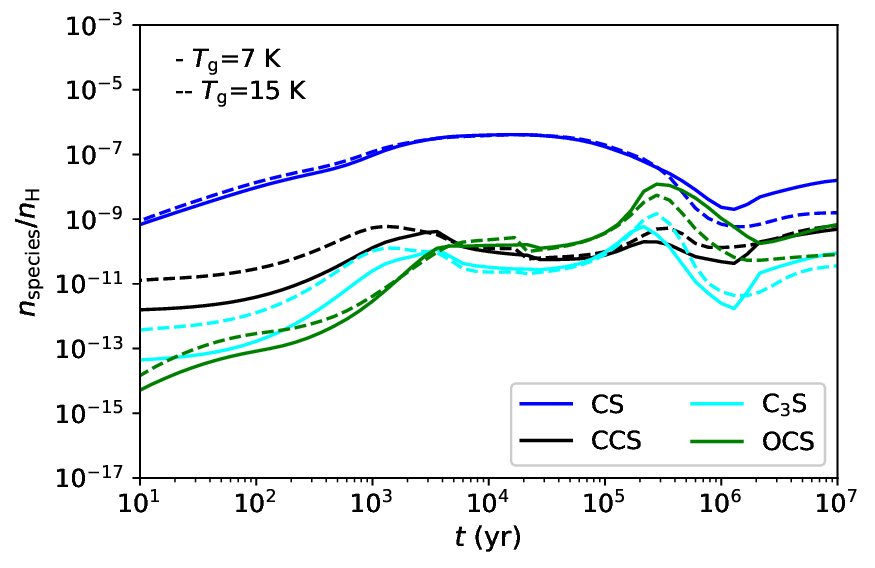}
  \includegraphics[width=0.4\textwidth]%
    {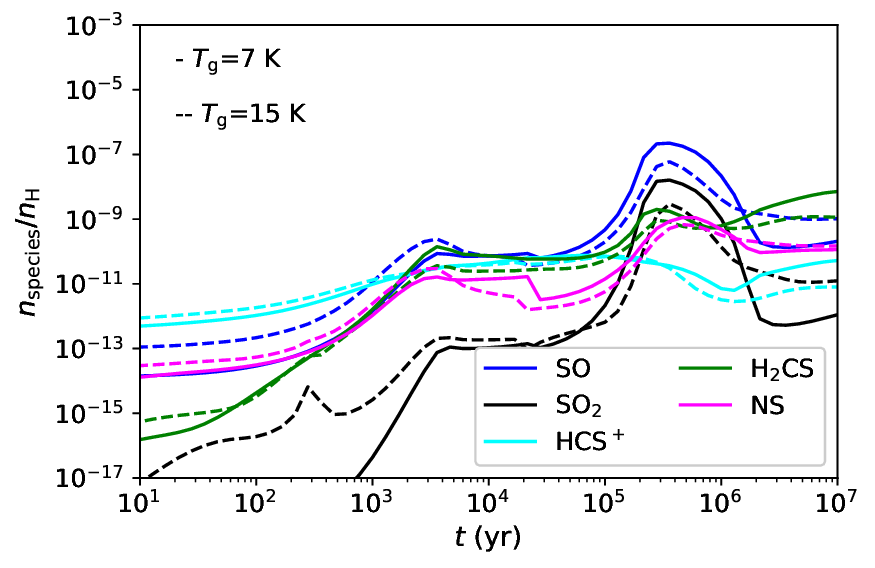}
  \label{figure:Tg_effect_plot1}
\end{SCfigure*}

\begin{SCfigure*}
  \centering
  \caption{Evolution of fractional abundances of CS, CCS, C$_3$S, OCS, SO, SO$_2$, HCS$^+$, H$_2$CS, and NS as a function of time for an initial sulphur abundance $S$$^{+}_{\mathrm{init}}$=1.5$\times$10$^{-6}$, a hydrogen number density $n$$_{\mathrm{H}}$=2$\times$10$^4$ cm$^{-3}$, $T$$_{\mathrm{gas}}$=7 K, and two values of CR ionisation rate $\zeta$=1.3$\times$10$^{-17}$ s$^{-1}$ (solid line) and $\zeta$=1.3$\times$10$^{-16}$ s$^{-1}$ (dashed line).}
  \includegraphics[width=0.4\textwidth]%
    {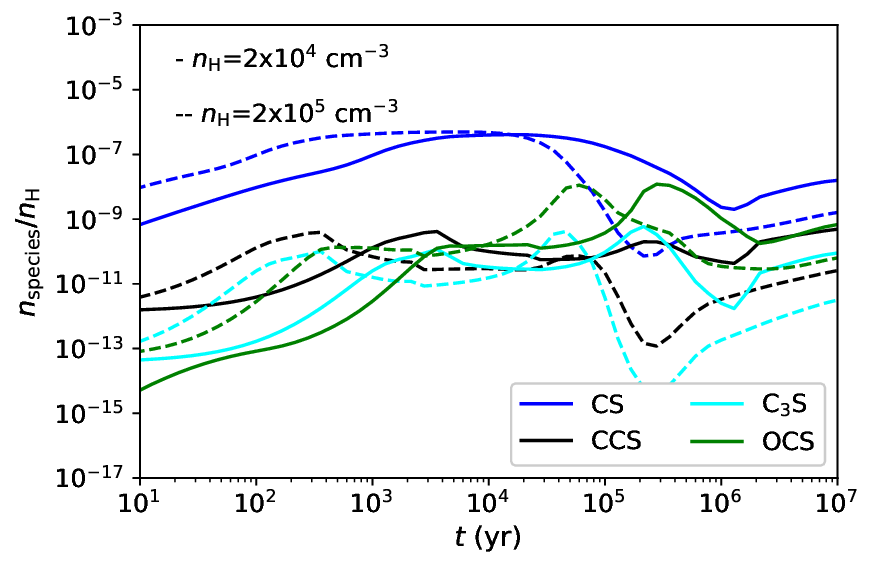}
  \includegraphics[width=0.4\textwidth]%
    {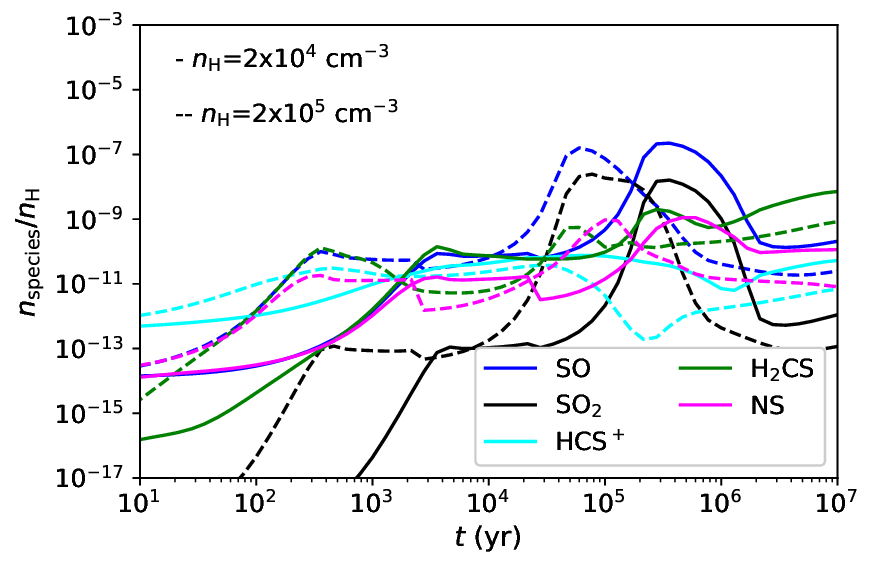}
    \label{figure:n_effect_plot1}
\end{SCfigure*}

\begin{SCfigure*}
  \centering
  \caption{Evolution of fractional abundances of CS, CCS, C$_3$S, OCS, SO, SO$_2$, HCS$^+$, H$_2$CS, and NS as a function of time for an initial sulphur abundance $S$$^{+}_{\mathrm{init}}$=1.5$\times$10$^{-6}$, a hydrogen number density $n$$_{\mathrm{H}}$=2$\times$10$^4$ cm$^{-3}$, $T$$_{\mathrm{gas}}$=7 K, and two values of CR ionisation rate $\zeta$=1.3$\times$10$^{-17}$ s$^{-1}$ (solid line) and $\zeta$=1.3$\times$10$^{-16}$ s$^{-1}$ (dashed line).}
  \includegraphics[width=0.4\textwidth]%
    {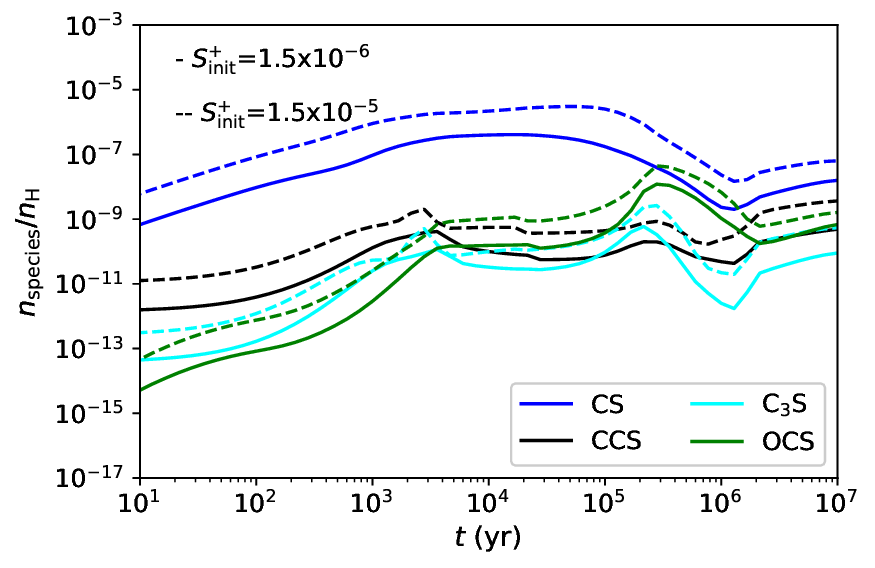}
  \includegraphics[width=0.4\textwidth]%
    {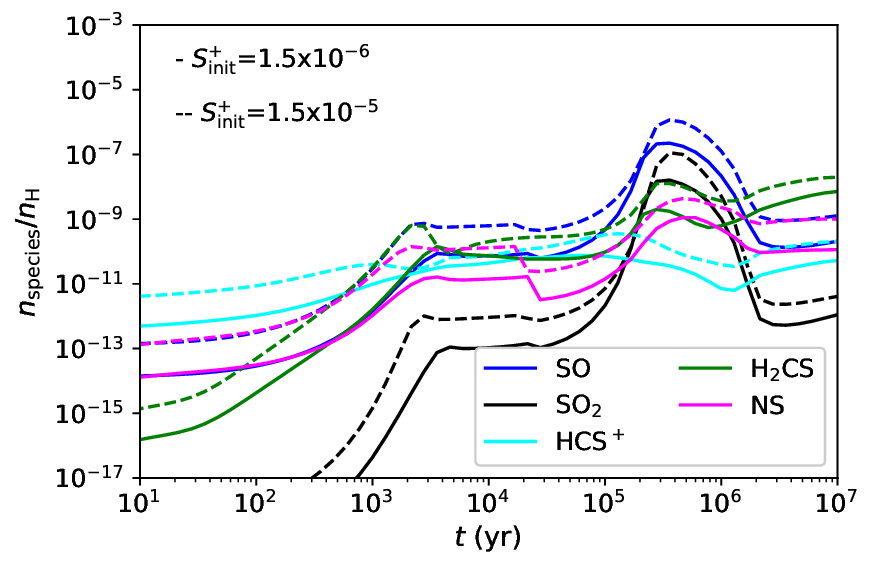}
    \label{figure:S+_effect_plot1}
\end{SCfigure*}

Figures \ref{figure:CR_effect_plot1}-\ref{figure:S+_effect_plot1} show the evolution of several S-bearing species along 10 Myr for different physical conditions of the cosmic-ray (CR) ionisation rate ($\zeta$$_{\mathrm{H_2}}$), gas temperature ($T$$_{\mathrm{g}}$), hydrogen number density ($n$$_{\mathrm{H}}$), and initial sulphur abundance ($S$$^{+}_{\mathrm{init}}$). Given that the CR ionisation rate is still uncertain and spreads over a range of values (\cite{vanderTak2000} determined as $\zeta$$_{\mathrm{H_2}}$$\sim$3$\times$10$^{-17}$ s$^{-1}$ in dense clouds, \cite{indriolo2012} obtained a range (1.7-10.6)$\times$$^{-16}$ s$^{-1}$ in a sample of diffuse clouds, and \cite{Neufeld2017} derived a $\zeta$$_{\mathrm{H_2}}$ of the order of a few 10$^{-16}$ s$^{-1}$ in the Galactic disk), we have considered two values for the cosmic ionisation rate $\zeta$$_{\mathrm{H_2}}$=1.3$\times$10$^{-16}$ s$^{-1}$ and 1.3$\times$10$^{-17}$ s$^{-1}$. For the gas temperature and density, given the results obtained in Sects. \ref{Rotational_diagrams} and \ref{column_densities}, we have run models considering $T$$_{\mathrm{g}}$=7 K and 15 K, and $n$$_{\mathrm{H}}$=2$\times$10$^4$ cm$^{-3}$ and 2$\times$10$^5$ cm$^{-3}$. For the case of the initial sulphur abundance, we have considered the solar elemental sulphur fractional abundance ($S$$^{+}_{\mathrm{init}}$=1.5$\times$10$^{-5}$) and also one factor of S depletion (1.5$\times$10$^{-6}$) since to reproduce observations in hot corinos one needs to assume a significant sulphur depletion of at least one order of magnitude lower than the solar elemental sulphur abundance as previously stated in Sect. \ref{section:Introduction}.

We have obtained these models using the Nautilus time-dependent chemical code \citep[]{Ruaud2016}. Nautilus is a three-phase model in which gas, surface icy mantle, bulk icy mantle, and their interactions are considered. Nautilus solves the kinetic equations for both the gas phase species and the surface species of interstellar dust grains and computes the evolution with time of chemical abundances for a given physical structure. 
The chemical network is based on the KInetic Database for Astrochemistry (KIDA\footnote{http://kida.obs.u-bordeaux1.fr/}). The used version here is detailed in \cite{Wakelam2021}. In particular, this Nautilus version considers gas-phase processes including neutral-neutral and ion-neutral reactions, direct cosmic ray ionisation or dissociation, ionisation or dissociation by UV photons, ionisation or dissociation produced by photons induced by cosmic-ray interactions with the medium \citep[]{Prasad1983}, and electronic recombinations.  
For species on the surfaces, there is a distinction between species in the most external layers (surface species) and species below these layers (mantle species). The species are adsorbed on the surface and become the mantle during the construction of the ices. Similarly, when the species desorb, only the species from the surface can desorb, but surface species are gradually replaced by the mantle species. Regarding surface reactions, the model includes thermal desorption, photo-desorption, chemical desorption, and cosmic-ray heating. See \cite{Wakelam2021} for extensive details about the surface parameters considered in each surface reaction.
In all models, we have adopted the initial abundances shown in Table \ref{table:abundances_Nautilus} and a visual extinction $A$$_{\mathrm{V}}$=15 mag\footnote{Derived from results in Table \ref{table:results_from_RD} and the expression
\begin{equation}
\label{conversor_abundances}
N_i = X_i\times A_V\times1.6\times10^{21},
\end{equation}

where $N$$_i$ is the column density of the species $i$, $X$$_i$ its abundance, $A$$_{\mathrm{V}}$ the visual extinction, and 1.6$\times$10$^{21}$ the hydrogen column density at 1 mag of extinction \citep[]{Bohlin1978}.}.

\begin{table}
\caption{Abundances with respect to total hydrogen nuclei considered in the chemical code Nautilus.
}             
\centering 
\begin{tabular}{l l l l l}     
\hline\hline       
Species &  Abundance   & Reference                       \\ 
\hline 
He      & 9.0$\times$10$^{-2}$   &  (1) \\
O       & 2.4$\times$10$^{-4}$   &  (2) \\
Si$^+$  & 8.0$\times$10$^{-9}$   &  (3) \\ 
Fe$^+$  & 3.0$\times$10$^{-9}$   &  (3) \\ 
S$^+$   & 1.5$\times$10$^{-5}$, 1.5$\times$10$^{-6}$   &  (1)  \\
Na$^+$  & 2.0$\times$10$^{-9}$   &  (3)  \\
Mg$^+$  & 7.0$\times$10$^{-9}$   &  (3)  \\
P$^+$   & 2.0$\times$10$^{-10}$  &  (3)  \\
Cl$^+$  & 1.0$\times$10$^{-9}$   &  (3)   \\
F$^+$   & 6.7$\times$10$^{-9}$   &  (4)  \\
N       & 6.2$\times$10$^{-5}$   &  (5) \\
C$^+$   & 1.7$\times$10$^{-4}$   &  (5) \\
\hline 
\label{table:abundances_Nautilus}                 
\end{tabular}
\tablefoot{
References: (1) Taken from \cite{Asplund2009} and \cite{Wakelam2008}. (2) Taken from \cite{Hincelin2011}. (3) Taken from \cite{Graedel1982}. (4) Taken from \cite{Neufeld2005}. (5) Taken from \cite{Jenkins2009}.\\
} 
\\
\end{table}

In general, we observe that the abundances of all the species considered in Figs. \ref{figure:CR_effect_plot1}-\ref{figure:S+_effect_plot1} increase with time up to $t$$\sim$5$\times$10$^{5}$ yr, moment at which the abundances of various species decrease abruptly. This is the case of CS, SO, and SO$_2$, whose abundances decrease by about 2-4 orders of magnitude between $t$$\sim$5$\times$10$^{5}$ yr and $t$$\sim$5$\times$10$^{6}$ yr since they react with other species to form more complex molecules. In particular, for $t$$>$10$^5$ yr, CS SO, and SO$_2$ are mainly destroyed by reacting with H$^{+}_{\mathrm{3}}$ and forming HCS$^+$, HSO$^+$, and HSO$^{+}_{\mathrm{2}}$, respectively. For the abundances of the rest of sulphur species, this decrease after reaching the maximum value is lower and not greater than two orders of magnitude. From these figures, we also deduce the formation time scales of S-species. In particular, we have observed that CS is one of the molecules reaching first its maximum abundance value in all the models at $t$$\sim$5$\times$10$^4$ yr, followed by HCS$^+$ with its maximum abundance reached at $t$$\sim$10$^5$ yr. Other molecules, such as H$_2$CS, can be considered as late molecules, since their maximum abundance values are found when the chemical evolution exceeds one million years.

Regarding the influence of the physical conditions (cosmic-ray ionisation rate, temperature, density, and initial sulphur abundance) on the S-abundances, the parameter with the highest impact is the hydrogen nuclei number density, $n$$_{\mathrm{H}}$, since varying $n$$_{\mathrm{H}}$ by one order of magnitude leads to abundance variations of up to three orders of magnitude for CS, CCS, OCS, HCS$^+$, SO\textbf{,} and SO$_2$ given a specific evolutionary time (see Fig. \ref{figure:n_effect_plot1}). In general terms, we observe in Fig. \ref{figure:n_effect_plot1} that the chemical evolution of each considered species is accelerated when the density increases since collisions are more frequent than in a low density regime. This results in the abundance peak of the species being reached at earlier times for larger densities, although the chemical behaviour along time is roughly maintained. In particular, for $t$$\sim$10$^4$ yr, we obtain that SO$_2$ is mainly formed by the reaction O+SO for $n$$_{\mathrm{H}}$=2$\times$10$^4$ and 2$\times$10$^5$ cm$^{-3}$, while for $t$$\sim$10$^5$ yr (when the SO$_2$ peak is reached) this molecule mainly forms by the reaction O+SO for $n$$_{\mathrm{H}}$=2$\times$10$^4$ cm$^{-3}$ and by OH+SO for $n$$_{\mathrm{H}}$=2$\times$10$^5$ cm$^{-3}$. The most responsible species of destroying SO$_2$ for both densities is C for $t$$\sim$10$^4$ yr, and H$^{+}_{\mathrm{3}}$  for $t$$\geq$10$^5$ yr. Regarding CS, for $t$$\sim$10$^4$ yr, it is mainly formed by the reaction HCS$^+$+e$^-$ for the low density value, while CS is mainly formed through the reaction between atomic S and C$_4$ for $n$$_{\mathrm{H}}$=2$\times$10$^5$ cm$^{-3}$. For $t$$\sim$10$^5$ yr (when the CS abundance decreases for any model), CS is mainly destroyed by its reaction with H$^{+}_{\mathrm{3}}$ and with HCO$^+$. 

Another parameter with a big impact on S-abundances is the cosmic-ray ionisation rate, $\zeta$, since the variation from $\zeta$=1.3$\times$10$^{-17}$ to 1.3$\times$10$^{-16}$ s$^{-1}$ changes by about two orders of magnitude the abundances of all the sulphur species considered in the sample (Fig. \ref{figure:CR_effect_plot1}). For the case of carbon-chains (CCS, C$_3$S), we observe that the larger $\zeta$, the larger the S-abundances for $t$$\lesssim$10$^5$ yr. Similar results are found for the species SO, SO$_2$, NS, H$_2$CS, and OCS. For $t$$\sim$10$^4$ yr, we find that CCS is mainly formed by C+HCS for the low $\zeta$ model, while it is mainly formed through HC$_3$S$^+$+e$^-$ in a more efficient way for a model with larger $\zeta$ (1.3$\times$10$^{-16}$ s$^{-1}$). In the case of SO (one of the most affected molecules by $\zeta$ between 10$^4$$\lesssim$$t$$\lesssim$10$^6$ yr), it is mostly formed by the reaction O+HS in a model with low $\zeta$, while the reaction between S+OH becomes more important to form SO$_2$ when increasing $\zeta$ by one order of magnitude.

On the other hand, changing the temperature (from $T$$_{\mathrm{gas}}$=7 K to 15 K) and especially the initial sulphur abundance (from $S$$^{+}_{\mathrm{init}}$=1.5$\times$10$^{-5}$ to 1.5$\times$10$^{-6}$) in the models leads to abundance variations smaller than two orders of magnitude for $t$$>$10$^4$ yr. In particular, we have found that an increase of the $S$$^{+}_{\mathrm{init}}$ value by one order of magnitude also enhances the S-abundances by one order of magnitude during the entire duration of core evolution.

For the particular case of HCS$^+$ and CS, the main physical parameters significantly influencing their abundances are the density (the lower density, the higher the HCS$^+$ and CS abundances for $t$$>$10$^4$ yr) and the cosmic-ray ionisation rate (the higher the $\zeta$, the higher the HCS$^+$ and CS abundances for $t$$>$5$\times$10$^5$ yr). By contrast, other physical parameters, such as the gas temperature, barely affect the HCS$^+$ and CS abundances for $t$$<$10$^6$ yr.

\begin{SCfigure*}
  \centering
  \caption{Evolution of fractional abundances of CS, CCS, C$_3$S, OCS, SO, SO$_2$, HCS$^+$, H$_2$CS, and NS as a function of time for an initial sulphur abundance $S$$^{+}_{\mathrm{init}}$=1.5$\times$10$^{-6}$, a hydrogen number density $n$$_{\mathrm{H}}$=2$\times$10$^4$ cm$^{-3}$, $T$$_{\mathrm{gas}}$=7 K, and two values of CR ionisation rate $\zeta$=1.3$\times$10$^{-17}$ s$^{-1}$ (solid line) and $\zeta$=1.3$\times$10$^{-16}$ s$^{-1}$ (dashed line).}
  \includegraphics[width=0.4\textwidth]%
    {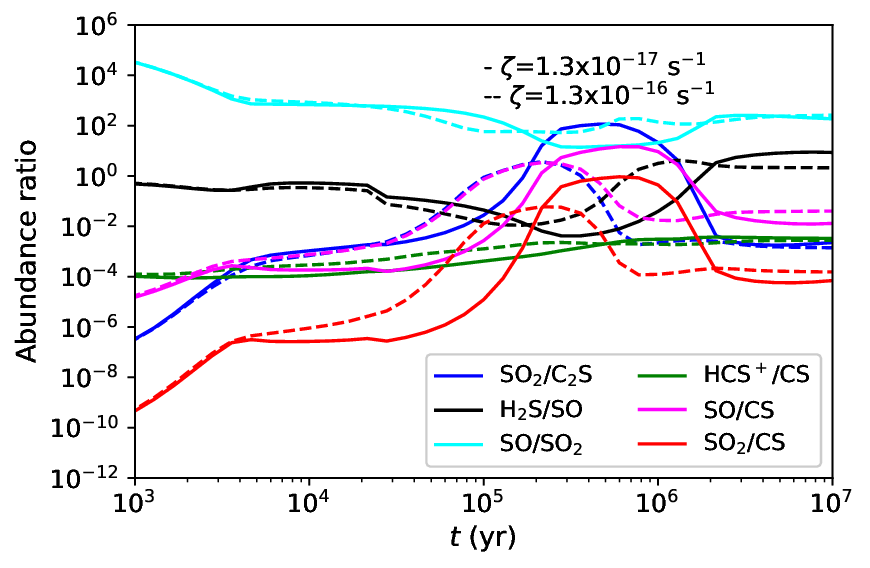}
  \includegraphics[width=0.4\textwidth]%
    {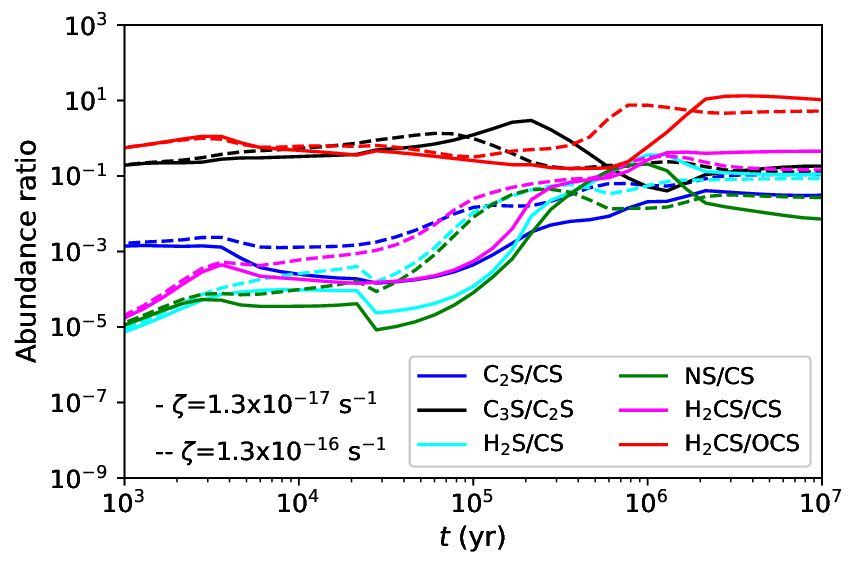}
   \label{figure:CR_effect_ratios}
\end{SCfigure*}

\begin{SCfigure*}
  \centering
  \caption{Evolution of fractional abundances of CS, CCS, C$_3$S, OCS, SO, SO$_2$, HCS$^+$, H$_2$CS, and NS as a function of time for an initial sulphur abundance $S$$^{+}_{\mathrm{init}}$=1.5$\times$10$^{-6}$, a hydrogen number density $n$$_{\mathrm{H}}$=2$\times$10$^4$ cm$^{-3}$, $T$$_{\mathrm{gas}}$=7 K, and two values of CR ionisation rate $\zeta$=1.3$\times$10$^{-17}$ s$^{-1}$ (solid line) and $\zeta$=1.3$\times$10$^{-16}$ s$^{-1}$ (dashed line).}
  \includegraphics[width=0.4\textwidth]%
    {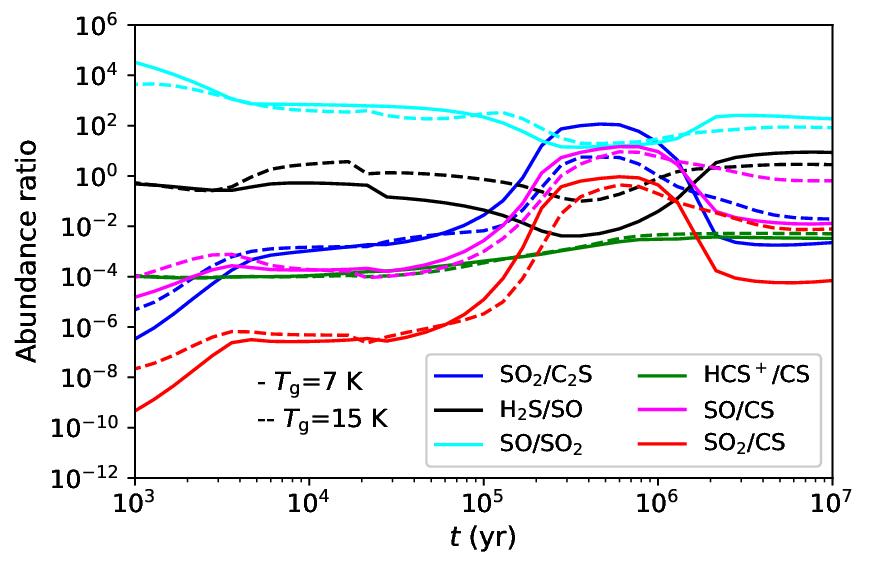}
  \includegraphics[width=0.4\textwidth]%
    {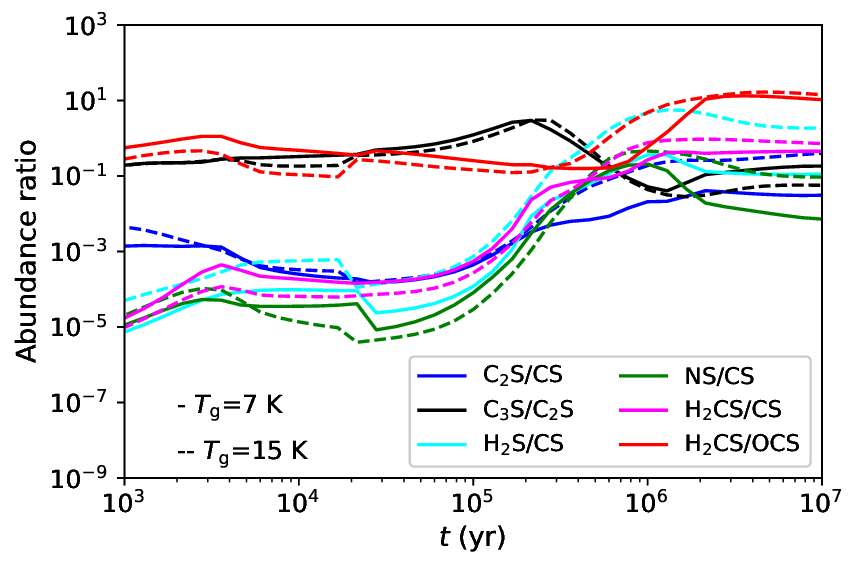}
   \label{figure:Tg_effect_ratios}
\end{SCfigure*}

\begin{SCfigure*}
  \centering
  \caption{Evolution of fractional abundances of CS, CCS, C$_3$S, OCS, SO, SO$_2$, HCS$^+$, H$_2$CS, and NS as a function of time for an initial sulphur abundance $S$$^{+}_{\mathrm{init}}$=1.5$\times$10$^{-6}$, a hydrogen number density $n$$_{\mathrm{H}}$=2$\times$10$^4$ cm$^{-3}$, $T$$_{\mathrm{gas}}$=7 K, and two values of CR ionisation rate $\zeta$=1.3$\times$10$^{-17}$ s$^{-1}$ (solid line) and $\zeta$=1.3$\times$10$^{-16}$ s$^{-1}$ (dashed line).}
  \includegraphics[width=0.4\textwidth]%
    {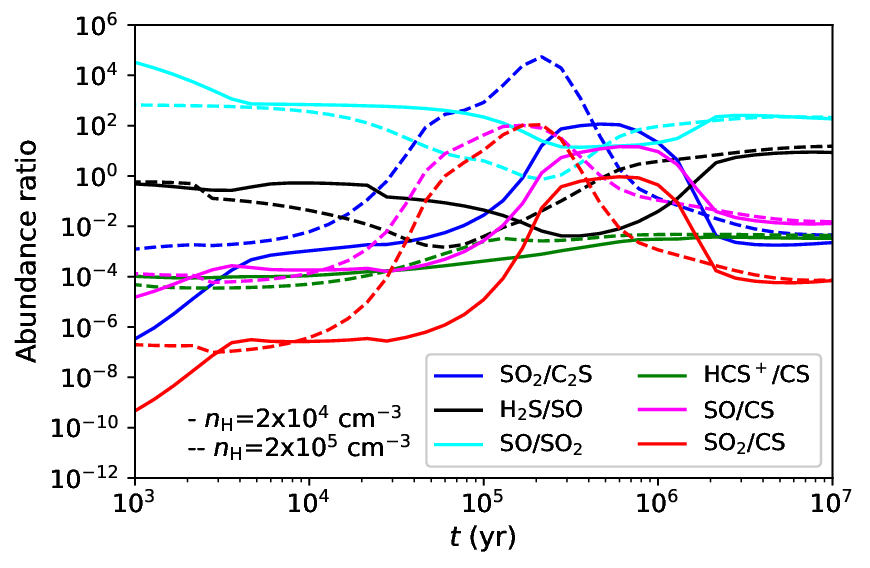}
  \includegraphics[width=0.4\textwidth]%
    {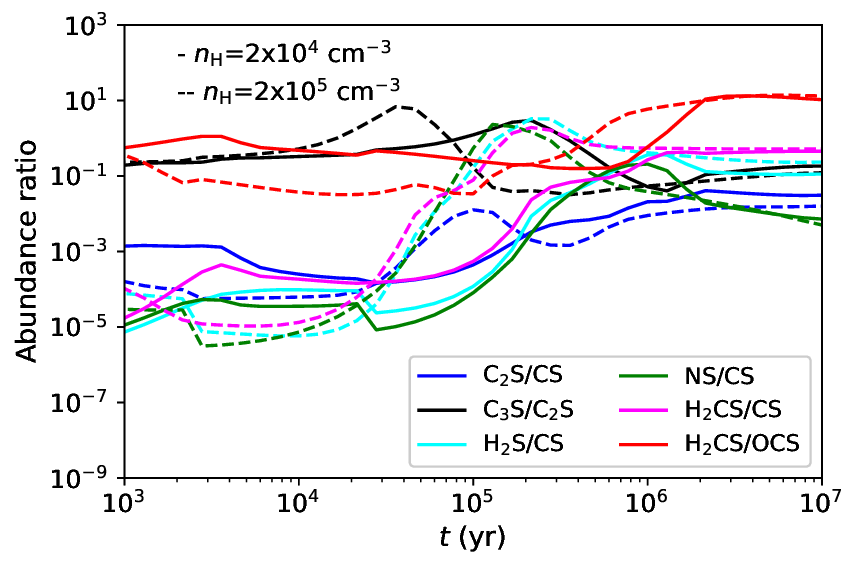}
   \label{figure:n_effect_ratios}
\end{SCfigure*}

\begin{SCfigure*}
  \centering
  \caption{Evolution of fractional abundances of CS, CCS, C$_3$S, OCS, SO, SO$_2$, HCS$^+$, H$_2$CS, and NS as a function of time for an initial sulphur abundance $S$$^{+}_{\mathrm{init}}$=1.5$\times$10$^{-6}$, a hydrogen number density $n$$_{\mathrm{H}}$=2$\times$10$^4$ cm$^{-3}$, $T$$_{\mathrm{gas}}$=7 K, and two values of CR ionisation rate $\zeta$=1.3$\times$10$^{-17}$ s$^{-1}$ (solid line) and $\zeta$=1.3$\times$10$^{-16}$ s$^{-1}$ (dashed line).}
  \includegraphics[width=0.4\textwidth]%
    {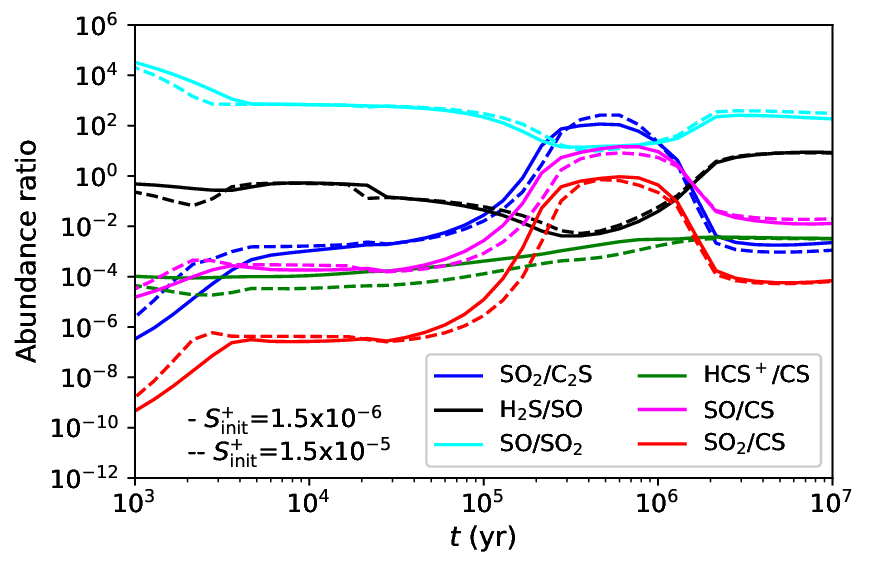}
  \includegraphics[width=0.4\textwidth]%
    {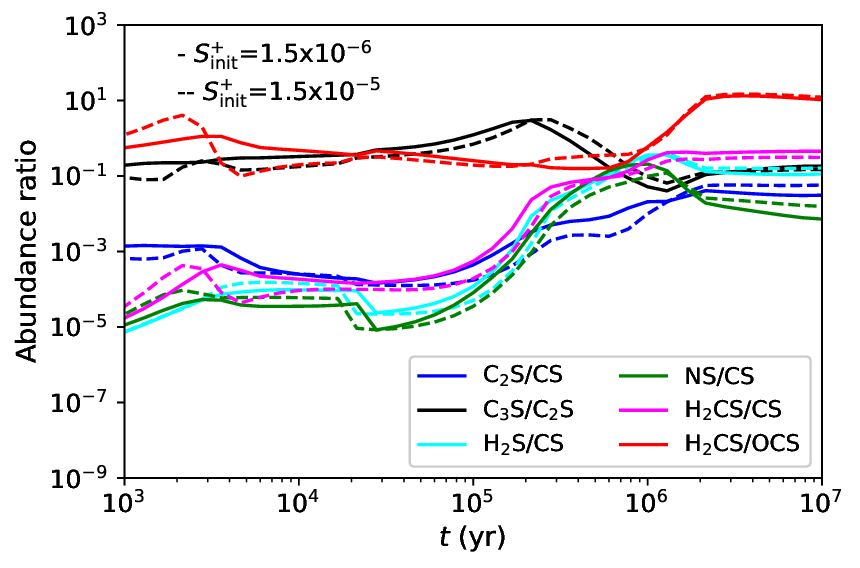}
   \label{figure:S+_effect_ratios}
\end{SCfigure*}

\subsection{\textbf{Comparison with observations}}
\label{comparison_with_observations}

We now compare the different models considered in the previous section with observational results to derive the current evolutionary stage of B335. Figures \ref{figure:model_vs_obs_CR}-\ref{figure:model_vs_obs_S+} show the ratio between the abundances obtained from the models ($X$$_{model}$) and from the observations ($X$$_{obs}$) for several sulphur-bearing species. In particular, we have considered those S-species for which we observe more than one emission line and those that are not close to the limit of detection. The results are shown for five specific evolution times ($t$=10$^4$, 5$\times$10$^4$, 10$^5$, 5$\times$10$^5$, and 10$^6$ yr). 
In order to find the cases with the closest values between the theoretical and observational results, we have considered for each model the largest number of S-species for which -0.5$\leq$log($X$$_{model}$/$X$$_{obs}$)$\leq$0.5 in order to only consider discrepancies between model and observational results smaller than a factor of $\sim$3. Results show that the evolutionary stages that better reproduce the observed abundances of the largest number of S-species in B\,335 are those between $t$=10$^4$-10$^5$ yr. This result obtained from a chemical approach agrees with the results obtained from \cite{Evans2015} who derived an age of $\sim$5$\times$10$^4$ yr for B\,335 through a dynamical perspective, by applying an inside-out collapse model that is compared with ALMA observations. Altogether this reveals, therefore, the particularly early evolutionary stage of B\,335, which is comparable to the ages of pre-stellar condensations \citep[e.g.][]{Caselli2012}. As B\,335 is an isolated globule, this may be due to the fact that the condensation has formed from very little dense gas ($\sim$10$^3$ cm$^{-3}$) without a dense pre-phase between the formation of the globule and the beginning of the collapse. This gives B\,335 a different chemistry compared to other young protostars that have form in dense molecular clouds, such as L\,483 and B1-b. Another plausible scenario is that material from the diffuse cloud surrounding B\,335 may currently be accreting onto the protostellar envelope, thus rejuvinating its chemistry. This accretion has been found \citep[e.g.][]{Pineda2020} in another Class 0 source embedded in the Perseus Molecular Cloud Complex. Interferometric observations are requested to distinguish between both scenarios.  

From Figs. \ref{figure:model_vs_obs_CR}-\ref{figure:model_vs_obs_S+}, we have also deduced that chemical models with values of $n$$_{H}$=2$\times$10$^4$ cm$^{-3}$, $T$$_{g}$=15 K, $S$$^{+}_{\mathrm{init}}$=1.5$\times$10$^{-6}$, and $\zeta$=1.3$\times$10$^{-17}$ s$^{-1}$ best reproduce the observations of the largest number of S-species for $t$=10$^4$-10$^5$ yr. This value of $\zeta$ agrees with results from \cite{Cabedo2023} who found a cosmic-ray ionisation rate $\lesssim$10$^{-16}$ s$^{-1}$ in the outer regions of B\,335, especially for an envelope radii$>$500 au. A sulphur depletion factor of 10 with respect to the sulphur cosmic elemental abundance is also consistent with results from \cite{Fuente2023} in low-mass star-forming regions, such as Taurus and Perseus.

\subsection{Sulphur chemical ratios}
\label{chemical_ratios}

We have considered several sulphur-bearing molecular ratios and have also analysed their evolution with time. Model results obtained with the Nautilus code are shown in Figures \ref{figure:CR_effect_ratios}-\ref{figure:S+_effect_ratios}.  
We have also changed some of the physical conditions (cosmic-ray ionisation rate, temperature, density, and initial sulphur abundance) in order to study their impact on the evolution of these ratios. We first notice that while there are some ratios, such as HCS$^+$/CS, that present small variations of up 1 order of magnitude between $t$=10$^4$-10$^6$ yr, there are other ratios that significantly change with time. In particular, the SO$_2$/C$_2$S, SO/CS, SO$_2$/CS, NS/CS, H$_2$S/CS, and H$_2$CS/CS ratios show variations of up to $\sim$5 orders of magnitude in that time range (10$^4$-10$^6$ yr). The physical parameter affecting sulphur ratios the most is the density, leading to differences of up to $\sim$4-5 orders of magnitude especially between 5$\times$10$^4$ and 10$^6$ yr. The cosmic-ray ionisation rate is the next physical parameter with the greatest effect on the sulphur ratios with differences of about one order of magnitude in their values. In particular, the SO$_2$/CS, SO/CS, SO$_2$/C$_2$S, and C$_2$S/CS ratios are among those mostly influenced by $\zeta$, while the SO/SO$_2$ ratio is one of the least affected. The SO/SO$_2$ ratio is, however, more sensitive to the density variation. We also observe a large density impact on the NS/CS ratio, finding that the higher the density, the larger the NS/CS for $t$$\sim$10$^5$-10$^6$ yr, with NS being mostly formed during that time range through the neutral-neutral reaction N+HS and, to a lesser extent, through the dissociative recombination of HNS$^+$. On the other hand, varying the gas temperature from 7 to 15 K (Fig. \ref{figure:Tg_effect_ratios}) mostly affects the H$_2$S/SO ratio between 10$^4$ and 10$^6$ yr, and the SO/CS, SO$_2$/CS, and H$_2$S/CS ratios at late times ($t$$\gtrsim$10$^6$ yr).
Regarding the only ion detected in B\,335, its HCS$^+$/CS ratio slightly increases when the initial sulphur abundance is decreased, and also when the cosmic-ray ionisation rate or the density increase.

\begin{figure}
\includegraphics[scale=0.65, angle=0]{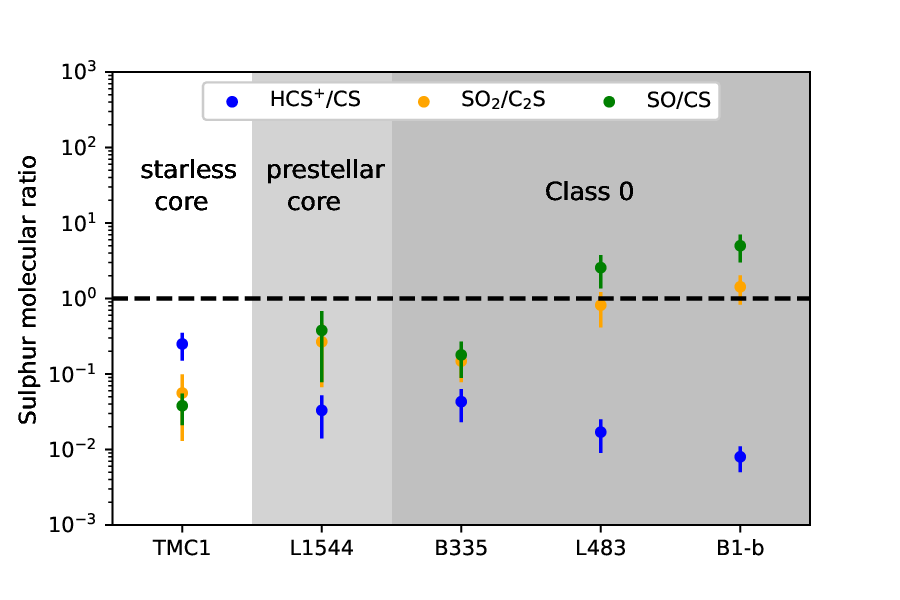}  \hspace{0.0cm}
\caption{Observational sulphur molecular ratios of potentially good chemical evolution tracers of the prestellar to protostellar transition for different objects (TMC1, L\,1544, B\,335, L\,483, and B1-b). Data for TMC1 were taken from \cite{Marcelino2007}, \cite{Cernicharo2012}, \cite{Cernicharo2021}, and \cite{Rodriguez-Baras2021}. Data for L\,1544 were taken from \cite{Vastel2018}. L\,483 and B1-b data were gathered from \cite{Agundez2019} and \cite{Fuente2016}, respectively.}
\label{figure:molecular_ratios}
\end{figure}

The use of molecular ratios as indicators of chemical evolution of young stellar sources has been previously approached from both observational and theoretical perspectives \citep[e.g.][]{PineauDesForets1993, Charnley1997, Hatchell1998, Codella1999, Gibb2000, Boogert2000, Wakelam2004}. Our models (Figs. \ref{figure:CR_effect_plot1}-\ref{figure:S+_effect_plot1}) show that SO and SO$_2$ constantly increase with time reaching both their peaks at $t$$\gtrsim$5$\times$10$^5$ yr, which prevents the use of their ratio as a chemical clock. 
Nevertheless, Figures \ref{figure:CR_effect_ratios}-\ref{figure:S+_effect_ratios} show that there are other sulphur ratios that could be used as evolutionary tracers since they maintain a constant trend over time. This is for instance the case of the SO$_2$/C$_2$S and SO/CS ratios, whose values constantly increase during the first $\sim$2$\times$10$^5$ yr. In order to compare these theoretical results with observations, we have considered a sample of sources that includes starless cores \citep[TMC1 and L\,1544 considered as starless and prestellar cores, respectively, see for example][]{Crapsi2005, Schnee2007, Lefloch2018, Agundez2019, Cernicharo2021} and Class 0 objects (B\,335, L\,483, and B1-b). Observational results are shown in Fig. \ref{figure:molecular_ratios}, where we have indeed found a clear trend where the SO$_2$/C$_2$S and SO/CS ratios increase with the age of the objects, suggesting that these sulphur ratios could be used as chemical clocks. We have also included in Fig. \ref{figure:molecular_ratios} observational results for the HCS$^+$/CS ratio
. In this case, we have obtained a negative trend in which the value of HCS$^+$/CS decreases with the age of the objects. 

Comparing the observational HCS$^+$/CS, SO$_2$/C$_2$S, and CO/CS ratios for B\,335 with the theoretical results (Figs. \ref{figure:CR_effect_ratios}-\ref{figure:S+_effect_ratios}), we observe that, for the temperature and density values used in Sect. \ref{column_densities} to derive observational results, the three ratios would be reproduced at 5$\times$10$^4$$\lesssim$$t$$\lesssim$10$^5$ yr when considering a high cosmic-ray ionisation rate ($\zeta$=1.3$\times$10$^{-16}$ s$^{-1}$) and a S depletion of at least a  factor 10 with respect to the solar elemental sulphur abundance. This depletion is also found by many authors \citep[e.g.][]{Wakelam2004, Crockett2014, Vastel2018, Bulut2021, Navarro-Almaida2021, Esplugues2022, Hily-Blant2022, Fuente2023} in other star-forming regions.

\section{Summary and conclusions}
\label{section:summary}

We have carried out a comprehensive observational and theoretical study of sulphur in the Class 0 object B\,335. This source is an isolated dense Bok globule located at a distance of $\sim$100 pc. 
We have detected 20 different sulphur-bearing species (including isotopologues) using Yebes 40m and IRAM 30m telescope observations.  
In order to derive column densities, we have applied the method of rotational diagrams. For this purpose, we have also used non-LTE radiative transfer codes given the low number of transitions detected for some species and the optically thick emission of some lines. Our results show low rotational temperatures ($\lesssim$15 K), suggesting that the gas is sub-thermally excited, and thus revealing a relatively low ($<$10$^5$ cm$^{-3}$) gas density. 
These low temperatures, the narrow line widths, and the relatively low density are also consistent with emission arising from a cold and extended region, suggesting that the detected S-emission does not trace the hot corino, but its cold envelope.

From the column densities of all the observed sulphur species, we have derived a total molecular sulphur abundance of 2.0$\times$10$^{-8}$, which is similar to that found in the envelopes of other Class 0 objects. Nevertheless, we have also found that the abundances of some specific S-molecules vary significantly across the sample of Class 0 objects (B\,335, L\,483, and B1-b) considered in this paper. In particular, we have derived high abundances of sulphur carbon-chain molecules in B\,335 compared to those found in other Class 0 objects, and also compared with those abundances of S-molecules containing oxygen or nitrogen. This points out the nature of B\,335 as a source especially rich in sulphur carbon chains, which could be due to the fact that B\,335 is an isolated source compared to other young protostars that have formed in dense molecular clouds, such as L\,483 and B1-b, or to accretion of material from the diffuse cloud surrounding B\,335 onto the protostellar envelope, thus rejuvenating its chemistry.
The comparison between envelopes of different Class 0 objects has also allowed to deduce a low SO abundance in B\,335 with respect to L\,483 and B1-b. SO is a well-known outflow tracer, which is thought to be enhanced in shocks. Its low abundance in B\,335 together with the non-detection of other shock tracers, such SO$^+$, suggests that its chemistry in B\,335 is not particularly influenced by shocks.

In this work, we have also used the time-dependent chemical code Nautilus to study the time evolution of fractional S-abundances, as well as the influence of several physical and chemical parameters (cosmic-ray ionisation rate, density, gas temperature, and initial sulphur abundance) on these abundances. The comparison between chemical model results and observations shows that the evolutionary stages that better reproduce the observed abundances of the largest number of S-species in B\,335 correspond to $t$=10$^4$-10$^5$ yr, in agreement with previous age estimations obtained by applying theoretical methods with the use of a dynamical model. The results from Nautilus have also revealed that the physical parameters with the highest influence  on the S-abundances are the density and the cosmic-ray ionisation rate whose variations by one order of magnitude lead to abundance variations of several orders of magnitude for different sulphur-bearing species such as SO, SO$_2$, CS, and CCCS. We have also used Nautilus to explore the behaviour of different S-molecular ratios with time, finding that there are some sulphur ratios that could be used as chemical evolution tracers since they maintain a constant trend over time. A comparison of the observational results of S-ratios in a sample of objects (including starless and prestellar cores, and Class 0 objects) characterised by different evolutionary stages confirms a clear trend where the SO$_2$/C$_2$S and SO/CS ratios increase with the age of the objects, while HCS$^+$/CS decreases with the evolutionary stage. We, therefore, conclude that these S-ratios could be used as good chemical evolutionary indicators tracing the prestellar to protostellar transition.

\begin{acknowledgements}

We thank the Spanish MICINN for funding support from PID2019-106235GB-I00. This project has been carried out with observations from the 30-m radio telescope of the Institut de radioastronomie millimétrique (IRAM) and from the 40-m radio telescope of the National Geographic Institute of Spain (IGN) at Yebes Observatory. Yebes Observatory thanks the ERC for funding support under grant ERC-2013-Syg-610256-NANOCOSMOS. A.F. is grateful to the European Research Council (ERC) for funding under the Advanced Grant project SUL4LIFE, grant agreement No101096293. D.N.A. acknowledges funding support from Fundación Ramón Areces through their international postdoc grant program. Á.S.M. acknowledges support from the RyC2021-032892-I grant funded by MCIN/AEI/10.13039/501100011033 and by the European Union 'Next GenerationEU'/PRTR, as well as the program Unidad de Excelencia María de Maeztu CEX2020-001058-M. M.N.D. acknowledges the support by the Swiss National Science Foundation (SNSF) Ambizione grant no. 180079, the Center for Space and Habitability (CSH) Fellowship, and the IAU Gruber Foundation Fellowship. We also thank the anonymous referee for valuable comments that improved the manuscript.

\end{acknowledgements}

\bibliographystyle{aa}
\bibliography{biblio}

\begin{appendix}

\pagebreak

\section{Tables and figures}
\label{tables}

\onecolumn
\begin{longtable}{lllllll}
\caption{Line parameters of sulphur-bearing molecules detected in B\,335.}\\
\hline\hline                      
\hline
\endfirsthead
\caption{continued.}\\
\hline\hline
Species    & Transition   & Frequency    & v$_{\mathrm{LSR}}$ & $\Delta$v    & $T$$_{\mathrm{MB}}$  & $\int$$T$$_{\mathrm{MB}}$dv  \\
           &              & (MHz)        & (km s$^{-1}$)        & (km s$^{-1}$)  & (K)                  & (K km s$^{-1}$)                \\
\hline
\endhead
\hline
\endfoot
Species    & Transition   & Frequency    & v$_{\mathrm{LSR}}$ & $\Delta$v    & $T$$_{\mathrm{MB}}$  & $\int$$T$$_{\mathrm{MB}}$ dv  \\
           &              & (MHz)        & (km s$^{-1}$)        & (km s$^{-1}$)  & (K)                  & (K km s$^{-1}$)                \\
\hline

CS         & 1-0          &  48990.98   & 8.27$\pm$0.07         & 1.20$\pm$0.08   & 1.29$\pm$0.50        & 1.64$\pm$0.03 \\
           & 2-1          &  97980.95   & 8.25$\pm$0.08         & 1.41$\pm$0.05   & 1.13$\pm$0.32        & 1.70$\pm$0.07 \\
           & 3-2          &  146969.02  & 8.01$\pm$0.06         & 1.05$\pm$0.03   & 1.50$\pm$0.17        & 1.68$\pm$0.04 \\
\hline
$^{13}$CS  & 1-0          &  46247.56   & 8.31$\pm$0.08         & 0.81$\pm$0.02   & 0.15$\pm$0.03        & 0.128$\pm$0.002 \\
           & 2-1          &  92494.27   & 8.34$\pm$0.04         & 1.34$\pm$0.07   & 0.06$\pm$0.01        & 0.095$\pm$0.007 \\
           & 3-2          &  138739.26  & 8.24$\pm$0.06         & 0.99$\pm$0.07   & 0.06$\pm$0.01        & 0.066$\pm$0.008 \\
\hline
C$^{34}$S  & 1-0          &  48206.97    & 8.43$\pm$0.02         & 0.97$\pm$0.06   & 0.39$\pm$0.05        & 0.403$\pm$0.004 \\
           & 2-1          &  96412.95    & 8.31$\pm$0.07         & 1.27$\pm$0.09   & 0.21$\pm$0.05        & 0.279$\pm$0.006 \\
           & 3-2          &  144617.10    & 8.28$\pm$0.07         & 0.84$\pm$0.08   & 0.18$\pm$0.06        & 0.165$\pm$0.005 \\
\hline
C$^{33}$S  & 1-0          &  48585.88   & 8.17$\pm$0.05         & 0.80$\pm$0.08   & 0.04$\pm$0.01        & 0.041$\pm$0.004 \\
           & 2-1          &  97172.06    & 9.07$\pm$0.04         & 0.86$\pm$0.05   & 0.02$\pm$0.01        & 0.020$\pm$0.006 \\
           & 3-2          &  145755.62   & 8.21$\pm$0.07         & 0.80$\pm$0.06   & 0.02$\pm$0.01        & 0.020$\pm$0.007 \\
\hline
CCS    & 2$_3$-1$_2$         &  33751.37   & 8.80$\pm$0.03         & 0.83$\pm$0.04   & 0.34$\pm$0.05        & 0.306$\pm$0.001 \\
       &                     &             & 7.39$\pm$0.03         & 1.10$\pm$0.05   & 0.15$\pm$0.02        & 0.185$\pm$0.001 \\ 
       & 3$_3$-2$_2$         &  38866.42   & 8.29$\pm$0.05         & 0.77$\pm$0.04   & 0.11$\pm$0.02        & 0.087$\pm$0.001 \\
       & 4$_3$-3$_2$         &  43981.02   & 8.39$\pm$0.08         & 0.55$\pm$0.08   & 0.16$\pm$0.03        & 0.095$\pm$0.001 \\
       & 3$_4$-2$_3$         &  45379.03   & 8.35$\pm$0.02         & 0.67$\pm$0.06   & 0.85$\pm$0.07        & 0.606$\pm$0.001 \\
       & 6$_6$-5$_5$         &  77731.71   & 8.13$\pm$0.04         & 1.24$\pm$0.07   & 0.08$\pm$0.02        & 0.100$\pm$0.010 \\
       & 6$_7$-5$_6$         &  81505.17   & 8.05$\pm$0.01         & 1.52$\pm$0.07   & 0.22$\pm$0.03        & 0.357$\pm$0.006 \\
       & 7$_6$- 6$_5$        &  86181.39   & 8.18$\pm$0.07         & 0.78$\pm$0.05   & 0.10$\pm$0.04        & 0.084$\pm$0.004 \\
       & 7$_7$- 6$_6$        &  90686.38   & 8.47$\pm$0.07         & 0.77$\pm$0.09   & 0.06$\pm$0.01        & 0.051$\pm$0.006 \\
       & 7$_8$- 6$_7$        &  93870.09   & 9.14$\pm$0.06         & 1.35$\pm$0.04   & 0.19$\pm$0.02        & 0.274$\pm$0.005 \\
       & 8$_7$- 7$_6$        &  99866.52   & 8.54$\pm$0.03         & 0.75$\pm$0.03   & 0.08$\pm$0.02        & 0.061$\pm$0.007 \\
       & 8$_8$- 7$_7$        &  103640.75  & 8.42$\pm$0.05         & 0.76$\pm$0.04   & 0.05$\pm$0.01        & 0.044$\pm$0.004 \\
       & 8$_9$- 7$_8$        &  106347.72  & 7.99$\pm$0.09         & 1.10$\pm$0.05   & 0.12$\pm$0.04        & 0.140$\pm$0.004 \\
       & 9$_8$- 8$_7$        &  113410.18  & 8.63$\pm$0.04         & 0.93$\pm$0.07   & 0.04$\pm$0.01        & 0.040$\pm$0.006 \\
       & 10$_{11}$-9$_{10}$  &  131551.96  & 8.38$\pm$0.03         & 0.72$\pm$0.04   & 0.05$\pm$0.01        & 0.038$\pm$0.007 \\
       & 11$_{10}$-10$_9$    &  140180.74  & 8.60$\pm$0.04         & 1.10$\pm$0.08   & 0.03$\pm$0.01        & 0.033$\pm$0.007 \\
       & 11$_{11}$-10$_{10}$ &  142501.69  & 8.55$\pm$0.06         & 0.95$\pm$0.04   & 0.03$\pm$0.01        & 0.027$\pm$0.006 \\
       & 11$_{12}$-10$_{11}$ &  144244.82  & 8.35$\pm$0.05         & 0.92$\pm$0.07   & 0.05$\pm$0.01        & 0.040$\pm$0.020 \\
       & 13$_{14}$-12$_{13}$ &  169753.46  & 8.14$\pm$0.07         & 0.84$\pm$0.08   & 0.03$\pm$0.01        & 0.027$\pm$0.005 \\
       & 13$_{13}$-12$_{12}$ &  168406.79  & 8.38$\pm$0.05         & 0.62$\pm$0.08   & 0.03$\pm$0.01        & 0.017$\pm$0.004 \\
\hline
CC$^{34}$S & 2$_3$-1$_2$     &  33111.83    & 8.97$\pm$0.05        & 0.55$\pm$0.06   & 0.010$\pm$0.001        & 0.070$\pm$0.003 \\
           &                 &              & 8.10$\pm$0.02        & 1.63$\pm$0.05   & 0.008$\pm$0.001        & 0.013$\pm$0.003 \\
           & 4$_3$-3$_2$     &  42918.18    & 8.37$\pm$0.04        & 0.60$\pm$0.05   & 0.008$\pm$0.001        & 0.030$\pm$0.002 \\
           & 3$_4$-2$_3$     &  44497.59    & 8.23$\pm$0.01        & 0.60$\pm$0.07   & 0.05$\pm$0.01        & 0.005$\pm$0.001 \\
\hline
CCCS       & 6-5             &  34684.36   & 8.78$\pm$0.03         & 0.66$\pm$0.08   & 0.12$\pm$0.02        & 0.090$\pm$0.001 \\
           &                 &             & 7.54$\pm$0.05         & 0.83$\pm$0.09   & 0.06$\pm$0.01        & 0.053$\pm$0.001 \\
           & 7-6             &  40465.01   & 8.32$\pm$0.04         & 0.60$\pm$0.05   & 0.24$\pm$0.04        & 0.158$\pm$0.001 \\
           & 8-7             &  46245.62   & 8.34$\pm$0.02         & 0.77$\pm$0.05   & 0.20$\pm$0.05        & 0.170$\pm$0.012 \\
           & 13-12           &  75147.92   & 8.60$\pm$0.02         & 1.20$\pm$0.06   & 0.04$\pm$0.01        & 0.063$\pm$0.012 \\
           & 14-13           &  80928.18   & 7.82$\pm$0.09         & 1.66$\pm$0.09   & 0.03$\pm$0.01        & 0.060$\pm$0.007 \\
           & 15-14           &  86708.37   & 8.30$\pm$0.06         & 0.90$\pm$0.03   & 0.07$\pm$0.01        & 0.071$\pm$0.005 \\
           & 16-15           &  92488.49   & 8.57$\pm$0.05         & 1.01$\pm$0.08   & 0.02$\pm$0.01        & 0.027$\pm$0.009 \\
\hline
CCC$^{34}$S & 6-5            &  33844.24   & 8.77$\pm$0.04         & 0.84$\pm$0.05   & 0.005$\pm$0.001        & 0.005$\pm$0.001 \\
            & 7-6            &  39484.87   & 8.36$\pm$0.03         & 0.55$\pm$0.05   & 0.012$\pm$0.03        & 0.007$\pm$0.001 \\
            & 8-7            &  45125.47   & 8.33$\pm$0.04         & 0.41$\pm$0.07   & 0.009$\pm$0.003        & 0.004$\pm$0.001 \\
\hline
OCS        & 3-2             &  36488.81   & 8.45$\pm$0.06         & 0.94$\pm$0.06   & 0.015$\pm$0.002        & 0.015$\pm$0.005 \\
           & 4-3             &  48651.60   & 8.11$\pm$0.07         & 1.10$\pm$0.08   & 0.023$\pm$0.002        & 0.027$\pm$0.001 \\
           & 7-6             &  85139.10   & 8.50$\pm$0.03         & 0.90$\pm$0.06   & 0.033$\pm$0.003        & 0.031$\pm$0.012 \\
           & 8-7             &  97301.20   & 8.40$\pm$0.04         & 1.45$\pm$0.09   & 0.017$\pm$0.002        & 0.026$\pm$0.012 \\
           & 9-8             &  109463.06  & 8.62$\pm$0.03         & 1.09$\pm$0.05   & 0.020$\pm$0.002        & 0.024$\pm$0.007 \\
\hline
HCS$^{+}$   & 1-0            &  42674.19   & 8.34$\pm$0.05         & 0.56$\pm$0.05   & 0.10$\pm$0.02        & 0.061$\pm$0.001 \\
            & 2-1            &  85347.87   & 8.25$\pm$0.03         & 0.75$\pm$0.07   & 0.11$\pm$0.04        & 0.089$\pm$0.004 \\
            & 4-3            &  170691.64  & 8.27$\pm$0.06         & 0.56$\pm$0.04   & 0.05$\pm$0.01        & 0.035$\pm$0.004 \\
\hline
HC$^{34}$S$^{+}$ & 1-0       &  41983.06   & 8.43$\pm$0.07         & 0.60$\pm$0.04   & 0.006$\pm$0.001        & 0.004$\pm$0.001 \\
                 & 4-3       &  167927.26  & 8.90$\pm$0.08         & 0.75$\pm$0.08   & 0.023$\pm$0.01        & 0.018$\pm$0.006 \\
\hline
SO         & 2$_3$-2$_2$     &  36202.04   & 8.77$\pm$0.09         & 0.51$\pm$0.05   & 0.005$\pm$0.001        & 0.003$\pm$0.001 \\
           & 2$_2$-1$_1$     &  86093.98   & 8.55$\pm$0.06         & 0.86$\pm$0.04   & 0.04$\pm$0.01        & 0.045$\pm$0.007 \\
           & 3$_2$-2$_1$     &  99299.90   & 8.35$\pm$0.07         & 0.91$\pm$0.07   & 0.56$\pm$0.04        & 0.550$\pm$0.006 \\
           & 2$_3$-1$_2$     &  109252.21  & 8.39$\pm$0.02         & 1.08$\pm$0.08   & 0.05$\pm$0.01        & 0.057$\pm$0.005 \\
           & 4$_3$-3$_2$     &  138178.68  & 8.32$\pm$0.02         & 0.81$\pm$0.06   & 0.37$\pm$0.02        & 0.318$\pm$0.006 \\
           & 3$_4$-2$_3$     &  158971.85  & 8.65$\pm$0.01         & 0.88$\pm$0.05   & 0.07$\pm$0.01        & 0.066$\pm$0.012 \\
           & 4$_4$-3$_3$     &  172181.40  & 8.38$\pm$0.05         & 1.00$\pm$0.08   & 0.06$\pm$0.01        & 0.071$\pm$0.009 \\
\hline
$^{34}$SO  & 2$_3$-1$_2$     &  97715.40   & 8.24$\pm$0.04         & 1.20$\pm$0.09    & 0.02$\pm$0.01        & 0.023$\pm$0.006 \\
\hline
SO$_2$        & 3$_{1}$$_{,}$$_{3}$-2$_{0}$$_{,}$$_{2}$  & 104029.41    & 8.63$\pm$0.03         & 1.15$\pm$0.07   & 0.02$\pm$0.01        & 0.028$\pm$0.005 \\
\hline
H$_2$S        & 1$_{1}$$_{,}$$_{0}$-1$_{0}$$_{,}$$_{1}$  & 168762.76    & 8.10$\pm$0.04         & 1.00$\pm$0.06   & 0.17$\pm$0.04        & 0.184$\pm$0.009 \\
              &                                          &              & 8.30$\pm$0.03         & 2.50$\pm$0.10   & 0.11$\pm$0.03        & 0.290$\pm$0.008       \\
\hline
H$_2$$^{34}$S & 1$_{1}$$_{,}$$_{0}$-1$_{0}$$_{,}$$_{1}$  & 167910.51    & 8.20$\pm$0.02         & 1.00$\pm$0.08   & 0.05$\pm$0.01        & 0.049$\pm$0.004 \\

\hline
o-H$_2$CS   & 1$_{0}$$_{,}$$_{1}$-0$_{0}$$_{,}$$_{0}$   &  34351.45    & 8.84$\pm$0.01         & 0.76$\pm$0.05   & 0.09$\pm$0.01        & 0.069$\pm$0.001 \\
            &                                           &              & 7.62$\pm$0.08         & 0.86$\pm$0.05   & 0.05$\pm$0.01        & 0.0423$\pm$0.001 \\
            & 3$_{0}$$_{,}$$_{3}$-2$_{0}$$_{,}$$_{2}$   &  103040.54   & 8.31$\pm$0.01         & 0.85$\pm$0.06   & 0.17$\pm$0.04        & 0.151$\pm$0.003 \\
            & 4$_{0}$$_{,}$$_{4}$-3$_{0}$$_{,}$$_{3}$   &  137371.29   & 8.52$\pm$0.02         & 0.81$\pm$0.03   & 0.11$\pm$0.04        & 0.095$\pm$0.007 \\
            & 5$_{0}$$_{,}$$_{5}$-4$_{0}$$_{,}$$_{4}$   &  171687.90   & 8.01$\pm$0.01         & 0.92$\pm$0.07   & 0.09$\pm$0.01        & 0.089$\pm$0.006 \\
\hline
p-H$_2$CS   & 3$_{1}$$_{,}$$_{3}$-2$_{1}$$_{,}$$_{2}$  &  101477.88   & 8.31$\pm$0.02         & 0.97$\pm$0.08   & 0.18$\pm$0.05        & 0.181$\pm$0.008 \\
            & 3$_{1}$$_{,}$$_{2}$-2$_{1}$$_{,}$$_{1}$  &  104617.11   & 8.48$\pm$0.03         & 0.68$\pm$0.08   & 0.23$\pm$0.07        & 0.171$\pm$0.003 \\
            & 4$_{1}$$_{,}$$_{4}$-3$_{1}$$_{,}$$_{3}$  &  135298.36   & 8.50$\pm$0.06         & 0.75$\pm$0.05   & 0.14$\pm$0.04        & 0.115$\pm$0.007 \\
            & 4$_{1}$$_{,}$$_{3}$-3$_{1}$$_{,}$$_{2}$  &  139483.69   & 8.38$\pm$0.04         & 0.60$\pm$0.07   & 0.13$\pm$0.04        & 0.080$\pm$0.005 \\
            & 5$_{1}$$_{,}$$_{5}$-4$_{1}$$_{,}$$_{4}$  &  169114.16   & 8.41$\pm$0.03         & 0.76$\pm$0.04   & 0.09$\pm$0.01        & 0.075$\pm$0.004 \\
            & 5$_{1}$$_{,}$$_{4}$-4$_{1}$$_{,}$$_{3}$  &  174345.22   & 8.35$\pm$0.05         & 0.72$\pm$0.04   & 0.07$\pm$0.01        & 0.055$\pm$0.006 \\
\hline
HSCN        & 4$_{0}$$_{,}$$_{4}$-3$_{0}$$_{,}$$_{3}$  & 45877.81     & 8.05$\pm$0.07         & 0.55$\pm$0.04   & 0.014$\pm$0.001        & 0.008$\pm$0.002 \\
\hline
HNCS        & 3$_{0}$$_{,}$$_{3}$-2$_{0}$$_{,}$$_{2}$  & 35187.11     & 8.40$\pm$0.06         & 0.60$\pm$0.05   & 0.005$\pm$0.001        & 0.003$\pm$0.001 \\
\hline
NS          & $J$=5/2-3/2, $\Omega$=1/2, $F$=5/2-3/2, $l$=$e$  &  115156.81   & 8.40$\pm$0.03         & 1.50$\pm$0.08   & 0.06$\pm$0.01        & 0.10$\pm$0.02 \\
            & $J$=5/2-3/2, $\Omega$=1/2, $F$=3/2-3/2, $l$=$f$  &  115524.60   & 8.75$\pm$0.05         & 0.95$\pm$0.06   & 0.08$\pm$0.01        & 0.08$\pm$0.02 \\
            & $J$=5/2-3/2, $\Omega$=1/2, $F$=7/2-5/2, $l$=$f$  &  115556.25   & 7.95$\pm$0.06         & 0.90$\pm$0.06   & 0.06$\pm$0.01        & 0.05$\pm$0.02 \\
            & $J$=7/2-5/2, $\Omega$=1/2, $F$=9/2-7/2, $l$=$e$  &  161297.24   & 8.26$\pm$0.01         & 0.73$\pm$0.05   & 0.03$\pm$0.01        & 0.020$\pm$0.004 \\
            & $J$=7/2-5/2, $\Omega$=1/2, $F$=7/2-5/2, $l$=$e$  &  161298.41   & 8.62$\pm$0.01         & 0.70$\pm$0.06   & 0.02$\pm$0.01        & 0.016$\pm$0.004 \\
            & $J$=7/2-5/2, $\Omega$=1/2, $F$=9/2-7/2, $l$=$f$  &  161697.25   & 7.95$\pm$0.03         & 0.65$\pm$0.07   & 0.02$\pm$0.01        & 0.014$\pm$0.004 \\
\label{table:line_parameters} 
\end{longtable}

\pagebreak

\begin{figure*}
\includegraphics[scale=0.65, angle=0]{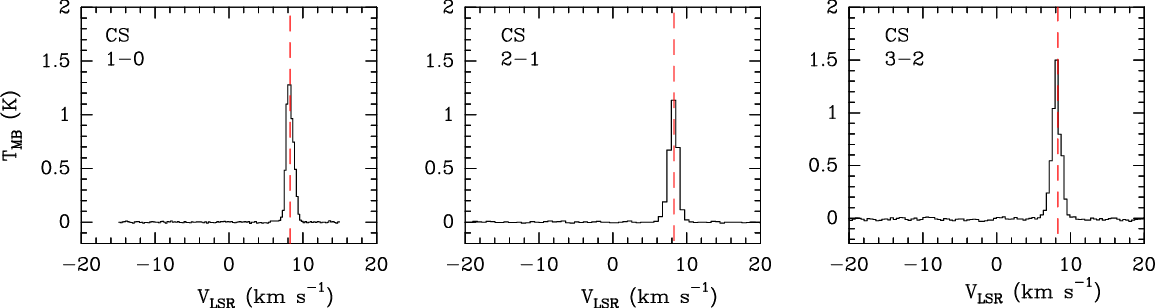}  \hspace{0.0cm}
\caption{Observed lines of CS in B\,335 (black) with rms$>$3$\sigma$. The red dashed line indicates the systemic velocity of the source v$_{\mathrm{LSR}}$=8.3 km s$^{-1}$.}
\label{figure:CS_lines}
\end{figure*}

\begin{figure*}
\includegraphics[scale=0.65, angle=0]{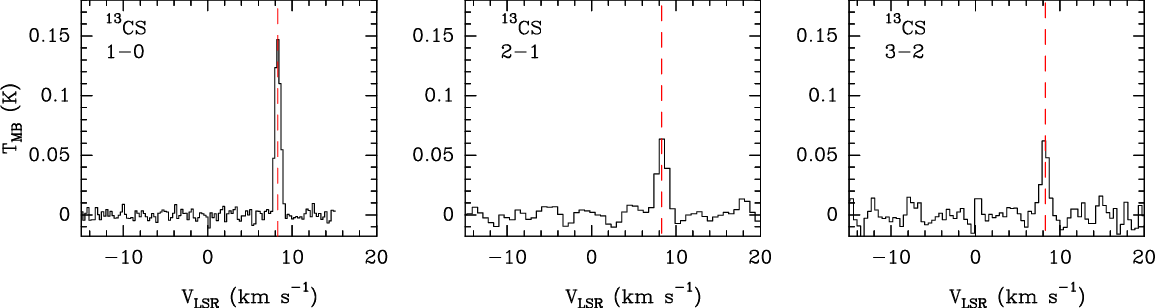}  \hspace{0.0cm}
\caption{Observed lines of $^{13}$CS in B\,335 (black) with rms$>$3$\sigma$. The red dashed line indicates the systemic velocity of the source v$_{\mathrm{LSR}}$=8.3 km s$^{-1}$.}
\label{figure:13CS_lines}
\end{figure*}

\begin{figure*}
\includegraphics[scale=0.65, angle=0]{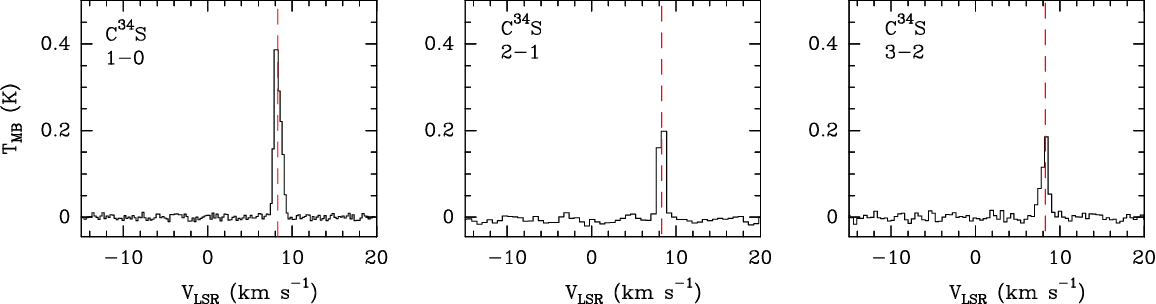}  \hspace{0.0cm}
\caption{Observed lines of C$^{34}$S in B\,335 (black) with rms$>$3$\sigma$. The red dashed line indicates the systemic velocity of the source v$_{\mathrm{LSR}}$=8.3 km s$^{-1}$.}
\label{figure:C34S_lines}
\end{figure*}

\begin{figure*}
\includegraphics[scale=0.65, angle=0]{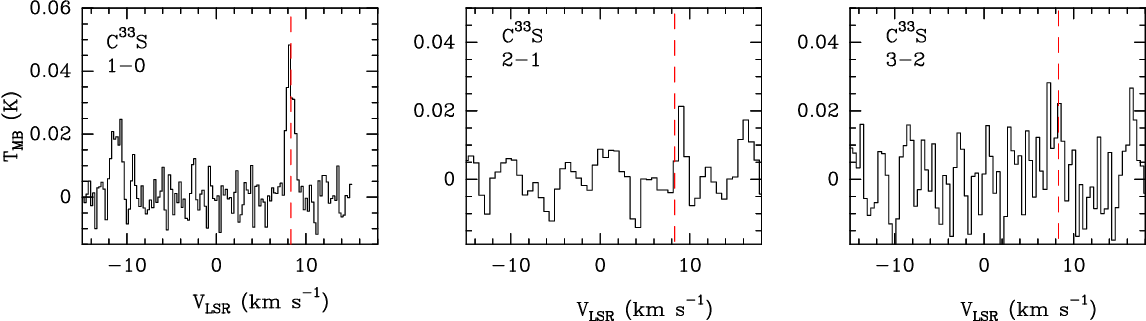}  \hspace{0.0cm}
\caption{Observed lines of C$^{33}$S in B\,335 (black) with rms$>$3$\sigma$. The red dashed line indicates the systemic velocity of the source v$_{\mathrm{LSR}}$=8.3 km s$^{-1}$.}
\label{figure:C33S_lines}
\end{figure*}

\begin{figure*}
\includegraphics[scale=0.65, angle=0]{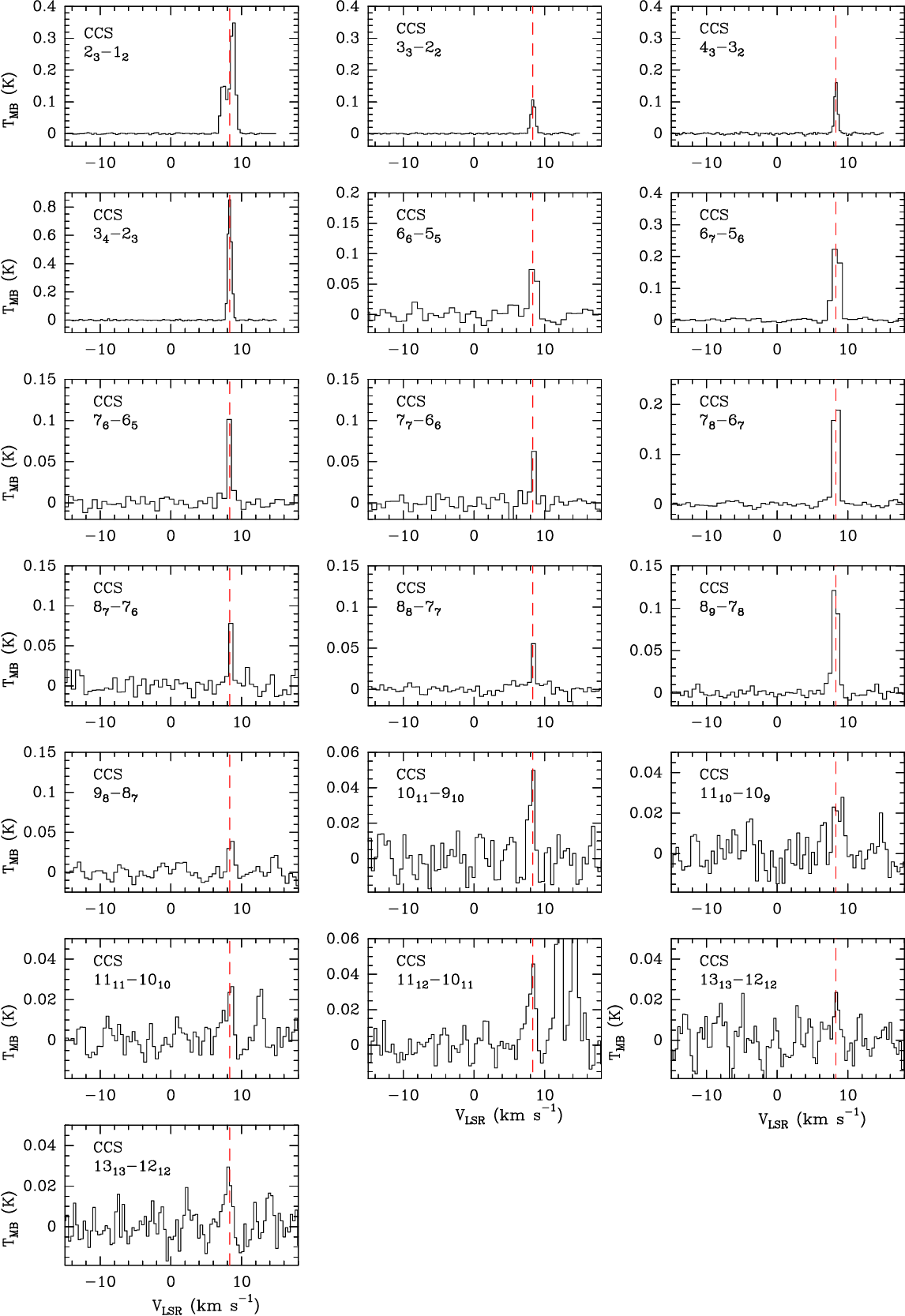}  \hspace{0.0cm}
\caption{Observed lines of CCS in B\,335 with rms$>$3$\sigma$. The red dashed line indicates the systemic velocity of the source v$_{\mathrm{LSR}}$=8.3 km s$^{-1}$.}
\label{figure:CCS_lines}
\end{figure*}

\begin{figure*}
\includegraphics[scale=0.65, angle=0]{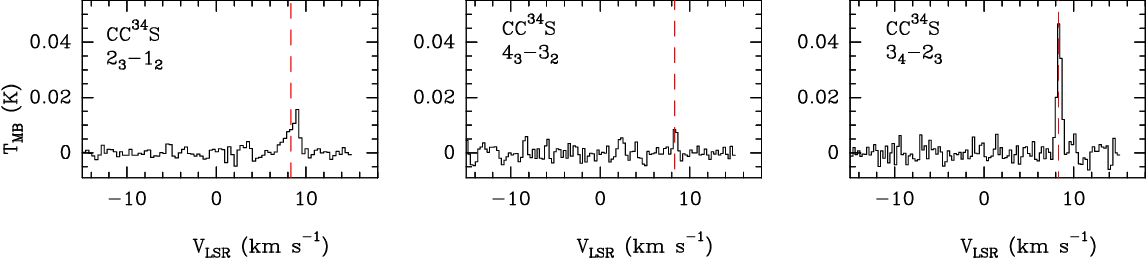}  \hspace{0.0cm}
\caption{Observed lines of CC$^{34}$S in B\,335 with rms$>$3$\sigma$. The red dashed line indicates the systemic velocity of the source v$_{\mathrm{LSR}}$=8.3 km s$^{-1}$.}
\label{figure:CC34S_lines}
\end{figure*}

\begin{figure*}
\includegraphics[scale=0.65, angle=0]{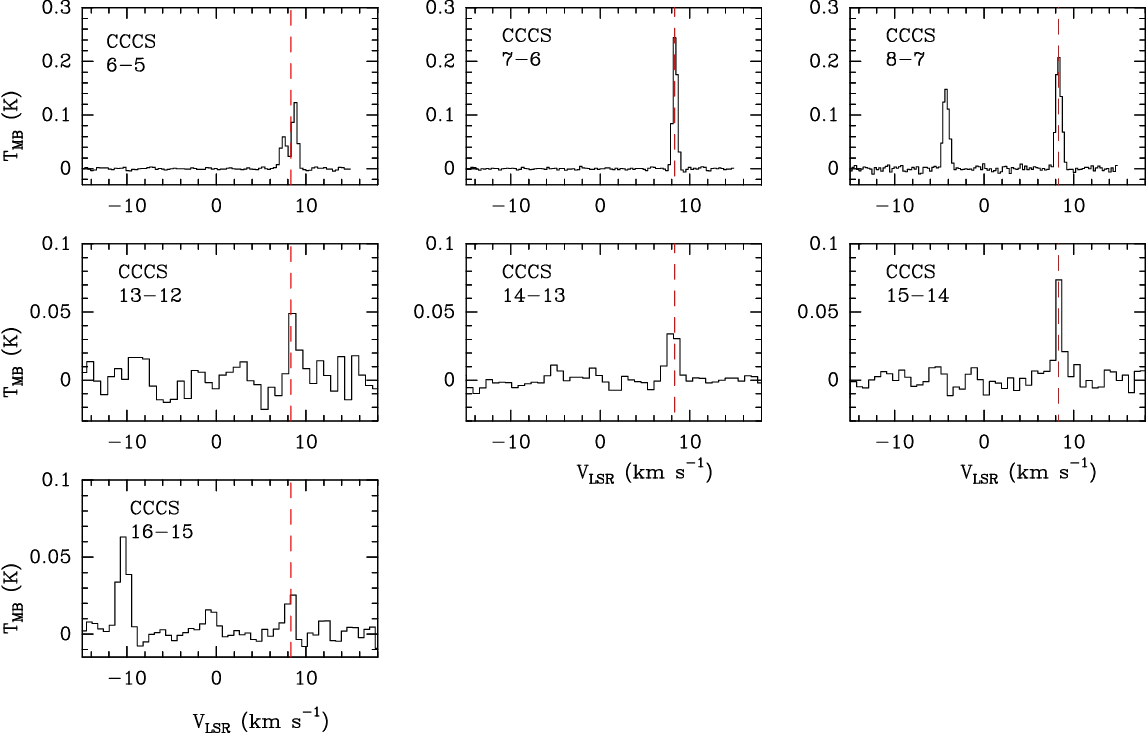}  \hspace{0.0cm}
\caption{Observed lines of CCCS in B\,335 with rms$>$3$\sigma$. The red dashed line indicates the systemic velocity of the source v$_{\mathrm{LSR}}$=8.3 km s$^{-1}$.}
\label{figure:CCCS_lines}
\end{figure*}

\begin{figure*}
\includegraphics[scale=0.65, angle=0]{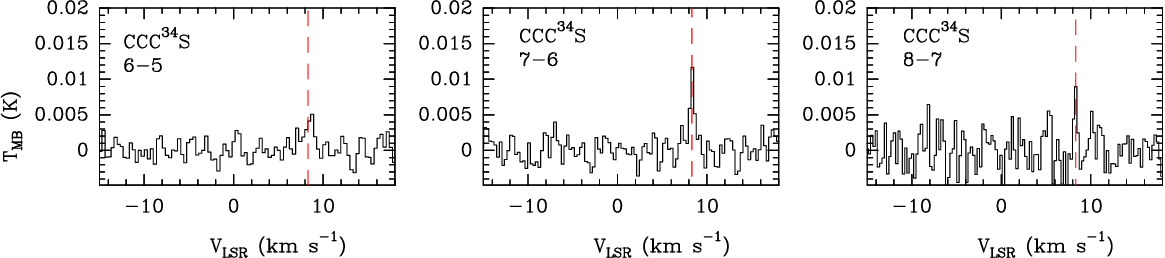}  \hspace{0.0cm}
\caption{Observed lines of CCC$^{34}$S in B\,335 with rms$>$3$\sigma$. The red dashed line indicates the systemic velocity of the source v$_{\mathrm{LSR}}$=8.3 km s$^{-1}$.}
\label{figure:CCC34S_lines}
\end{figure*}

\begin{figure*}
\includegraphics[scale=0.65, angle=0]{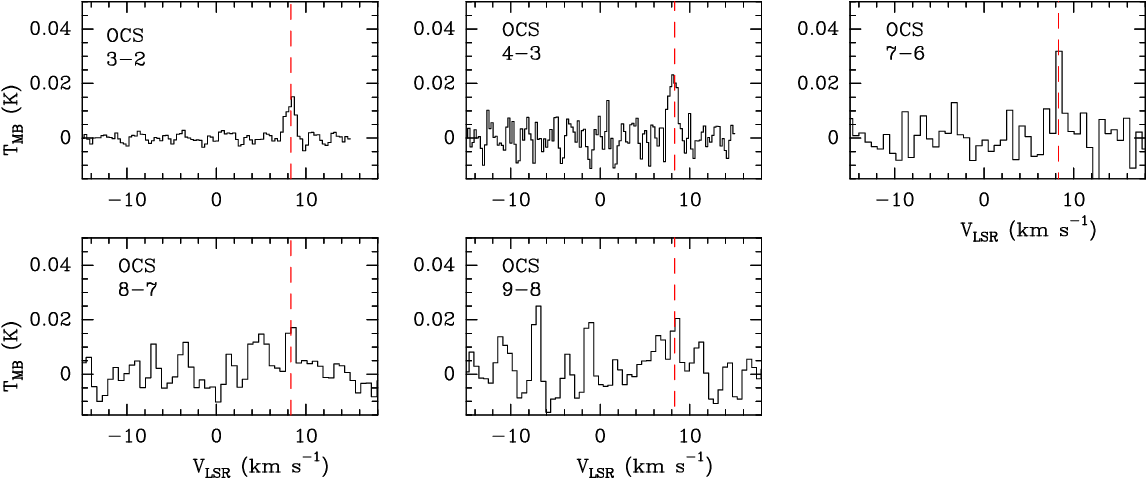}  \hspace{0.0cm}
\caption{Observed lines of OCS in B\,335 with rms$>$3$\sigma$. The red dashed line indicates the systemic velocity of the source v$_{\mathrm{LSR}}$=8.3 km s$^{-1}$.}
\label{figure:OCS_lines}
\end{figure*}

\begin{figure*}
\includegraphics[scale=0.65, angle=0]{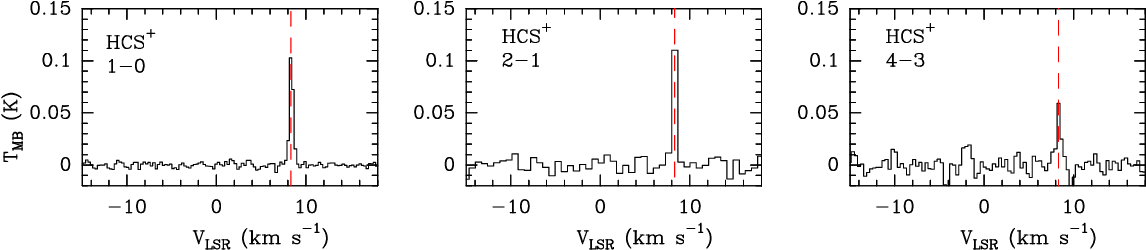}  \hspace{0.0cm}
\caption{Observed lines of HCS$^+$ in B\,335 with rms$>$3$\sigma$. The red dashed line indicates the systemic velocity of the source v$_{\mathrm{LSR}}$=8.3 km s$^{-1}$.}
\label{figure:HCS+_lines}
\end{figure*}

\begin{figure*}
\includegraphics[scale=0.65, angle=0]{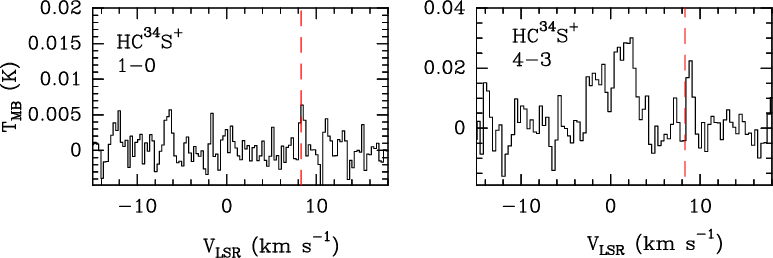}  \hspace{0.0cm}
\caption{Observed lines of HC$^{34}$S$^+$ in B\,335 with rms$>$3$\sigma$. The red dashed line indicates the systemic velocity of the source v$_{\mathrm{LSR}}$=8.3 km s$^{-1}$.}
\label{figure:HC34S+_lines}
\end{figure*}

\begin{figure*}
\includegraphics[scale=0.65, angle=0]{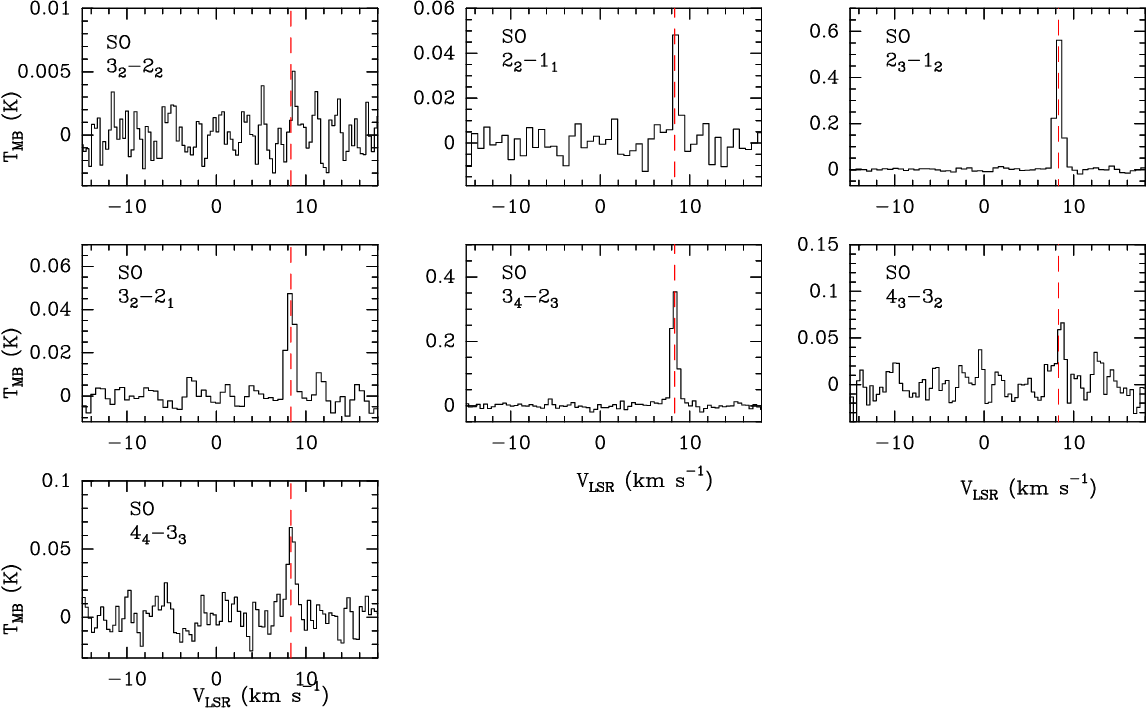}  \hspace{0.0cm}
\caption{Observed lines of SO in B\,335 with rms$>$3$\sigma$. The red dashed line indicates the systemic velocity of the source v$_{\mathrm{LSR}}$=8.3 km s$^{-1}$.}
\label{figure:SO_lines}
\end{figure*}

\begin{figure*}
\includegraphics[scale=0.65, angle=0]{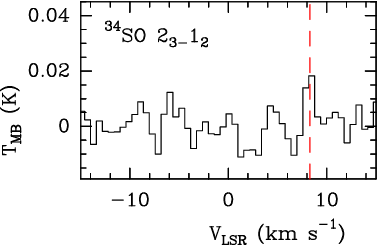}  \hspace{0.0cm}
\caption{Observed lines of $^{34}$SO in B\,335 with rms$>$3$\sigma$. The red dashed line indicates the systemic velocity of the source v$_{\mathrm{LSR}}$=8.3 km s$^{-1}$.}
\label{figure:34SO_lines}
\end{figure*}

\begin{figure*}
\includegraphics[scale=0.65, angle=0]{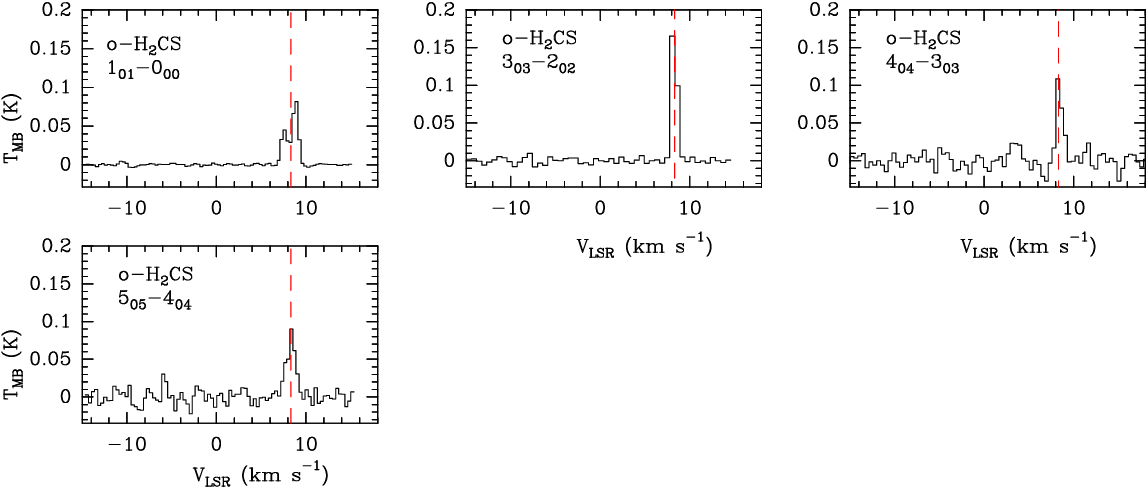}  \hspace{0.0cm}
\caption{Observed lines of o-H$_2$CS in B\,335 with rms$>$3$\sigma$. The red dashed line indicates the systemic velocity of the source v$_{\mathrm{LSR}}$=8.3 km s$^{-1}$.}
\label{figure:o-H2CS_lines}
\end{figure*}

\begin{figure*}
\includegraphics[scale=0.65, angle=0]{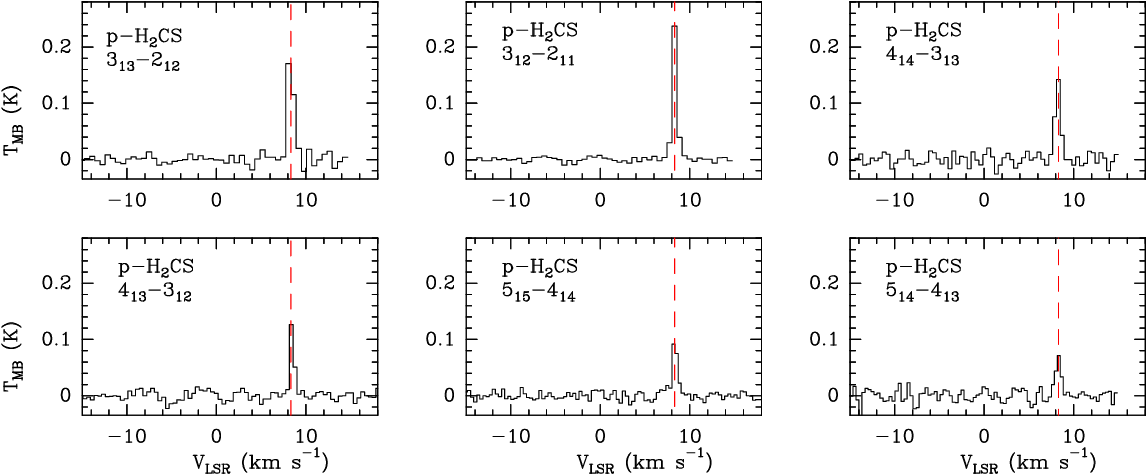}  \hspace{0.0cm}
\caption{Observed lines of p-H$_2$CS in B\,335 with rms$>$3$\sigma$. The red dashed line indicates the systemic velocity of the source v$_{\mathrm{LSR}}$=8.3 km s$^{-1}$.}
\label{figure:p-H2CS_lines}
\end{figure*}

\begin{figure*}
\includegraphics[scale=0.65, angle=0]{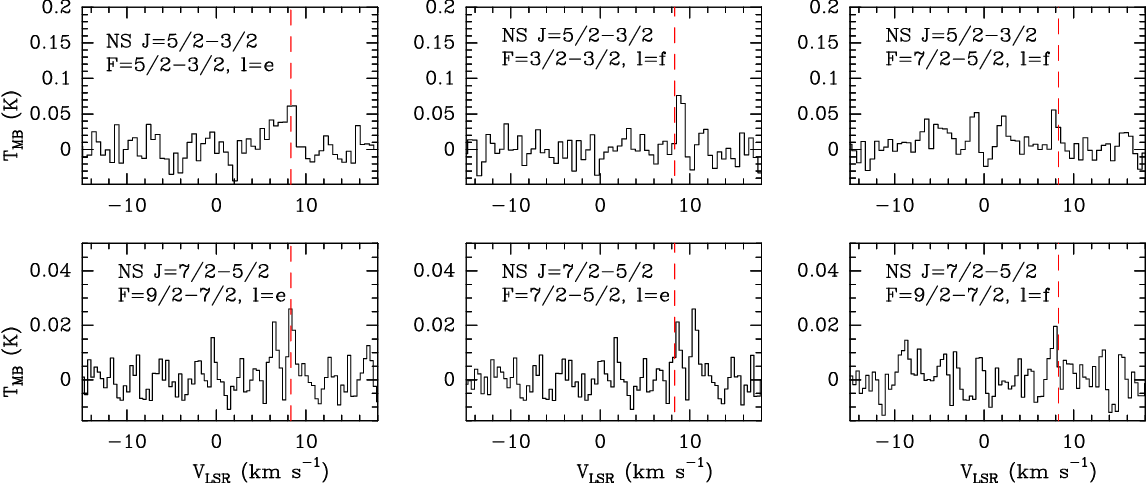}  \hspace{0.0cm}
\caption{Observed lines of NS in B\,335 with rms$>$3$\sigma$. The red dashed line indicates the systemic velocity of the source v$_{\mathrm{LSR}}$=8.3 km s$^{-1}$.}
\label{figure:NS_lines}
\end{figure*}

\begin{figure}
\includegraphics[scale=0.65, angle=0]{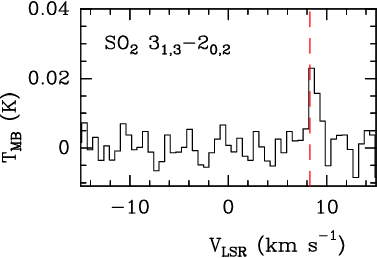}  \hspace{0.0cm}
\caption{Observed line of SO$_2$ in B\,335 with rms$>$3$\sigma$. The red dashed line indicates the systemic velocity of the source v$_{\mathrm{LSR}}$=8.3 km s$^{-1}$.}
\label{figure:SO2_lines}
\end{figure}

\begin{figure}
\includegraphics[scale=0.65, angle=0]{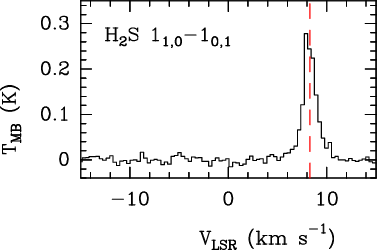}  \hspace{0.0cm}
\caption{Observed line of H$_2$S in B\,335 with rms$>$3$\sigma$. The red dashed line indicates the systemic velocity of the source v$_{\mathrm{LSR}}$=8.3 km s$^{-1}$.}
\label{figure:H2S_lines}
\end{figure}

\begin{figure}
\includegraphics[scale=0.65, angle=0]{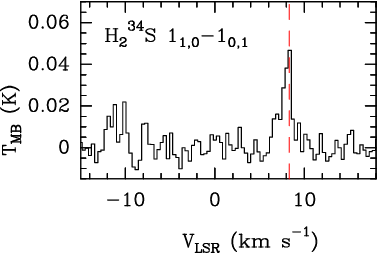}  \hspace{0.0cm}
\caption{Observed line of H$_2$$^{34}$S in B\,335 with rms$>$3$\sigma$. The red dashed line indicates the systemic velocity of the source v$_{\mathrm{LSR}}$=8.3 km s$^{-1}$.}
\label{figure:H234S_lines}
\end{figure}

\begin{figure}
\includegraphics[scale=0.65, angle=0]{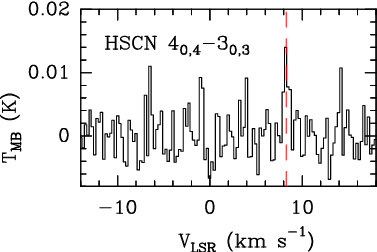}  \hspace{0.0cm}
\caption{Observed line of HSCN in B\,335 with rms$>$3$\sigma$. The red dashed line indicates the systemic velocity of the source v$_{\mathrm{LSR}}$=8.3 km s$^{-1}$.}
\label{figure:HSCN_lines}
\end{figure}

\begin{figure}
\includegraphics[scale=0.65, angle=0]{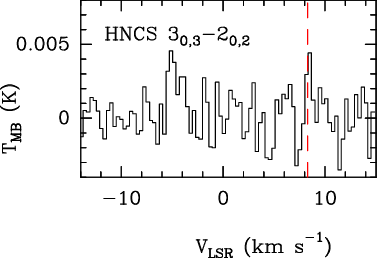}  \hspace{0.0cm}
\caption{Observed line of HNCS in B\,335 with rms$>$3$\sigma$. The red dashed line indicates the systemic velocity of the source v$_{\mathrm{LSR}}$=8.3 km s$^{-1}$.}
\label{figure:HSCN_lines}
\end{figure}
\pagebreak

\begin{figure*}
\centering
\hspace{-0.5cm}
\includegraphics[scale=0.26, angle=0]{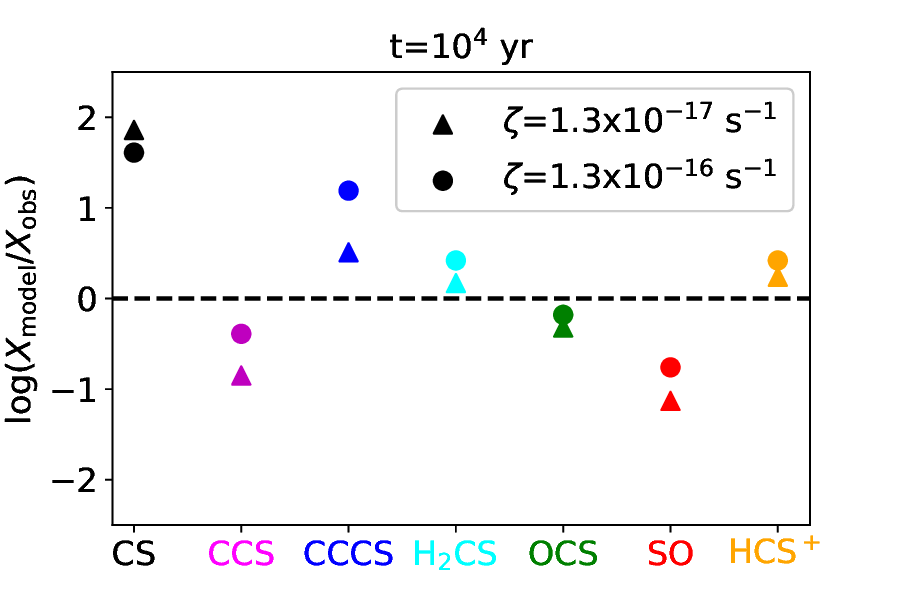}  
\hspace{-0.5cm}
\includegraphics[scale=0.26, angle=0]{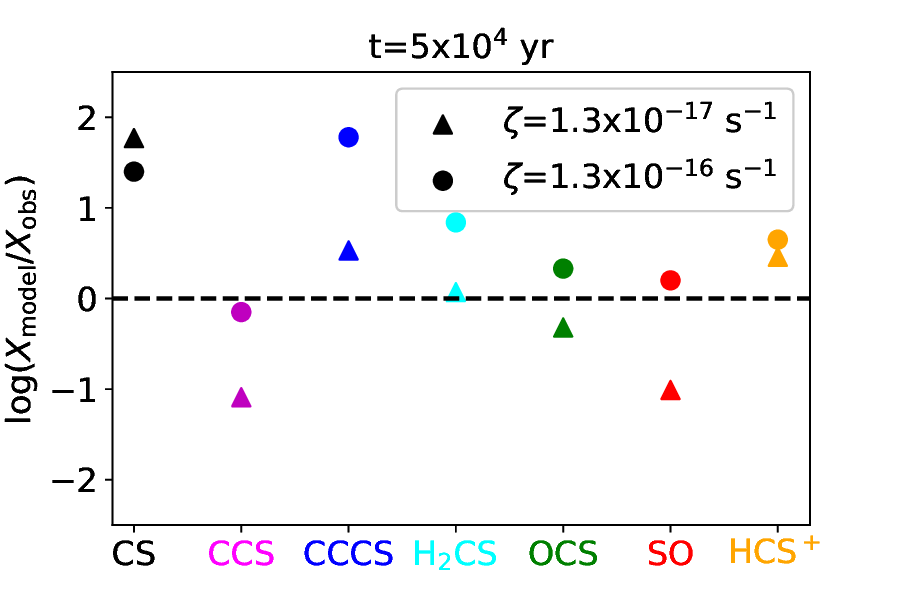} 
\hspace{-0.5cm}
\includegraphics[scale=0.26, angle=0]{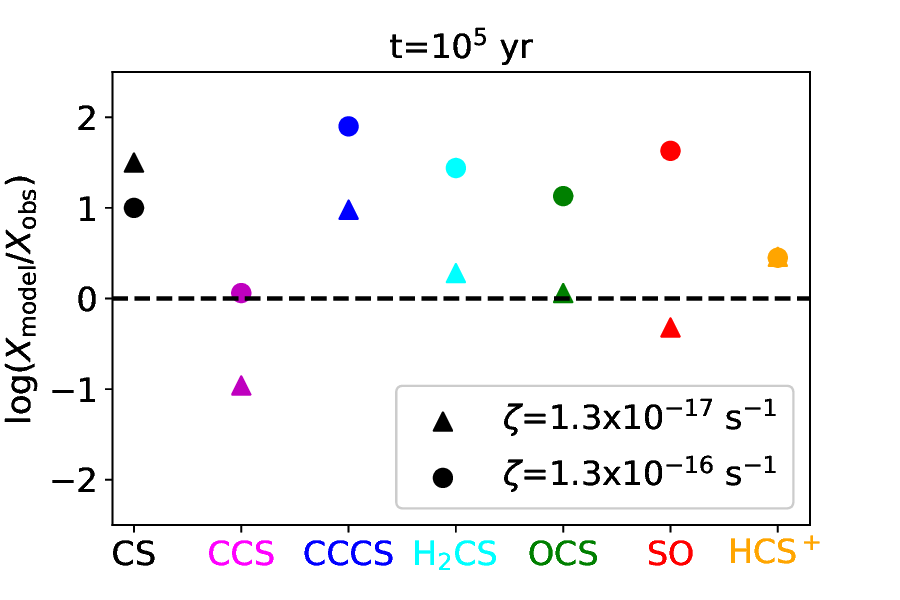} 
\hspace{-0.5cm}
\includegraphics[scale=0.26, angle=0]{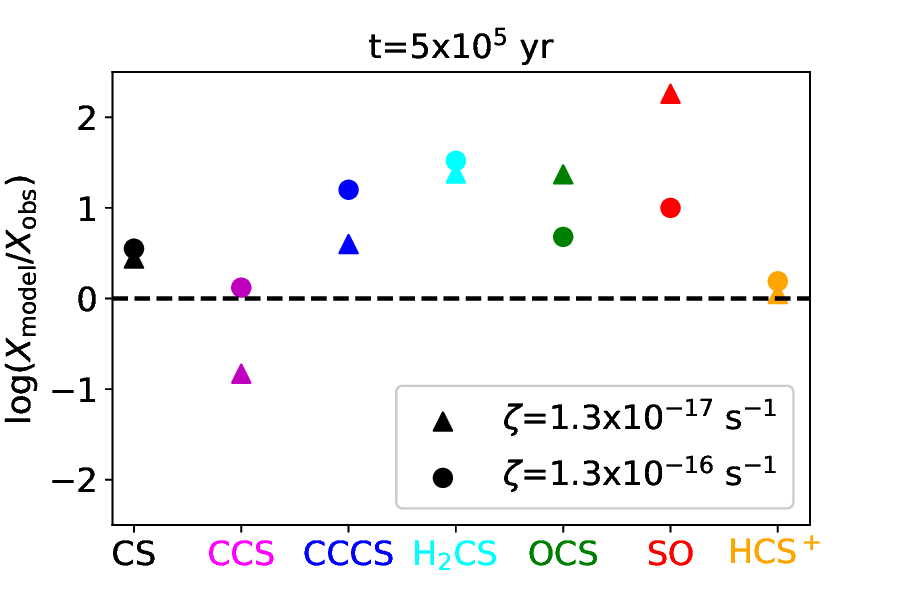} 
\hspace{-0.5cm}
\includegraphics[scale=0.26, angle=0]{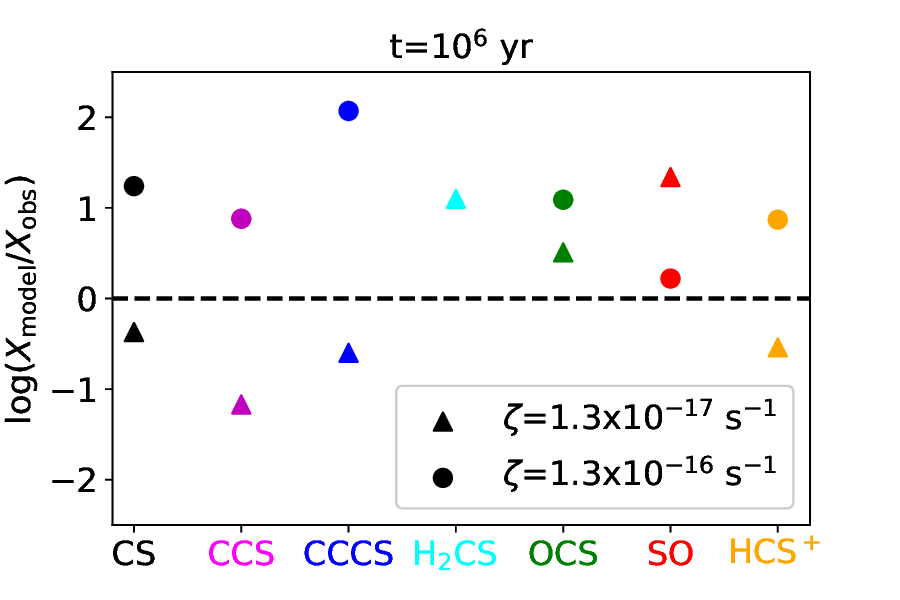} 
\hspace{-0.5cm}
\\
\caption{Ratios between the abundances obtained from the models, $X$$_{\mathrm{model}}$,  and the observations, $X$$_{\mathrm{obs}}$, for different sulphur species at specific evolution times ($t$=10$^4$, 5$\times$10$^4$, 10$^5$, 5$\times$10$^5$, and 10$^6$ yr), when varying the cosmic-ray rate ($\zeta$).}
\label{figure:model_vs_obs_CR}
\end{figure*}

\begin{figure*}
\centering
\hspace{-0.5cm}
\includegraphics[scale=0.26, angle=0]{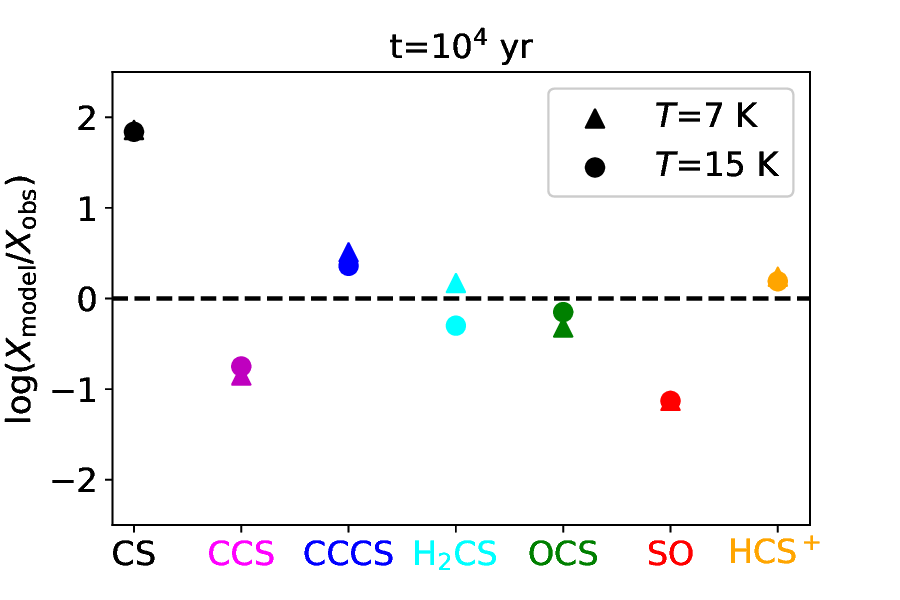}  
\hspace{-0.5cm}
\includegraphics[scale=0.26, angle=0]{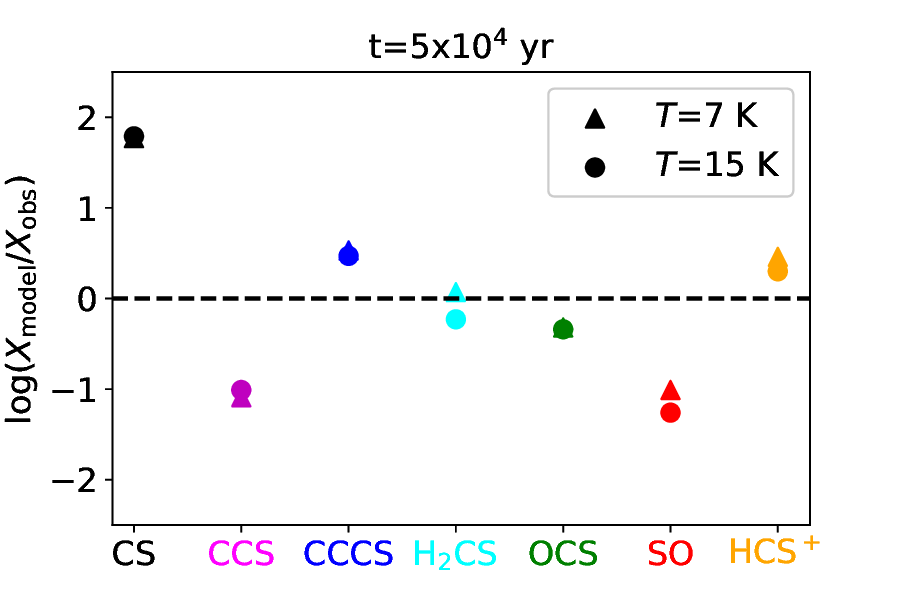} 
\hspace{-0.5cm}
\includegraphics[scale=0.26, angle=0]{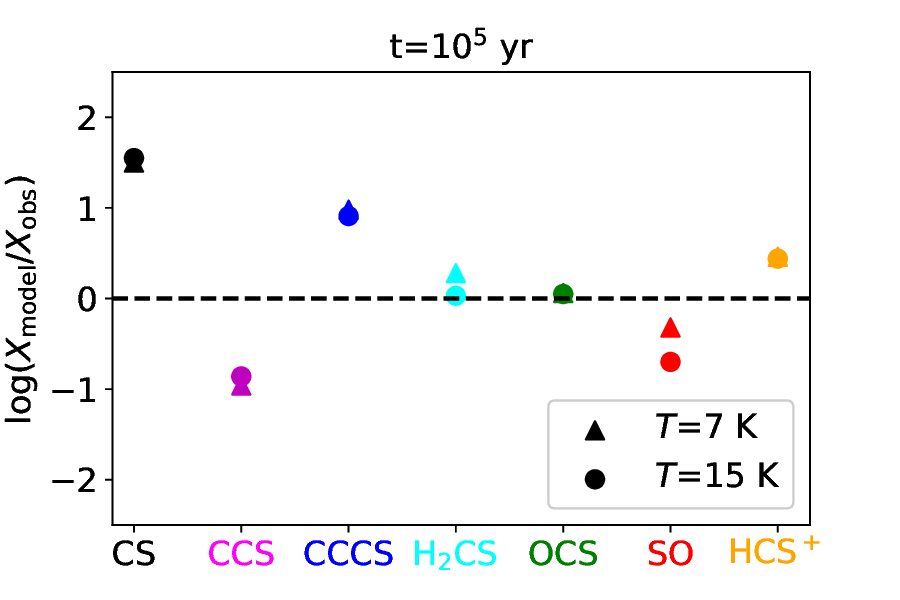} 
\hspace{-0.5cm}
\includegraphics[scale=0.26, angle=0]{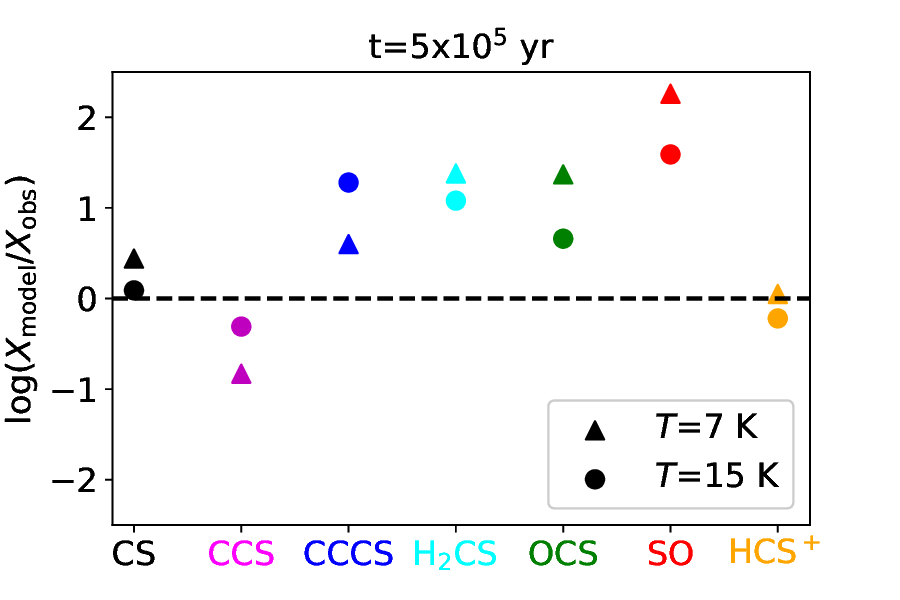} 
\hspace{-0.5cm}
\includegraphics[scale=0.26, angle=0]{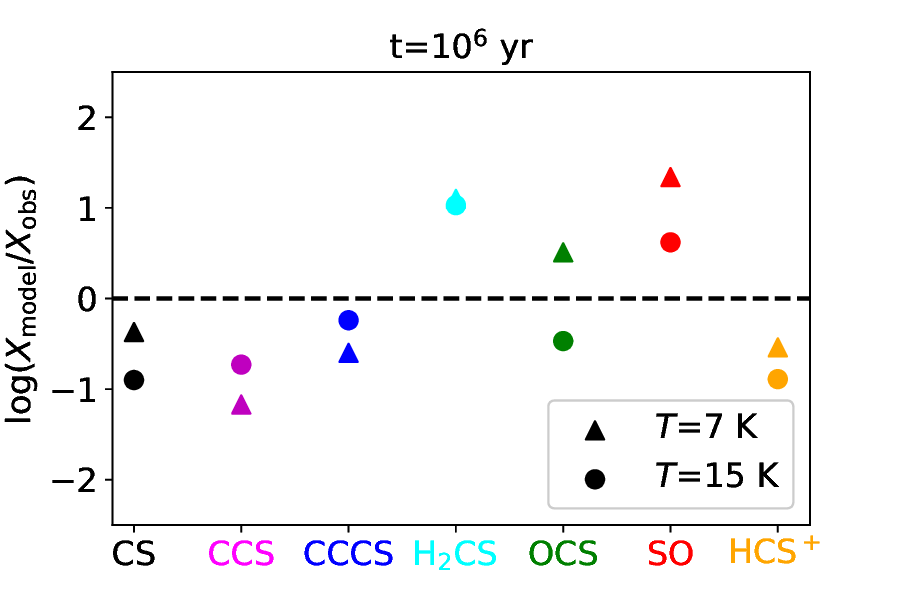} 
\hspace{-0.5cm}
\\
\caption{Ratios between the abundances obtained from the models, $X$$_{\mathrm{model}}$,  and the observations, $X$$_{\mathrm{obs}}$, for different sulphur species at specific evolution times ($t$=10$^4$, 5$\times$10$^4$, 10$^5$, 5$\times$10$^5$, and 10$^6$ yr), when varying the gas temperature.}
\label{figure:model_vs_obs_Tg}
\end{figure*}

\begin{figure*}
\centering
\hspace{-0.5cm}
\includegraphics[scale=0.26, angle=0]{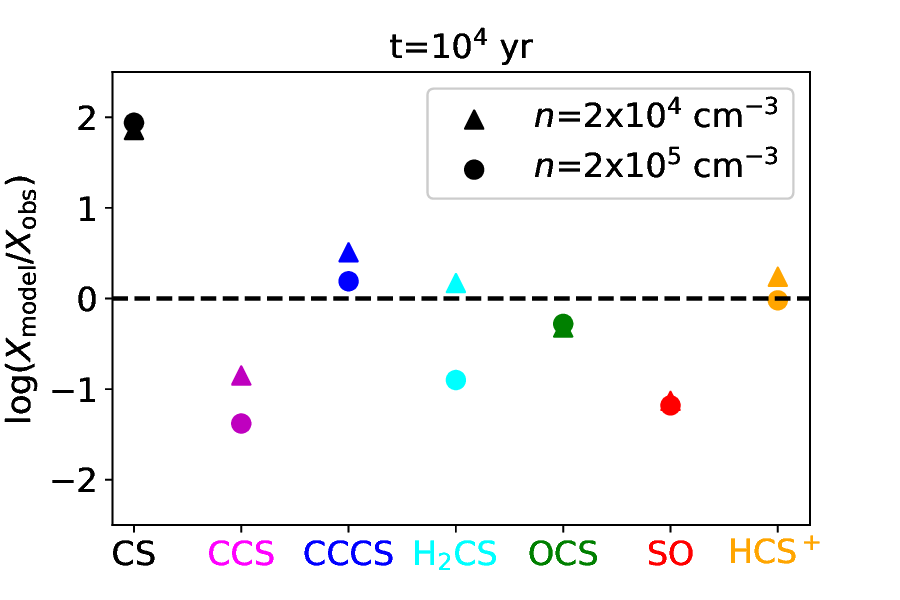}  
\hspace{-0.5cm}
\includegraphics[scale=0.26, angle=0]{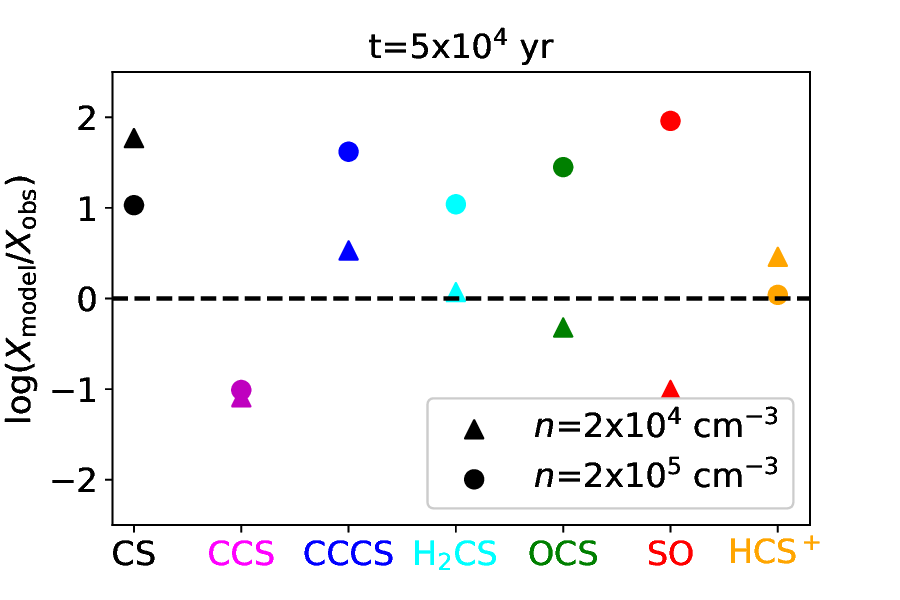} 
\hspace{-0.5cm}
\includegraphics[scale=0.26, angle=0]{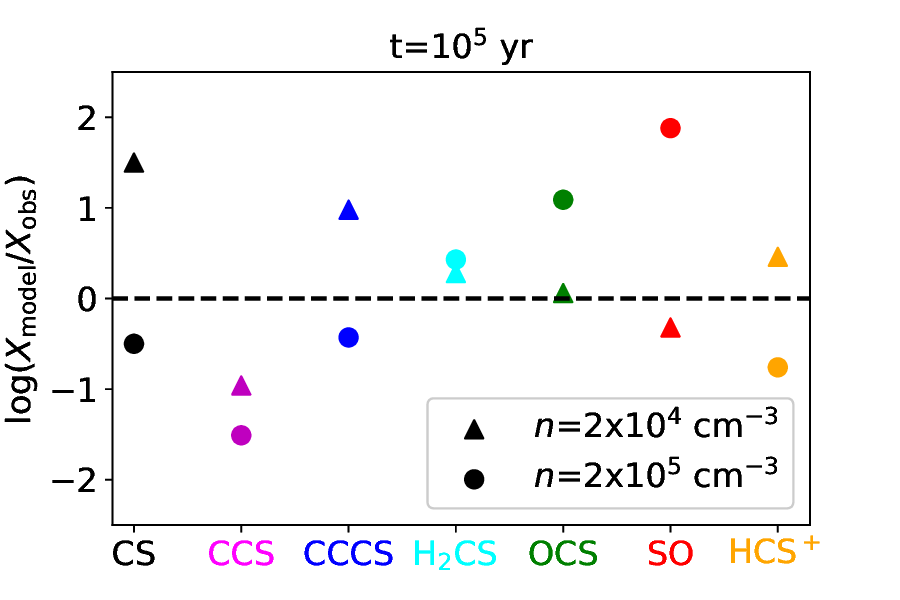} 
\hspace{-0.5cm}
\includegraphics[scale=0.26, angle=0]{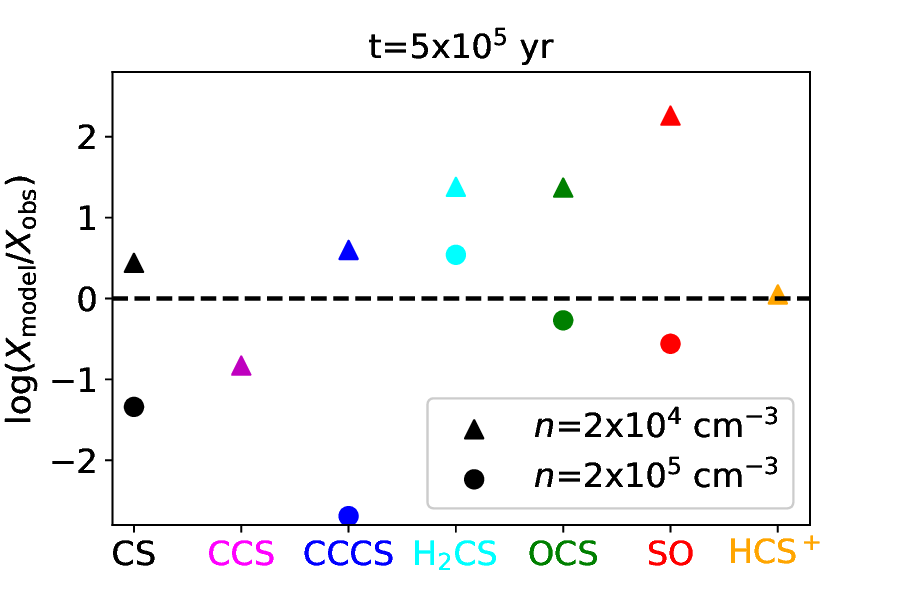} 
\hspace{-0.5cm}
\includegraphics[scale=0.26, angle=0]{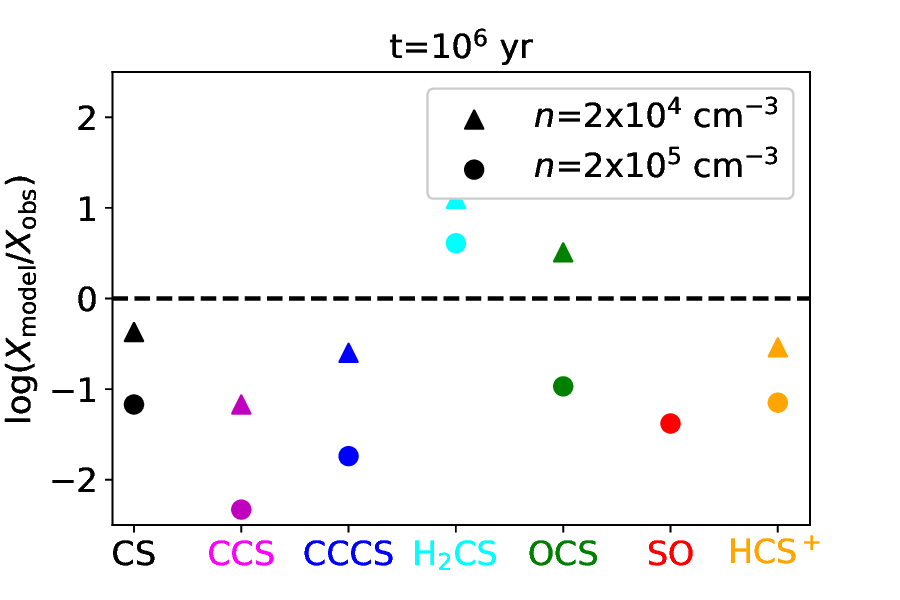} 
\hspace{-0.5cm}
\\
\caption{Ratios between the abundances obtained from the models, $X$$_{\mathrm{model}}$,  and the observations, $X$$_{\mathrm{obs}}$, for different sulphur species at specific evolution times ($t$=10$^4$, 5$\times$10$^4$, 10$^5$, 5$\times$10$^5$, and 10$^6$ yr), when varying the hydrogen number density ($n$$_{\mathrm{H}}$).}
\label{figure:model_vs_obs_n}
\end{figure*}

\begin{figure*}
\centering
\hspace{-0.5cm}
\includegraphics[scale=0.26, angle=0]{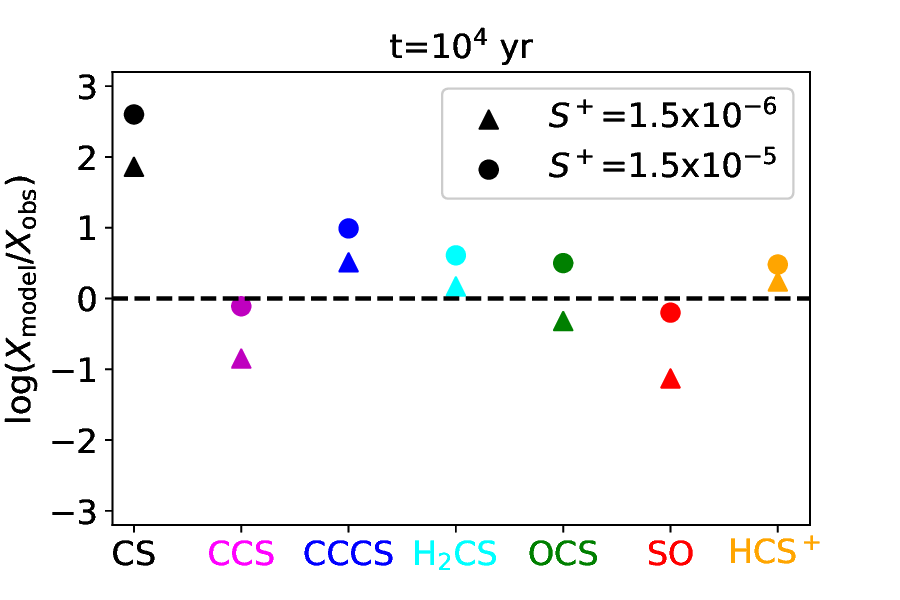}  
\hspace{-0.5cm}
\includegraphics[scale=0.26, angle=0]{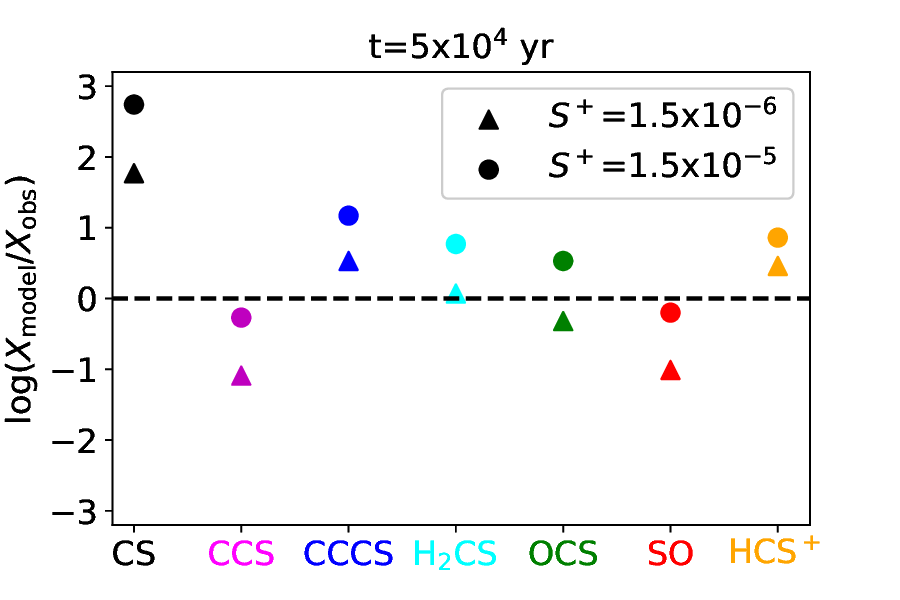} 
\hspace{-0.5cm}
\includegraphics[scale=0.26, angle=0]{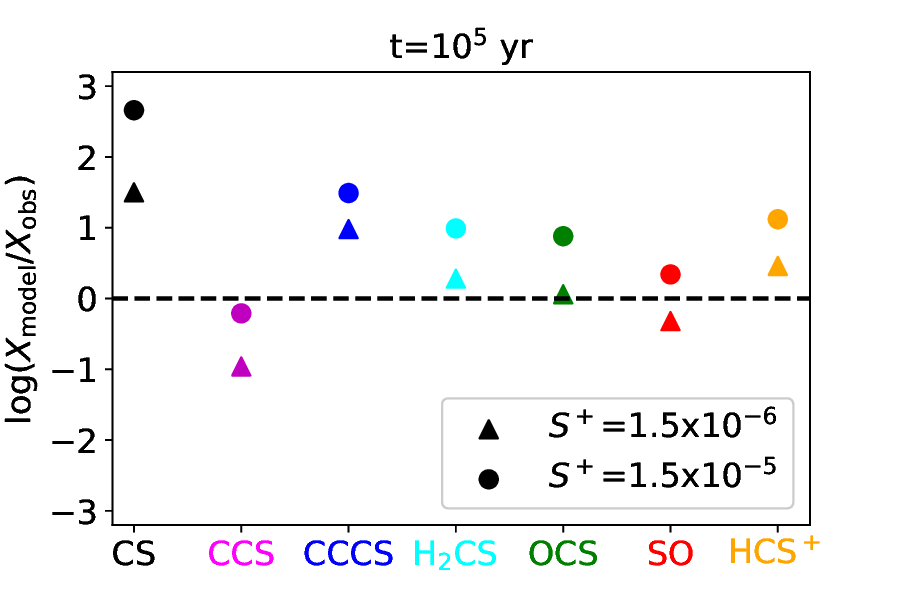} 
\hspace{-0.5cm}
\includegraphics[scale=0.26, angle=0]{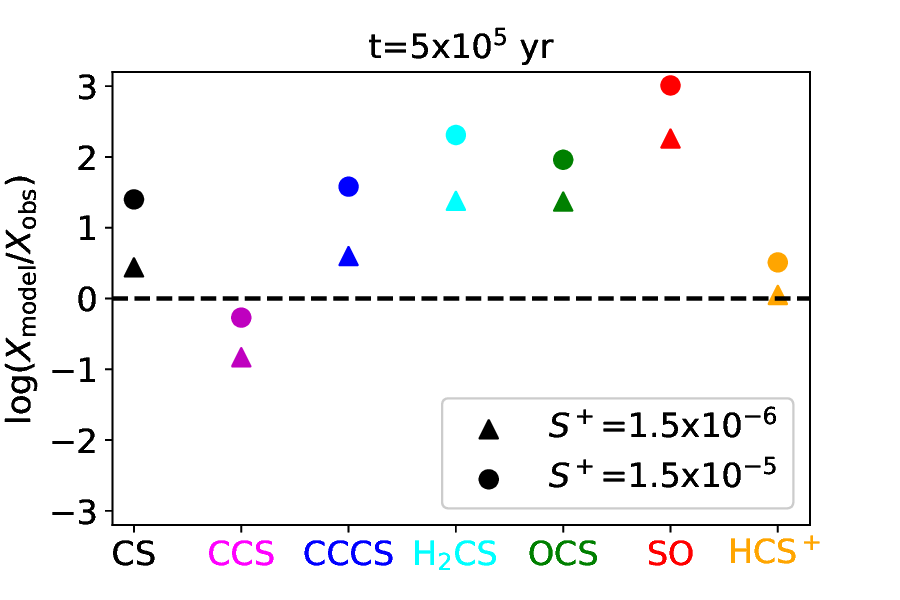} 
\hspace{-0.5cm}
\includegraphics[scale=0.26, angle=0]{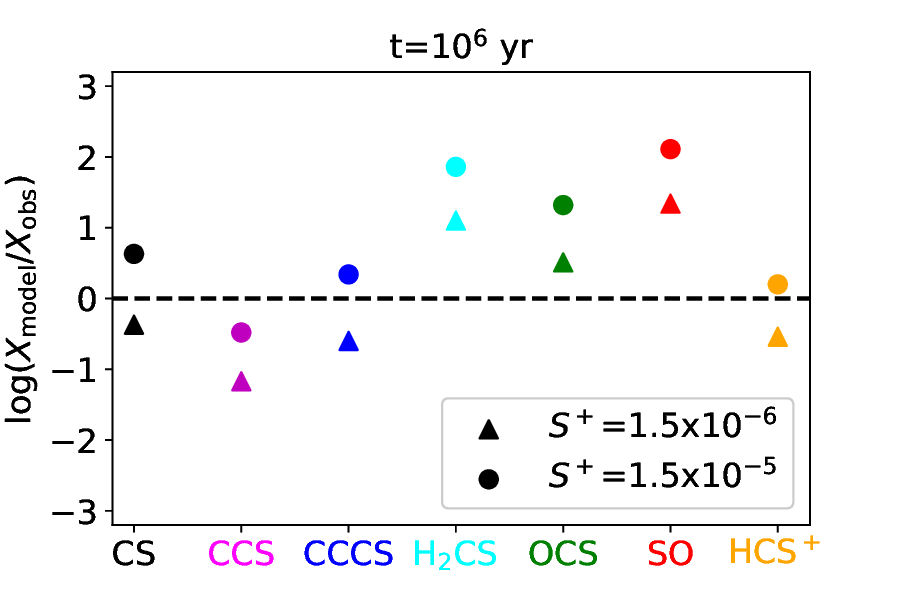} 
\hspace{-0.5cm}
\\
\caption{Ratios between the abundances obtained from the models, $X$$_{\mathrm{model}}$,  and the observations, $X$$_{\mathrm{obs}}$, for different sulphur species at specific evolution times ($t$=10$^4$, 5$\times$10$^4$, 10$^5$, 5$\times$10$^5$, and 10$^6$ yr), when varying the initial sulphur abundance ($S$$^{+}_{\mathrm{init}}$).}
\label{figure:model_vs_obs_S+}
\end{figure*}

\pagebreak
\end{appendix}

\end{document}